\RequirePackage{lineno}
\documentclass[a4paper,11pt,aps,final, reprint,showkeys,showpacs,superscriptaddress,floatfix,nofootinbib,amsmath,amssymb]{article}
\pdfoutput=1 
\usepackage{jcappub} 

\usepackage[T1]{fontenc} 

\usepackage[utf8]{inputenc}

\usepackage{hyperref} 
\usepackage{url}
\usepackage{graphicx}
\usepackage{esvect}
\usepackage{graphics}
\usepackage{natbib}
\usepackage{mathrsfs}
\usepackage{blindtext}

\usepackage[table]{xcolor}
\usepackage{multirow}

\usepackage[normalem]{ulem}
\usepackage{color}

\usepackage{soul}

\usepackage{supertabular}
\usepackage{tabularx}

\def\units#1{~\hbox{$\,{\rm #1}$}}

\title{\fontsize{22}{20} Implications of current nuclear cross sections on secondary cosmic rays with the upcoming {\tt DRAGON2} code}

\author[a,b]{P.~De~La~Torre~Luque}
\emailAdd{pedro.delatorreluque@ba.infn.it}
\author[a]{M.~N.~Mazziotta}
\emailAdd{mazziotta@ba.infn.it}
\author[a,b]{F.~Loparco}
\author[a]{F.~Gargano}
\author[a,b]{D.~Serini}

\affiliation[a]{Istituto Nazionale di Fisica Nucleare, Sezione di Bari, via Orabona 4, I-70126 Bari, Italy}
\affiliation[b]{Dipartimento di Fisica ``M. Merlin" dell'Universit\`a e del Politecnico di Bari, via Amendola 173, I-70126 Bari, Italy}

\date{\today}

\abstract{Current measurements of cosmic-ray fluxes have reached unprecedented accuracy thanks to the new generation of experiments, and in particular the AMS-02 mission. At the same time, significant progress has been made in the propagation models of galactic cosmic rays. These models include several propagation parameters, which are usually inferred from the ratios of secondary to primary cosmic rays, and which depend on the cross sections describing the collisions among the various species of cosmic-ray nuclei. At present, our knowledge of these cross sections in the energy range where cosmic-ray interactions occur is limited, and this is a source of uncertainties in the predicted fluxes of secondary cosmic-ray nuclei.
In this work we study the impact of the cross section uncertainties on the fluxes of light secondary nuclei (Li, Be, B) using a preliminary version of the upcoming DRAGON2 code.
We first present a detailed comparison of the secondary fluxes computed by implementing different parameterizations for the network of spallation cross sections. Then, we discuss the use of secondary-over-secondary cosmic-ray flux ratios as a tool to improve the consistency of cross sections parameterizations and give insight of the overall uncertainties coming from the cross sections parametrisations. We show that the uncertainties inferred from the cross section data are enough to explain the discrepancies in the Be and Li fluxes with respect to the AMS-02 data, with no need of a primary component in their spectra. In addition, we show that the fluxes of B, Be and Li can be simultaneously reproduced by rescaling their cross sections within the experimental uncertainty.
Finally, we also revisit the diffusive estimation of the halo size, obtaining good agreement with previous works and a best fit value of $6.8 \pm 1 \units{kpc}$ from the most updated cross sections parametrisations.
}

\keywords{Cosmic rays, diffusion, propagation, spallation, cross sections, magnetic halo}

\begin{document}
\maketitle
\flushbottom

\section{Introduction}
\label{sec:intro}

Propagation of Galactic cosmic rays (CRs) is governed by the magnetic collisionless interactions they suffer with the interstellar plasma waves. These interactions make them wander inside the Galaxy following a random walk that can be studied as a diffusive motion~\cite{Ginz&Syr,berezinskii1990astrophysics}. From the evaluation of the amount of matter traversed by CRs, we know that their diffusion is not limited to the disc of the Galaxy~\cite{stephens1998cosmic}, but extends up to a few\units{kpc} in the so-called magnetic halo. 

CRs are accelerated inside astrophysical sources (presumably supernova remnants) with an energy spectrum typically following a power law of the form $Q(E) = KE^{-\gamma}$. Acceleration of CRs at sources is explained by the diffusive shock acceleration (DSA) model~\cite{krymskii1977regular,bell1978acceleration, axford1982structure,blandford1978particle}. CRs produced at these sources are known as primary CRs and their spectra are usually called ``injection spectra''~\cite{blasi2013origin}. Nevertheless, the spectra of CRs detected at Earth are modified due to their diffusive propagation, resulting in power laws of the form $J(E) \propto E^{-(\gamma + \delta)}$, where the diffusion parameter $\delta$ is related to the time spent by CRs in the Galaxy  $\tau_{prop}(E)$ (propagation time).

On the other hand, during their journey, galactic CRs can eventually interact with the interstellar gas (which consists mainly of hydrogen with about a $11\%$ of helium and traces of heavier elements~\cite{asplund2004solar}) producing lighter nuclei (secondary CRs) that otherwise would be found in tiny proportions (mainly boron, beryllium, lithium and the so-called sub-Fe nuclei), since they are hardly produced in stellar fusion or other thermonuclear processes~\cite{biswas1965composition}. The spectra of secondary CRs produced from these interactions (spallation reactions) have the form: $J_{k}(E)\propto\rho_g \sum_{i}J_{i}(E)\sigma_{i \rightarrow k}(E) $. Here $\rho_g$ is the density of the interstellar medium (ISM) and $\sigma_{i \rightarrow k}(E)$ is the inclusive cross section of production of the CR species $k$ from the interaction of the CR species $i$ with the ISM. 

The study of secondary CRs can therefore provide valuable information about the interactions that CRs suffer during their journey. The relevant quantity used to describe CR interactions is the ``grammage'' $ X(E) = \rho_g l(E) \propto \rho_g \tau_{prop}(E)$, related to the diffusion coefficient used to describe the CR diffusive motion ($D \propto \tau_{prop}^{-1}$). Here $l(E)$ is the effective length traversed by CRs, which depends on the interaction cross sections and on the particle kinetic energy per nucleon $E$.  Nonetheless, the current experimental data on inclusive cross sections are based on a few data points that hardly reach energies of tens of\units{GeV/n}. These cross sections are crucial, since the best way to constrain the propagation parameters (mainly the diffusion coefficient) is by means of the secondary-over-primary CR flux ratios~\cite{reinert2018precision,derome2019fitting}.

In this work, we have implemented different publicly available cross sections data sets in the preliminary version of the {\tt DRAGON2} cosmic-ray propagation code \cite{DRAGON2-1, DRAGON2-2} to study how the choice of cross sections affects the predictions on the fluxes of the light secondary CRs boron, beryllium and lithium. In section~\ref{sec:model} the set-up of the simulations is illustrated. The results of the simulations are presented in section~\ref{sec:secondaries}, where the fluxes of these secondary species computed with different cross sections parametrisations are shown, and some relevant conclusions are drawn. The effects of variations on the cross sections are then tested in detail by studying secondary-over-secondary ratios in section~\ref{sec:uncert}. 
These ratios are extremely sensitive to the cross sections and, therefore, represent a useful tool to test the validity of the cross sections used in the propagation code. This allows us to evaluate the uncertainties in the predicted fluxes of B, Be and Li associated to their production cross sections.
In this section we also perform a simultaneous fit of the high-energy part of the flux ratios among secondary B, Be and Li by adjusting their production cross sections within the experimental uncertainties. In addition, these secondary-over-secondary ratios will be used to look for possible imprints of any extra primary source producing them. Then, in section~\ref{sec:size} the effect of the halo size on the ratios involving $^{10}$Be is explored with the different cross section models. In this section we also get a more robust estimate of the value of the halo size. Finally, the conclusions are drawn in section~\ref{sec:conc}.

\section{Propagation setup and cross section data sets}
\label{sec:model}

\begin{figure}[!t]
\centering
\includegraphics[width=0.65\columnwidth,height=0.23\textheight,clip]{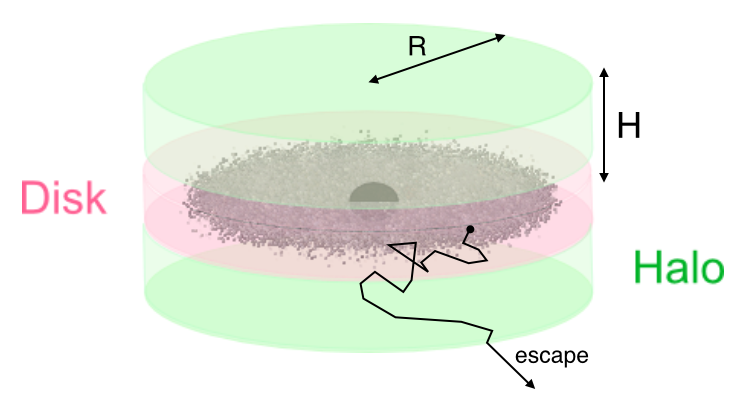}
\caption{Scheme of the 2D model used for the Galaxy structure. Taken from the lecture \url{https://w3.iihe.ac.be/~aguilar/PHYS-467/PA3.html}}
\label{fig:2D_model}
\end{figure}

Simulations of CR propagation are performed using the 2-dimensional model of the Galaxy offered by the DRAGON code~\cite{evoli2008cosmic}~\footnote{The software can be downloaded from \mbox{\url{https://github.com/cosmicrays/DRAGON}}}, where cylindrical symmetry is assumed. The Galaxy is described as a thin disc with radius $R \sim 20\units{kpc}$ and height $h \sim 100\units{pc}$ with the Sun on the Galactic plane at a distance of $8.3 \units{kpc}$ from the center. The disc is surrounded by the halo, which is a cylinder with the same radius as the disc and with height $2H$ of a few\units{kpc} (two-zone model). Gas and CR sources are distributed within the disc. An illustration of the model is shown in Fig.~\ref{fig:2D_model}.

A diffusion-reacceleration model is used in the CR propagation, without implicit advection, since recent studies of the AMS-02 data with similar diffusion coefficient parameterizations as we use here demonstrate to be compatible with no or negligible convection~\cite{Weinrich_combined, Niu:2017qfv}. These models have successfully reproduced the shape of the ratios of secondaries over primaries fluxes at low energy, as for example in refs.~\cite{Jaupart,Mazziot,Jones2001}. The inclusion of stochastic reacceleration seems to be necessary to naturally explain the shapes of these ratios.

The general formulae describing this propagation model are given by the following set of equations:
\begin{equation}
\label{eq:caprate}
\begin{split}
 \vec{\nabla}\cdot(\vec{J}_i - \vec{v}_{\omega}N_i) + \frac{\partial}{\partial p} \left[p^2D_{pp}\frac{\partial}{\partial p} \left( \frac{N_i}{p^2}\right) \right] = Q_i + \frac{\partial}{\partial p}  \left[\dot{p} N_i - \frac{p}{3}\left( \vec{\nabla}\cdot \vec{v}_{\omega} N_i \right)\right] \\
  - \frac{N_i}{\tau^f_i} + \sum \Gamma^s_{j\rightarrow i}(N_j) - \frac{N_i}{\tau^r_i} + \sum \frac{N_j}{\tau^r_{j\rightarrow i}} .
\end{split}
\end{equation}

In the previous equations $\vec{J_i}$ indicates the CR diffusive flux of the $i$-th species and $N_i$ is the density per unit momentum. The term $\vec{v}_{\omega}$ represents the advection speed, which we assume to be null. The second term in the left-hand side accounts for the diffusion in momentum space. The first term in the right-hand side, $Q_i$, represents the distribution and energy spectra of particle sources (injection spectra); the second term describes the momentum losses; finally, the remaining four terms describe losses and gains due to decays and fragmentations. These equations are numerically solved with a new version of the DRAGON code~\footnote{\label{note1}Now available in \url{https://github.com/cosmicrays/DRAGON2-Beta\_version}}. More information about the code can be found on refs.~\cite{DRAGON2-1} and~\cite{DRAGON2-2}.

The flux terms $\vec{J_i}$ are related to the densities $N_i$ by Fick's law, $\vec{J}_{i} = -\vec{\vec{D}} \vec{\nabla} N_i$, where $\vec{\vec{D}}$ is the spatial diffusion tensor~\cite{DRAGON2-1}.
In this study, we assume that the spatial diffusion coefficient parallel to the regular magnetic field lines vanishes because of the azimuthal symmetry adopted for the Galaxy structure~\cite{evoli2008cosmic} and we used a general form of the spatial diffusion coefficient perpendicular to the magnetic field lines, which is assumed to be homogeneous:

\begin{equation}
 D(R) = D_0 \beta^{\eta}\left(\frac{R}{R_0} \right)^{\delta} .
\label{eq:indepdiff}
\end{equation}
Here $\beta = v/c$, $R$ is the rigidity of the particle, $D_0$ is the constant diffusion coefficient at the reference rigidity $R_0$ (it is set to $4 \units{GV}$). The $\eta$ parameter describes the complex physical effects that may play a major role at low energy, as the dissipation of Alfven waves~\cite{ptuskin2006dissipation}. In this work $\eta$ will be treated as a free parameter, since also negative values are theoretically motivated~\cite{ptuskin2006dissipation}.

Finally, the relation between the spatial diffusion coefficient $D(R)$ and the diffusion coefficient in momentum space $D_{pp}$ depends on the Alfven velocity $v_A$ and, following refs.~\cite{osborne1987cosmic} and~\cite{seo1994stochastic}, it is given by:

\begin{equation}
    D_{pp} = \frac{4}{3}\frac{1}{\delta (4 - \delta^2)(4 - \delta)} \frac{v_A^2 p^2}{D(R)} . 
    \label{dpp} 
\end{equation}
Hereafter, when speaking of  ``diffusion parameters'', we will refer to the set of parameters in the spatial diffusion coefficient and to the Alfven velocity. 

The injection spectra of primary nuclei are parametrized with a doubly broken power law:  
\begin{equation}
Q = \begin{cases}
     K_1 \times\left(\frac{R}{R_0}\right)^{\gamma_1} & \text{for $R < R_{1}$} \\
     K_2 \times\left( \frac{R}{R_0}\right)^{\gamma_2} & \text{for $R_{1} < R < R_{2}$} \\
     K_3 \times\left( \frac{R}{R_0}\right)^{\gamma_3} & \text{for $R > R_{2}$}
  \end{cases} 
\label{eq:powerlaw} 
\end{equation}
where $Q$ is the differential energy flux in units of $\units{m^{-2} s^{-1} sr^{-1} GV^{-1}}$, $R_{1,2}$ are the rigidity breaks, $\gamma_{1,2,3}$ are the logarithmic slopes below and above each break and the parameters $K_{i}$ set the normalization of the flux. The low-energy break was set to $R_{1}=8 \units{GV}$ for all nuclei, while the high-energy break was set to $R_{2}=335 \units{GV}$ for protons and $R_{2}=200 \units{GV}$ for the heavier nuclei. 
In the present work we injected $^1$H, $^{4}$He, $^{12}$C, $^{14}$N, $^{16}$O, $^{20}$Ne, $^{24}$Mg and $^{28}$Si as primary nuclei with the spatial distribution of sources following the model in ref.\cite{Lorimer_2006}. The injection spectra are tuned such that these primary CRs reproduce the AMS-02 spectra (Figure~\ref{fig:primFit}).

In the last years CR experiments have reached unprecedented precision in the flux measurements of CRs, making possible the study of several features unexplored in the past. The accuracy showed in the last experimental results of the AMS-02 collaboration is of the order of $1-5\%$ (for the main nuclei involved in the creation of light secondary CRs, as C, N and O). However, the uncertainties on the cross sections reach levels of $20-50\%$ in some channels (see~\cite{Genolini}), which makes clear the need of new cross sections data with better accuracy to improve the precision of the predictions on CR fluxes. Even the exact determination of the uncertainties is complicated for most of the channels, since sometimes data from different experiments are difficult to reconcile. In addition, there are many channels with no data or just with a few data at low energies (below $10 \units{GeV/n}$), which makes the parametrisations at high energies not straightforward at all.
Also the so-called ghost-nuclei (i.e. those unstable nuclei whose lifetime is so small that they are not propagated and their contribution is directly added up to their daughter nuclei) have a sizeable effect in the estimation of the production of secondary nuclei.

Currently there are several parametrisations of the nuclei cross sections publicly available. The first parametrisations came from the early measurements of B.~Webber, published in a series of papers during the 80s (see~\cite{webber1990total} and references therein) that ultimately lead to the semi-empirical WNEW code~\cite{webber1990formula} with a last update in 2003~\cite{webber2003updated}. Then, there were important efforts to expand the known experimental measurements to other channels, turning out in the semi-empirical parametrisations by Silberberg and Tsao~\cite{tsao1993scaling,silberberg1973partial,silberberg1985improved} that converged in the YIELDX code~\cite{tsao1998partial,silberberg1998updated}. More recently, the {\tt GALPROP} team\footnote{\url{https://galprop.standford.edu}} developed a set of routines combining semi-empirical formulae~\cite{moskalenko2005propagation,moskalenko2001new,strong2007cosmic}. Nevertheless, although these routines have remained the state of the art, it is well known that with the current available cross section data they show serious shortcomings that limit the precision of CR studies~\cite{Tomassetti:2015nha}.

A couple of years ago, a new set of cross sections derived from a different parametrisation of data for all individual channels was presented as the default option for the incoming {\tt DRAGON2} code~\cite{DRAGON2-2, Evoli:2019wwu}. They have been successfully used in other studies as~\cite{CarmeloBeB} and are fully available in the github repository of the new DRAGON version \textsuperscript{\ref{note1}}. 

Due to uncertainties in cross section, the discrepancies found when comparing the predicted secondary CR fluxes to data might drive to misleading conclusions. It is therefore clear the need of comparing different cross section data sets and their performance in reproducing the  measurements.

In the present simulation we are propagating nuclei up to $Z=14$, which implies we can fully describe the generation of secondary nitrogen and boron (which is considered to be fully secondary, as Be and Li). Nevertheless, the missing iron (mainly $^{56}Fe$, but also its isotopes) and, in very low proportion, $^{32}S$ make us underestimate the total amount of Li and Be in a 3.22\% and 3.7\% in average, respectively (see tables $IV$ and $V$ in~\cite{Genolini}). In order to save computational time for the computations required for the rest of nuclei (up to $Z=14$), we compensate for the missing source terms (primary Fe and S) by adding just this extra $\sim3\%$ to the Li and Be spectra.

Finally, another important parameter affecting the low-energy region of the CR spectra is the solar modulation, which is modeled using the Parker equations and the force-field approximation 
\cite{forcefield}. This parameter depends on the solar activity and, therefore, on the epoch in which the experimental data were taken.

The variable used to characterize the solar modulation is the Fisk potential $\phi$. The flux of CRs reaching a detector at Earth is related to that in the Local Interstellar medium (LIS) by the following equation:

\begin{equation}
\Phi_{obs} (E_{obs})=  \left( \frac{2 m E_{obs} + E_{obs}^2}{2 m E_{LIS} + E_{LIS}^2}\right) \Phi_{LIS}(E_{LIS}) ,
\label{eq:potmod} 
\end{equation}
with $E_{LIS} = E_{obs} + e \phi |Z|/A$. In eq.~\ref{eq:potmod}, $E_{LIS}$ and $E_{obs}$ are the kinetic energies per nucleon in the LIS and at Earth, respectively, while $m$ indicates the proton mass.

In this work we have chosen for the solar modulation potential the value of $0.61 \units{GV}$, consistent with the data from the NEWK \footnote{\url{http://www01.nmdb.eu/station/newk/}} neutron monitor experiment (see \cite{ghelfi2016non,ghelfi2017neutron}), which allows us to reproduce the Voyager-1~\cite{stone2013voyager,cummings2016galactic} and AMS-02 data in the period 2011-2016 and is similar to the value found in previous studies~\cite{Mazziot}.

\section{Fluxes of the secondary CRs B, Be and Li}
\label{sec:secondaries}

In this section we show and discuss the predicted fluxes of B, Be and Li obtained in our analyses. These predictions are performed using the cross sections parameterizations\footnote{The {\tt GALPROP} cross sections are available as ASCII files at \url{https://dmaurin.gitlab.io/USINE/input\_xs\_data.html\#nuclei-xs-nuclei} and the {\tt DRAGON2} 
and {\tt Webber} cross sections at \url{https://github.com/cosmicrays/DRAGON2-Beta_version/tree/master/data}} from the upcoming {\tt DRAGON2} code, {\tt GALPROP} and a combination of the WNEW03 code and YIELDX code (only for the Li production) called here {\tt Webber}, and selecting the set of diffusion parameters
that reproduce the boron-over-carbon data (see Fig.~\ref{fig:BCComp} in appendix~\ref{sec:appendixB}) of AMS-02 \cite{aguilar2018observation} and using the optimal value for the halo size found from the procedure explained in Sec.~\ref{sec:size}. These parameters are in agreement with previous analyses~\cite{Boschini:2019gow, Weinrich_combined} and are summarized in Table~\ref{tab:diff_params}. Here, we do not perform a dedicated best-fit study of the diffusion parameters, but a scan over these parameters to fit the B/C ratio which is terminated when the condition  $\chi^2/n_{points} < 1.05$ is fulfilled, since this value corresponds to a p-value $\sim 50\%$. Additionally, we adjust the parameters of the source spectra in Eq.~\ref{eq:powerlaw} to reproduce the fluxes of individual CR species measured by AMS-02 \cite{aguilar2017observation, AMS_Ne}. 
As an example, Figure~\ref{fig:primFit} in appendix~\ref{sec:appendixB} shows the observed spectra at Earth of the main primary CR nuclei (from carbon to silicon) compared with the predictions from our simulation with the {\tt DRAGON2} set of cross sections.
We remark again that we are using the new released data of AMS-02 for the Mg, Si and Ne fluxes for testing the production of secondary CRs, since the collaboration reported fluxes on these CR species around $10-20$\% larger than previous experiments~\cite{AMS_Ne}, which lead up to $\sim2$\% changes in the flux of these secondary CRs.
We then proceed to study the spectra of secondary CRs and compare their fluxes to the available experimental data. The employed cross sections parametrisations for the channels of production of B, Be and Li from C and O projectiles (the main channels for their production) are compared to the available data in appendix~\ref{sec:appendixA} (see ref.~\cite{Genolini} for other rarer channels). 

Very few works have studied the spectra of Li and Be in order to determine the diffusion parameters (see~\cite{Weinrich_combined} for a recent work), while nearly all authors just use boron and its ratios to a primary CR species (usually C) to develop their models~\cite{AMS01}. This is due to the fact that, before AMS-02, the experimental data on Li and Be fluxes were poor and the uncertainty on the predicted fluxes from the cross sections parametrisations in the B channels is expected to be the smallest~\cite{LiBeB}. Typical uncertainties on the predicted Li and Be fluxes around $20-30\%$ and $15-25\%$ respectively are usually quoted~\cite{LiBeB}, while the uncertainties on the predicted B flux are around $10\%$~\cite{Genolini}. The reason is that the production of B is mainly regulated by the C and O reaction channels, while other very poorly constrained cross sections channels (mainly those of Mg, Ne and Si) become important for the production of Li and Be, contributing to $\sim 45\%$ of their flux.

\begin{table}[!t]
\centering
\begin{tabular}{|l|c|c|c|}
  \multicolumn{4}{c}{\hspace{0.3cm}\large} \\ \hline  & \textbf{ Webber}  & \hspace{0.2 cm}\textbf{GALPROP} & \hspace{0.2 cm}\textbf{DRAGON2}\\ 
  \hline
{$D_0$} ($10^{28}$ cm$^{2}$ s$^{-1}$) & 2.3 & 6.65 & 7.1\\
{$v_A$} (km/s) & 29.9 & 25.5 & 27.7\\
{$\eta$} & -0.25 & -0.55 & -0.6\\      
{$\delta$}  & 0.42 & 0.44 & 0.42\\   
{H} (kpc)  &  2.07 & 6.93 & 6.76\\  
\hline
\end{tabular}

\vspace{0.1cm}
\caption{Diffusion parameters used in the CR propagation with the different cross section parameterizations. 
The values have been obtained from the fit of the B/C data from AMS-02~\cite{aguilar2018observation}, \cite{aguilar2017observation} and of the $^{10}$Be/$^9$Be data from various experiments (see sec.~\ref{fig:sizes}), assuming the different cross section models.}
\label{tab:diff_params}
\end{table}

\begin{figure*}[!ht]
\hskip -0.3 cm
\includegraphics[width=1.0175\textwidth,height=0.6\textheight,clip] {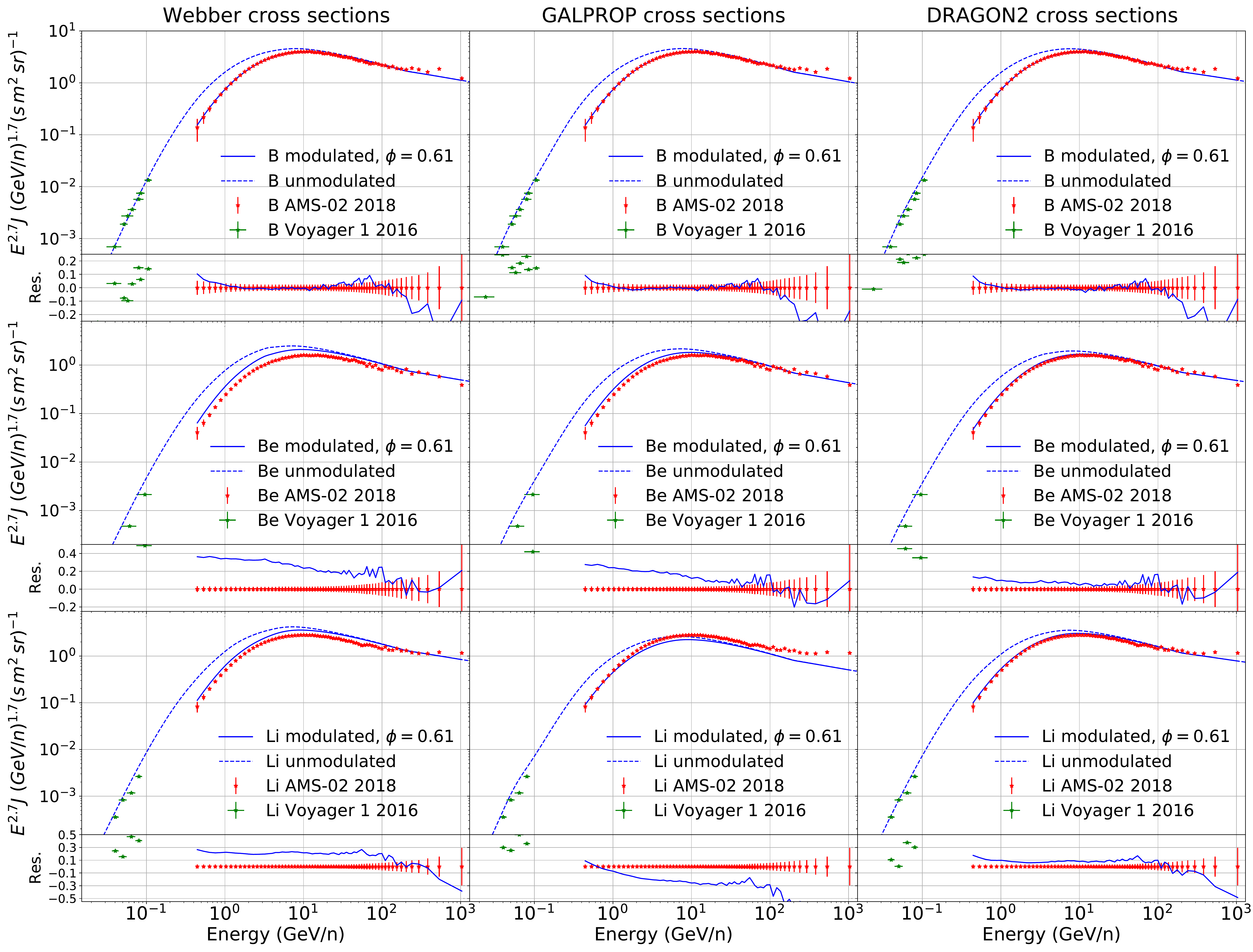} 
\rightskip 2cm


\caption{Spectra of the light secondary nuclei obtained with the diffusion parameters fitting the B/C spectrum using the {\tt Webber} cross sections (left column), the {\tt GALPROP} parametrisations (middle column) and the {\tt DRAGON2} model (right column). The residuals (defined as (model-data)/model throughout all the paper) are also shown to have an idea about how large discrepancies may be for different cross sections models. Experimental data from CR experiments were taken from \url{//https://lpsc.in2p3.fr/crdb/} \cite{Maurin_db1, Maurin_db2} and \url{https://tools.ssdc.asi.it/CosmicRays/} \cite{ssdc}. }
\label{fig:secfit_Webber}
\end{figure*} 

Figure~\ref{fig:secfit_Webber} shows the spectra of B, Be and Li for the three cross section parametrisations adopted in the present work.
However, in the very high-energy region, above $200\units{GeV/n}$, we see that the measured fluxes of Li, Be and B are slightly higher (harder) than the predicted ones for all the three cross section parametrisations. This feature suggests the need of introducing a break in the energy dependence of the diffusion coefficient (see, e.g., ref.~\cite{genolini2017indications}) rather than in the CR injection spectra, but this does not affect the conclusions of this article because the flux ratios among secondary cosmic rays mitigate the effects of features in the diffusion coefficient, as discussed later.

In the case of the {\tt Webber} cross sections, the Li and Be fluxes follow a similar trend, being overestimated with respect to the AMS-02 data with residuals generally below $25\%$. In turn, for the {\tt GALPROP} parametrisations, we see an opposite behaviour for Li, with the simulation underestimating the experimental data of about $25\%$, while in the case of Be the prediction is closer to the experimental data, with discrepancies less than $15\%$ above $10 \units{GeV/n}$. The shape of the residuals found for Be with the {\tt Webber} and {\tt GALPROP} parametrisations may also be related to the adjustment of the halo size value (since the flux of the unstable isotope $^{10}$Be strongly depends on this parameter, as explained in sec.\ref{sec:size}). Finally, the {\tt DRAGON2} default cross sections seem to reproduce all the secondary CR fluxes at the same time within $10\%$ discrepancies in the full energy range, with differences just in their normalization.

\begin{figure*}[!t]
\hskip -0.3 cm
\includegraphics[width=1.059\textwidth,height=0.6\textheight,clip] {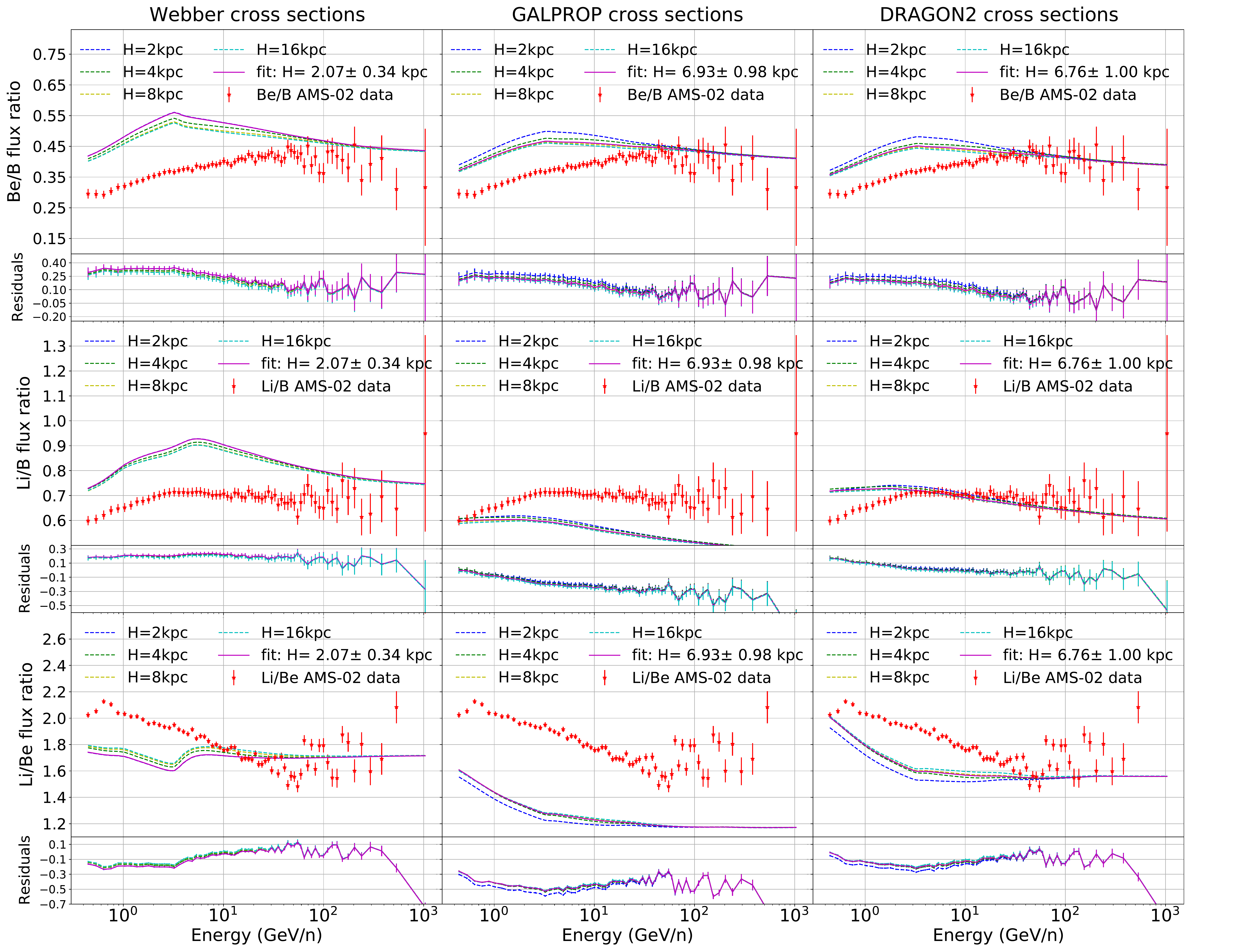} 
\rightskip 2cm
%
%
\caption{Secondary-to-secondary ratios of the light secondary CRs for the {\tt Webber}, {\tt GALPROP} and {\tt DRAGON2} cross sections models. The residuals are also shown to better illustrate how large are the discrepancies between simulations and data. These plots include the simulated spectra for various halo sizes, since the presence of $^{10}$Be and its beta decay to $^{10}$B modify the shape of the spectra at low energies. The simulated spectrum for the halo size that best fits the flux ratios of Be isotopes, as explained in section \ref{sec:size}, is also included. Data taken from \url{//https://lpsc.in2p3.fr/crdb/} \cite{Maurin_db1, Maurin_db2} and \url{https://tools.ssdc.asi.it/CosmicRays/} \cite{ssdc}.}
\label{fig:secsec_Webber}
\end{figure*} 

Comparing the fluxes of Li, Be and B shown in the panels of Fig.~\ref{fig:secfit_Webber}, we see that the main difference arise in the predicted Li flux, which is the nucleus that suffers more from missing cross sections data and poorly known reaction channels. Nevertheless, the differences among the predictions obtained with the three cross section parametrisations are subject to the choice of diffusion coefficient used, limiting a direct study of their production cross sections. In fact, the discrepancies between cross sections sets for the predicted flux of a secondary CR, (i.e. comparing the figures in the same column of Fig.~\ref{fig:secfit_Webber}), would change if we had used the same diffusion parameters for the three cross sections parametrisations. This strong dependence on the diffusion parameters employed makes the fluxes of individual secondary CRs to be an insufficient tool for discriminating among cross section models. On the contrary, the flux ratios among secondary CRs are almost unaffected by the diffusion coefficient, making this observable much more sensitive to the spallation cross sections than the fluxes of individual species.

Figure~\ref{fig:secsec_Webber} shows the Be/B, Li/B and Li/Be flux ratios predicted using the three cross section parametrisations, assuming the halo sizes reported in Table~\ref{tab:diff_params}, compared to the experimental data in the energy range from $500\units{MeV/n}$ up to $1\units{TeV/n}$. We see that the differences among the predictions obtained with the three parametrisations are very similar to those observed between the different individual fluxes in Fig.~\ref{fig:secfit_Webber}. However, the advantage is that these ratios are roughly unaffected by the parametrisation of the diffusion coefficient in the energy region above a few tens of GeV/n and mainly dependent on their production cross sections and on the spectra of primary CRs (set always to fit the AMS-02 data), as shown in appendix~\ref{sec:appendixC}. 

A parameter which can significantly affect the fluxes of secondary CRs and their ratios, at low energy, is the size of the galactic halo. To show the effect of the halo size on these ratios, we have performed an additional set of simulations changing the halo size from $2$ to $16 \units{kpc}$. From Figure~\ref{fig:secsec_Webber} we see that, in the case of the Li/B and Li/Be ratios, variations of the halo size do not yield large changes at any energies (the variations of the ratios are less than $5\%$). On the other hand, in the case of the Be/B ratio, variations of the halo size yield variations in the spectra up to $10\%$ in the low energy region. As mentioned above, different unstable CR species have different decay lengths and, depending on the path length they travel until reaching the Earth, different fractions of unstable nuclei can decay, thus influencing the secondary fluxes and their ratios. In particular, the Be/B ratio is highly sensitive to the halo size due to the presence of the radioactive isotope $^{10}$Be, which can decay into $^{10}$B (see~\cite{CarmeloBeB}, where the authors discuss the halo size repercussions on the Be/B ratio at low energy). This dependence is particularly evident below $10 \units{GeV/n}$, given the short lifetime of this isotope at low energies.  On top of this, we point out that, as the $^{10}$Be decay length at low energies is of the order of a few hundred$\units{pc}$, the Be flux at low energies can also depend on the local gas density distribution in the Galaxy~\cite{Donato:2001eq, Donato:2003va}. Changing this distribution could therefore change the predictions on the secondary flux ratios involving Be. However, a detailed study of this effect is beyond the goal of the present paper. 
Other uncertainties in the secondary CR fluxes are related to total inelastic cross sections and are expected to have negligible effects in comparison ($\mathcal{O}$(2\%) \cite{derome2019fitting}).

From Figure~\ref{fig:secsec_Webber}, we can see that the largest residuals are usually found at low energies, as expected. The shape of the flux ratios at low energy is not well reproduced when using the {\tt Webber} cross sections, with residuals up to $\sim 30\%$. On the other hand, the shapes of the flux ratios obtained from the {\tt GALPROP} and {\tt DRAGON2} parametrisations are very similar, with different normalizations. The smaller residuals with respect to data are found with the {\tt DRAGON2} parametrisation and are less than $10\%$ above $5 \units{GeV/n}$. 

In conclusion, we find that the secondary-over-secondary spectra are mainly related to their production cross sections, and above $\sim 10-20 \units{GeV/n}$ they have very little dependence on all the other discussed effects. Hence, these ratios represent an extremely useful tool to constrain the parametrisations of the inclusive cross sections used in CR propagation codes. 
In the next section we will use the flux ratios among B, Be and Li in order to evaluate the uncertainties associated to their production cross sections and achieve a simultaneous fit from a rescaling on their production cross sections within the experimental uncertainties.

\section{Cross section uncertainties assessment}
\label{sec:uncert}

Given the different predictions from the different parametrisations, each cross section model may lead to a different interpretation of the CR data. As an example, from Fig.~\ref{fig:secfit_Webber} we see that the {\tt Webber} cross section parametrisation yields a $\sim20\%$ excess in the Be and Li fluxes in comparison to the B flux. One could correct this discrepancy either by adding a primary component of boron (injecting boron from the source) and reducing the total grammage traversed by CRs (in order to fit the B/C ratio) or by rescaling the cross sections of boron production (which would have the same effect on the grammage necessary to fit the B/C ratio), such that the three fluxes will reproduce AMS-02 data at the same time. On the other hand, the {\tt GALPROP} parametrisation yields a $\sim 25\%$ deficit in the Li flux, which could be explained, for example, adding an additional component of Li generated at the source (see ref.~\cite{Boschini:2019gow}). Nevertheless, the {\tt GALPROP} parametrisations could also be tuned to reproduce the flux ratios by a proper rescaling of the production cross sections (i.e. renormalizing some channels), within the experimental uncertainties.

To investigate the effects of the uncertainties on the cross section parametrisations we have defined two bracketing models from the {\tt GALPROP} parametrisation, in which the spallation cross sections for each interaction channel with $^{12}$C and $^{16}$O as projectiles have been shifted up or down using a scaling factor corresponding to the average uncertainties on the cross sections experimental data at $\pm 1 \sigma$. This is motivated by the fact that the energy dependence of the cross sections is supposed to be well known, while their normalization is not precisely determined (see ref.~\cite{CarmeloBlasi}). A couple of channels were differently rescaled, to better contain the cross sections data. In the case of the Li production channels from $^{16}$O and $^{12}$C, extra shifts of $12\%$ and $7\%$ respectively were applied to the upper model, since a few data points exhibit larger excursions with respect to the nominal model. In the channels of $^{16}$O producing $^{9}$Be and $^{10}$Be the shift was taken to be half of the experimental uncertainties. 

\begin{figure*}[!bt]
\centering
\includegraphics[width=0.48\textwidth,height=0.23\textheight,clip] {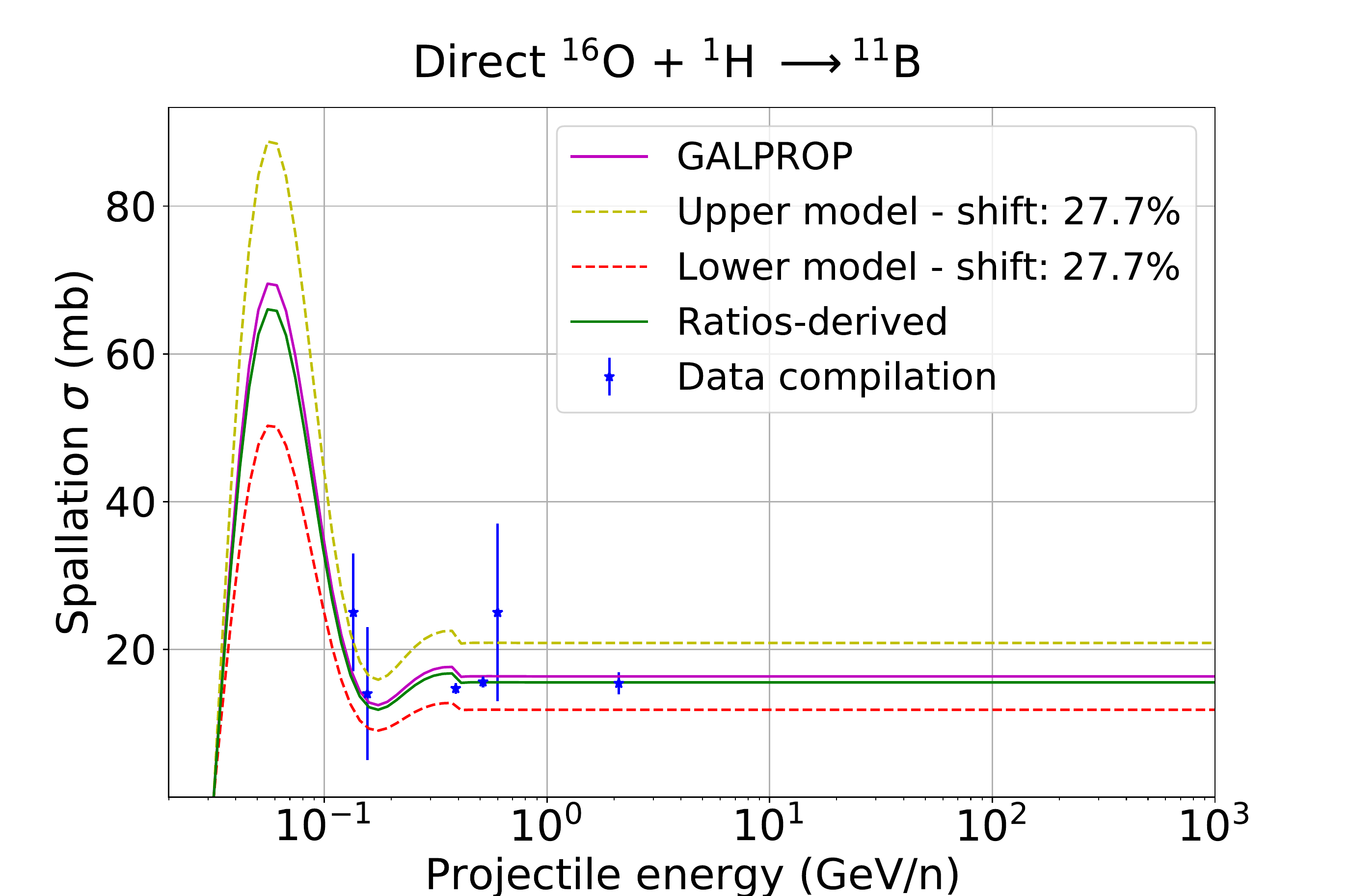} 
\includegraphics[width=0.48\textwidth,height=0.23\textheight,clip] {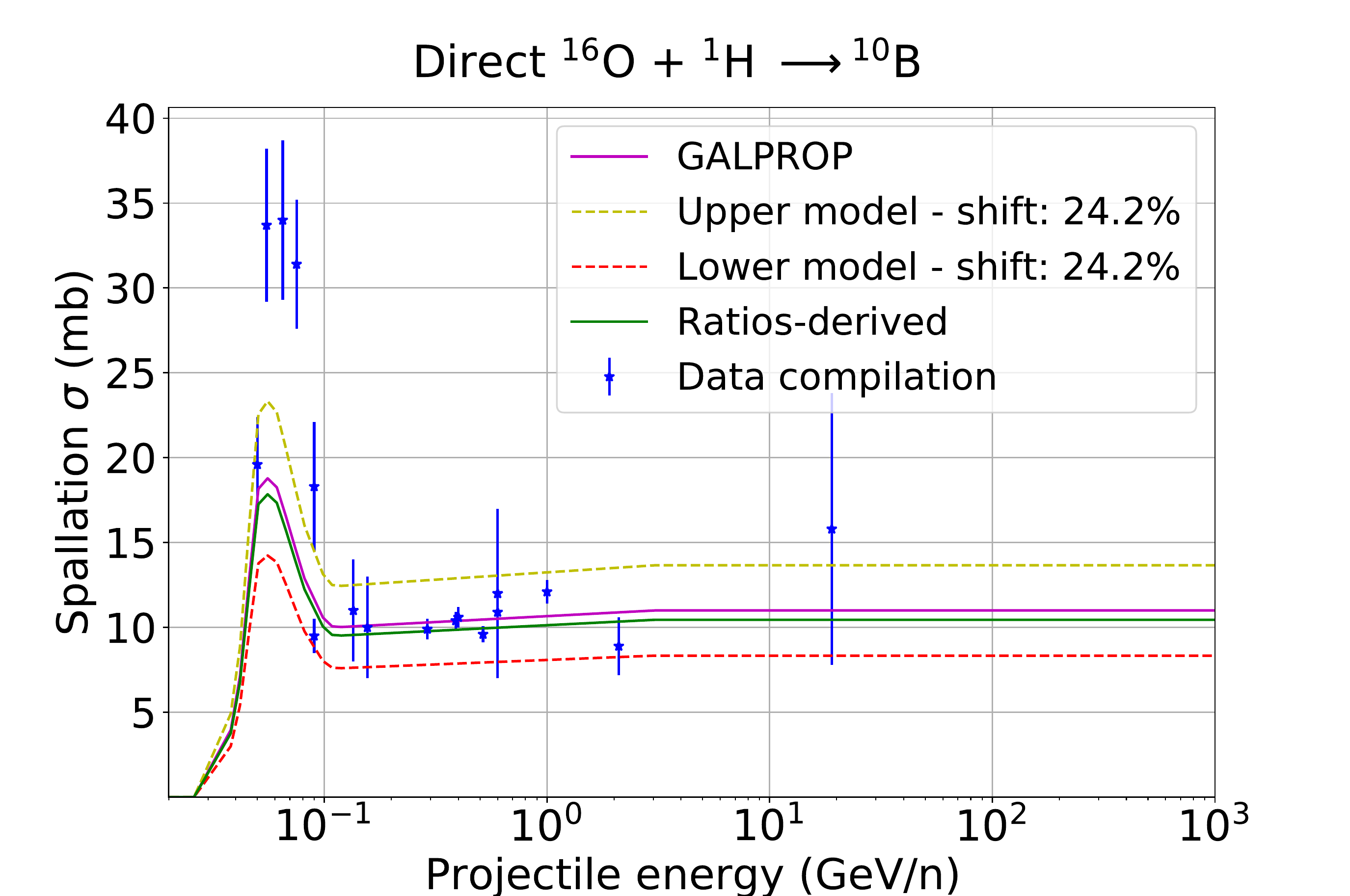} 
\includegraphics[width=0.48\textwidth,height=0.23\textheight,clip] {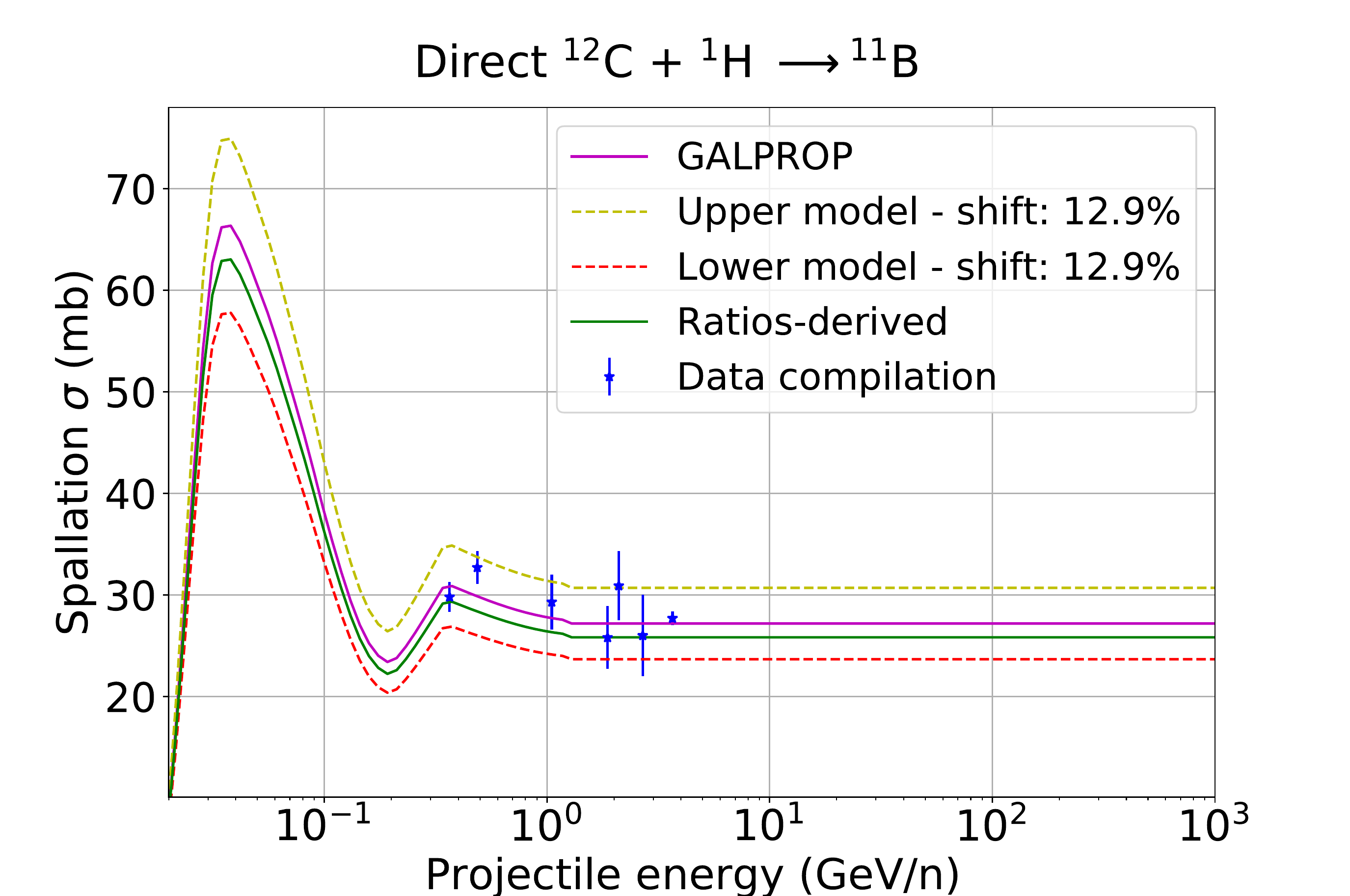} 
\includegraphics[width=0.48\textwidth,height=0.23\textheight,clip] {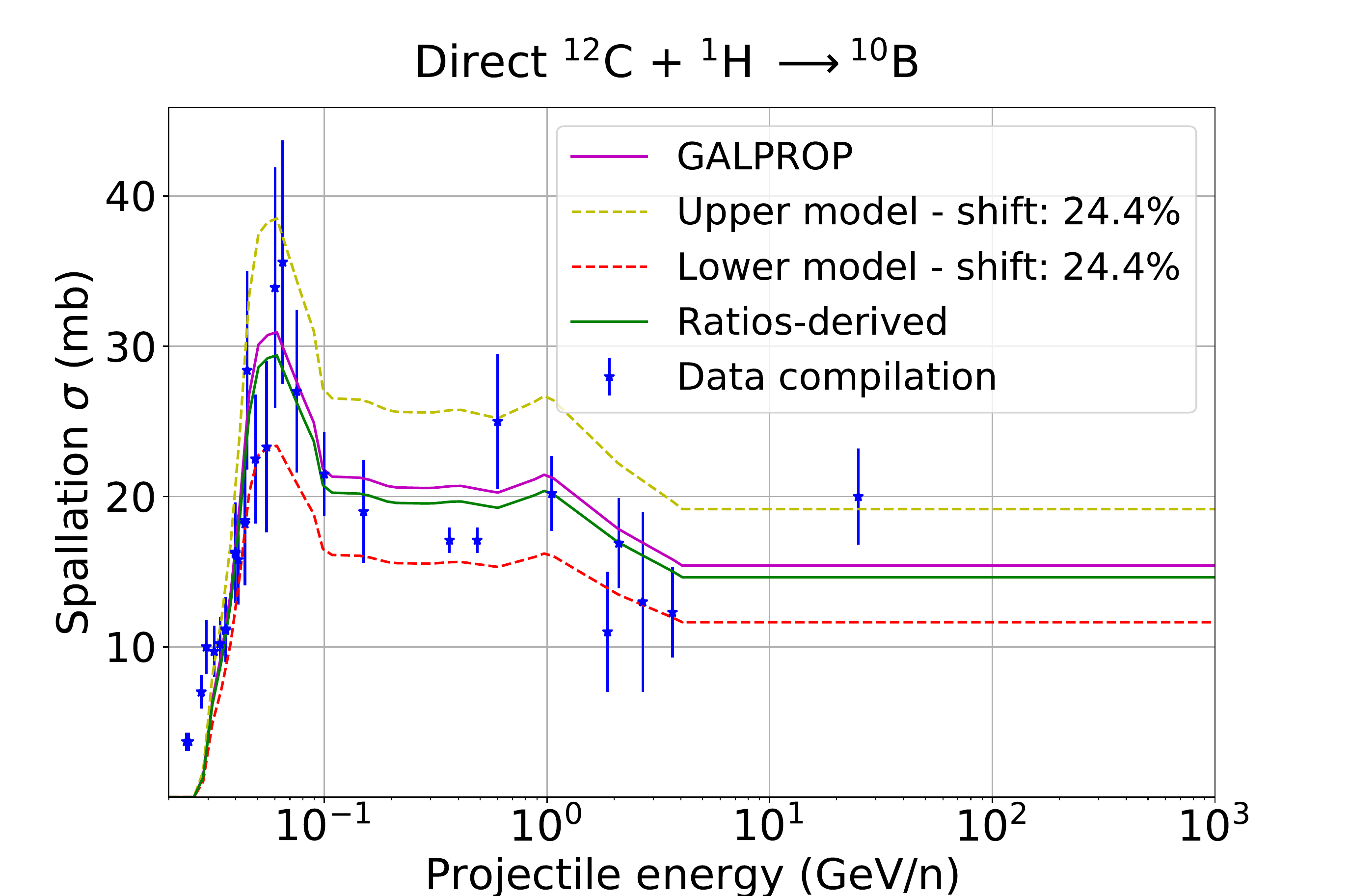} 
\caption{Cross sections describing the production of $^{11}$B and $^{10}$B isotopes from $^{16}$O and $^{12}$C. The upper and lower bracketing cross sections are shown and the percentage of renormalization is indicated in the legends. The cross sections derived by the fit of the secondary-over-secondary flux ratios are also shown and compared with the {\tt GALPROP} cross sections.}
\label{fig:LUmodelsB}
\end{figure*}

\begin{figure*}[!bt]
\centering

\includegraphics[width=0.48\textwidth,height=0.23\textheight,clip] {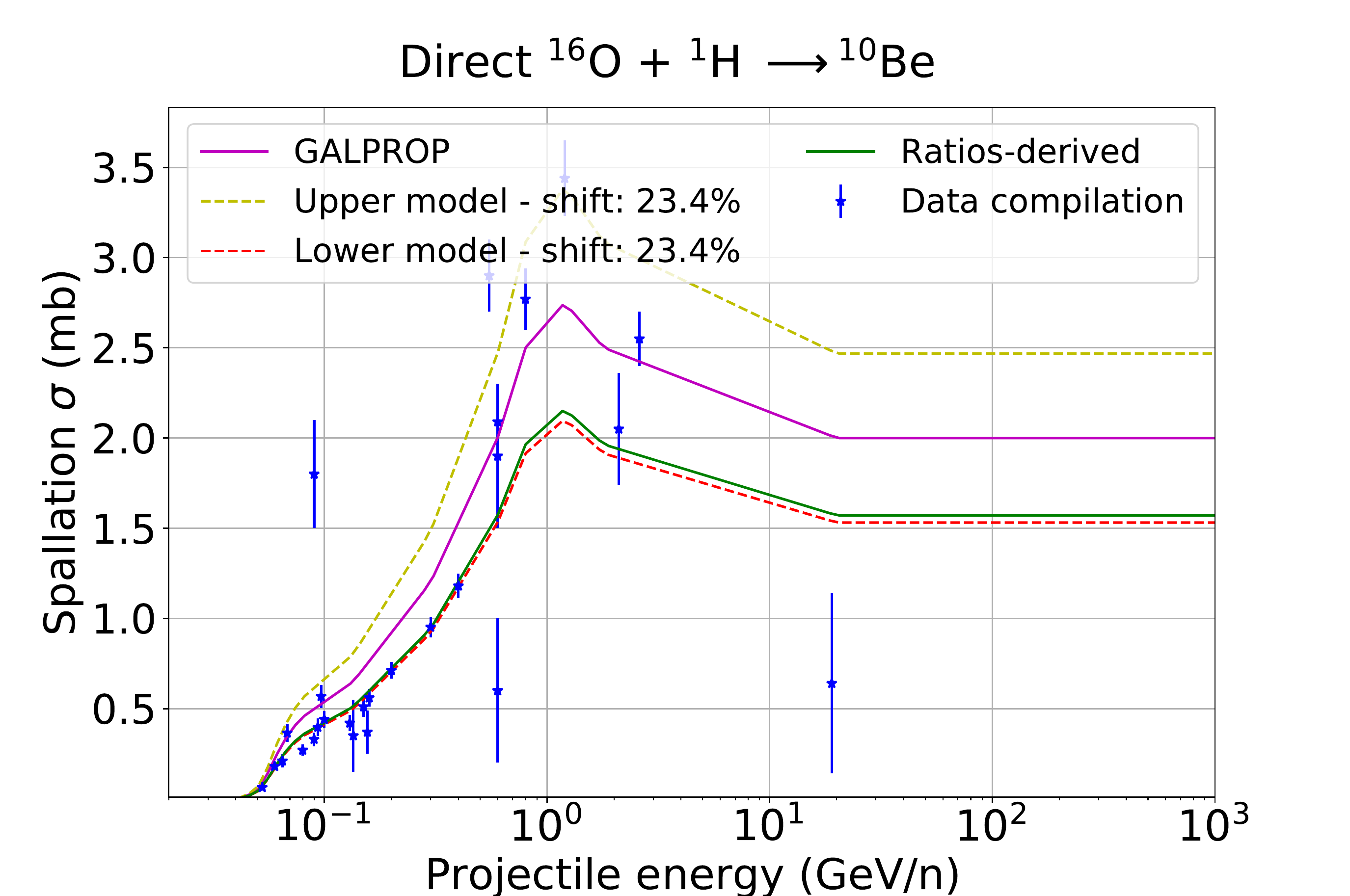} 
\includegraphics[width=0.48\textwidth,height=0.23\textheight,clip] {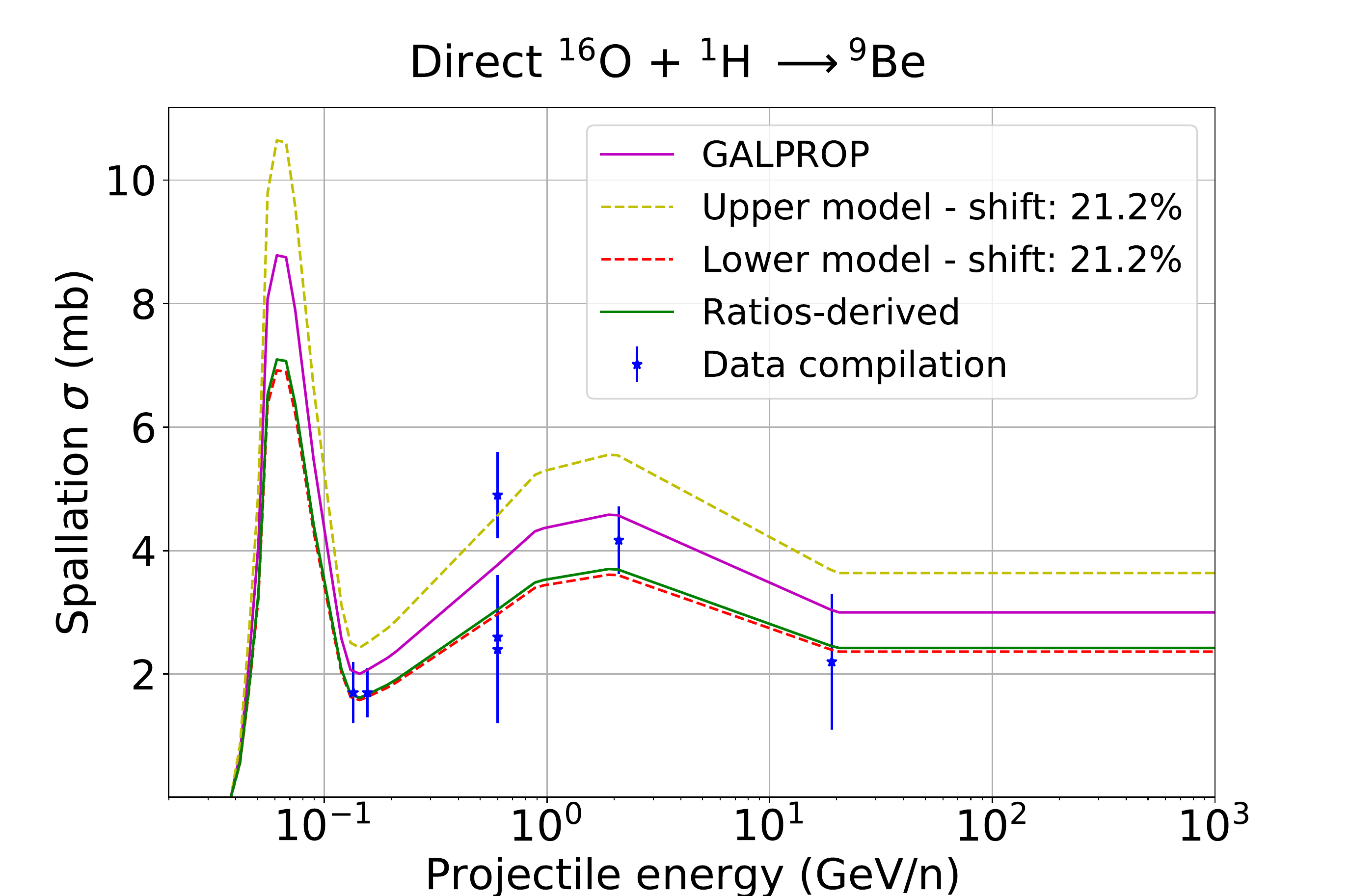}
\includegraphics[width=0.48\textwidth,height=0.23\textheight,clip] {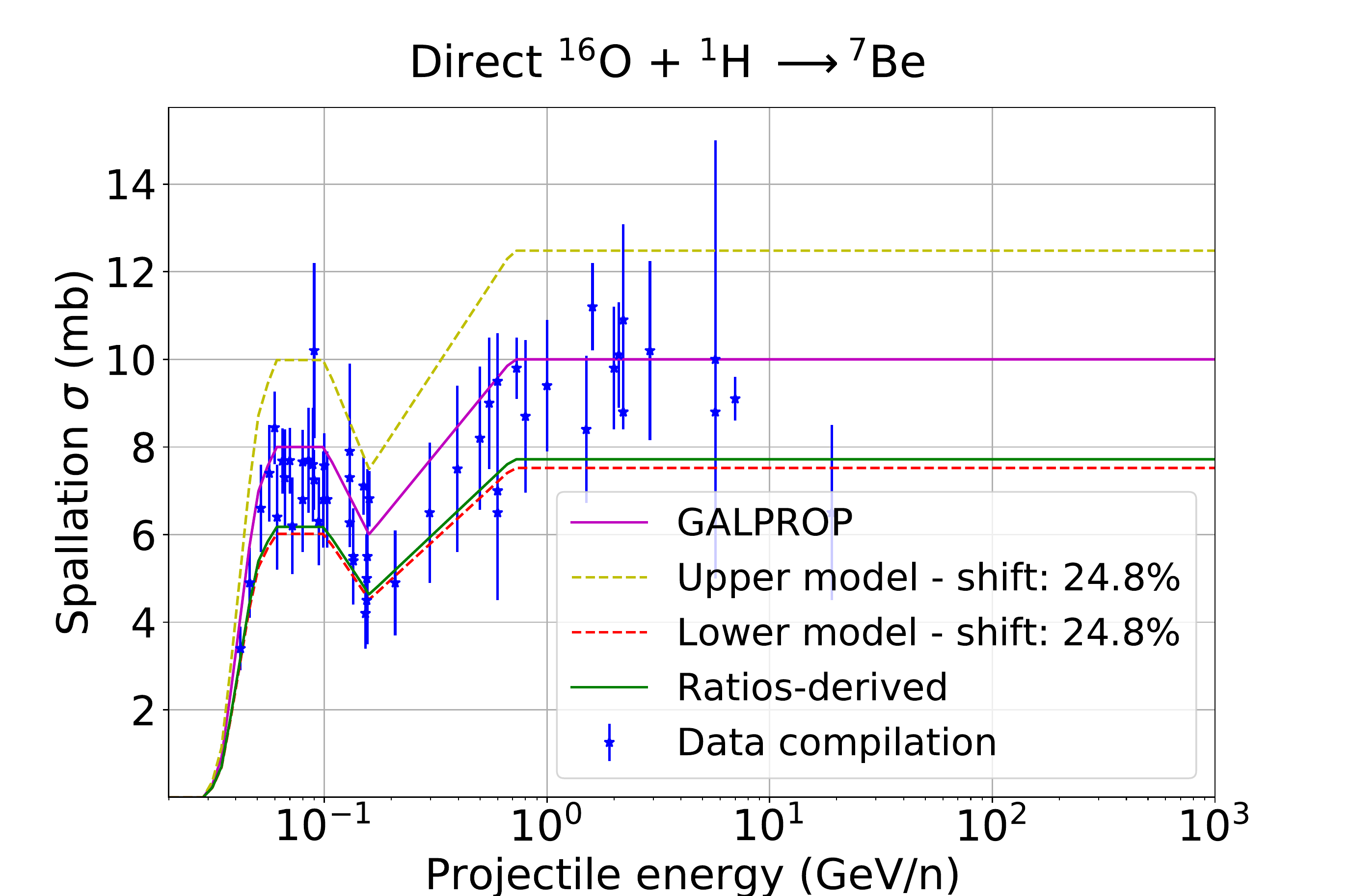}
\includegraphics[width=0.48\textwidth,height=0.23\textheight,clip] {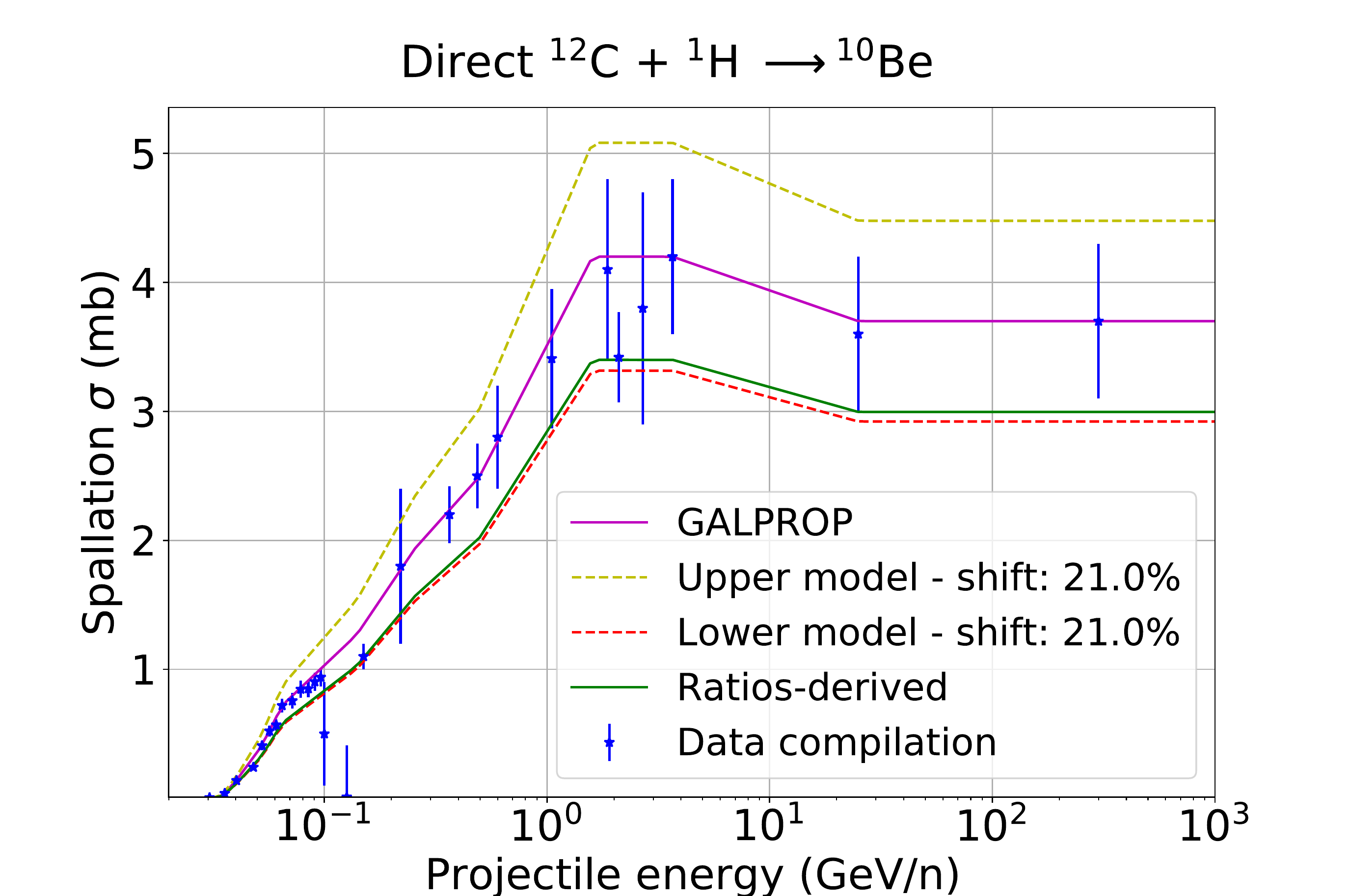} 
\includegraphics[width=0.48\textwidth,height=0.23\textheight,clip] {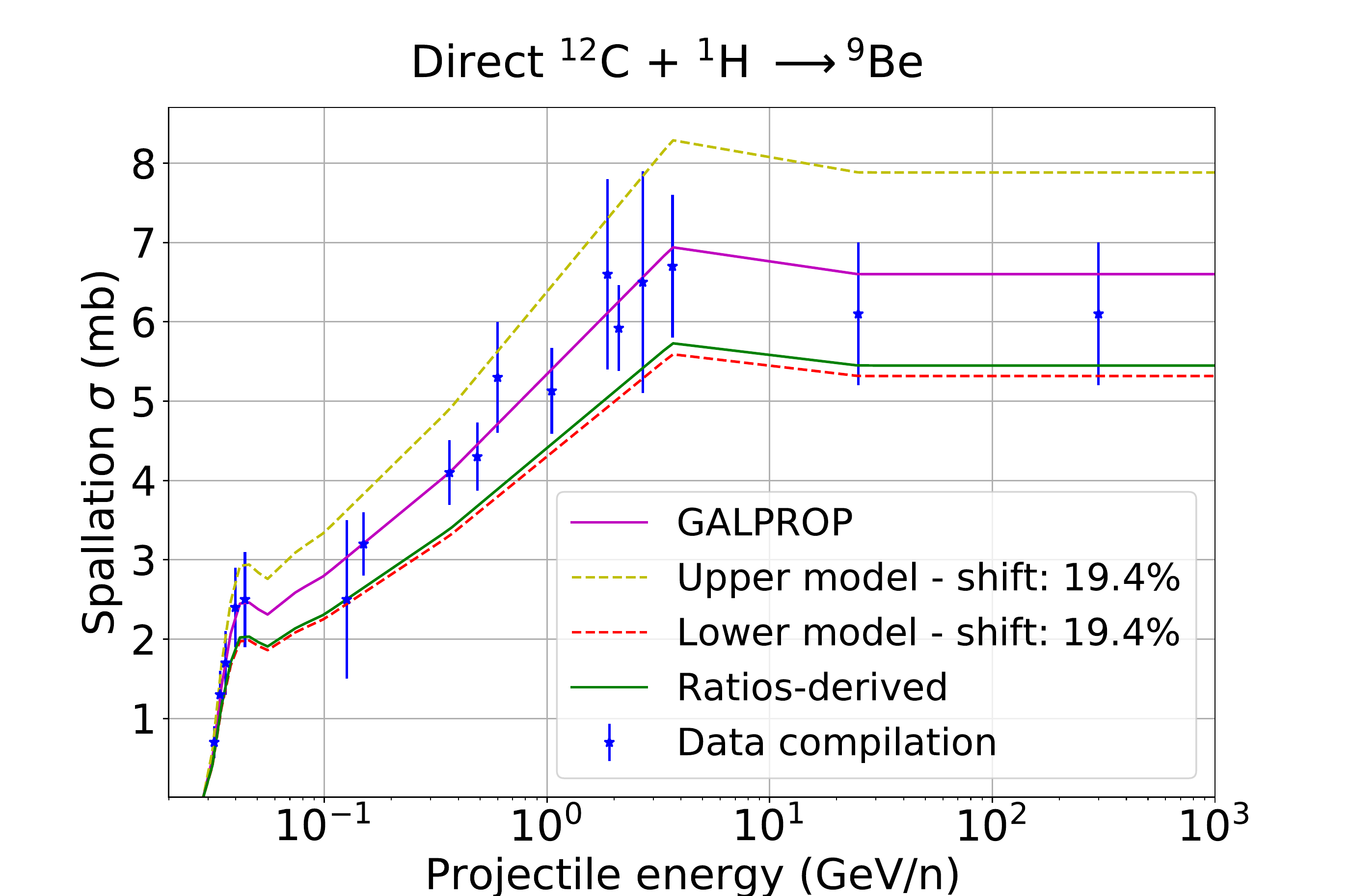}
\includegraphics[width=0.48\textwidth,height=0.23\textheight,clip] {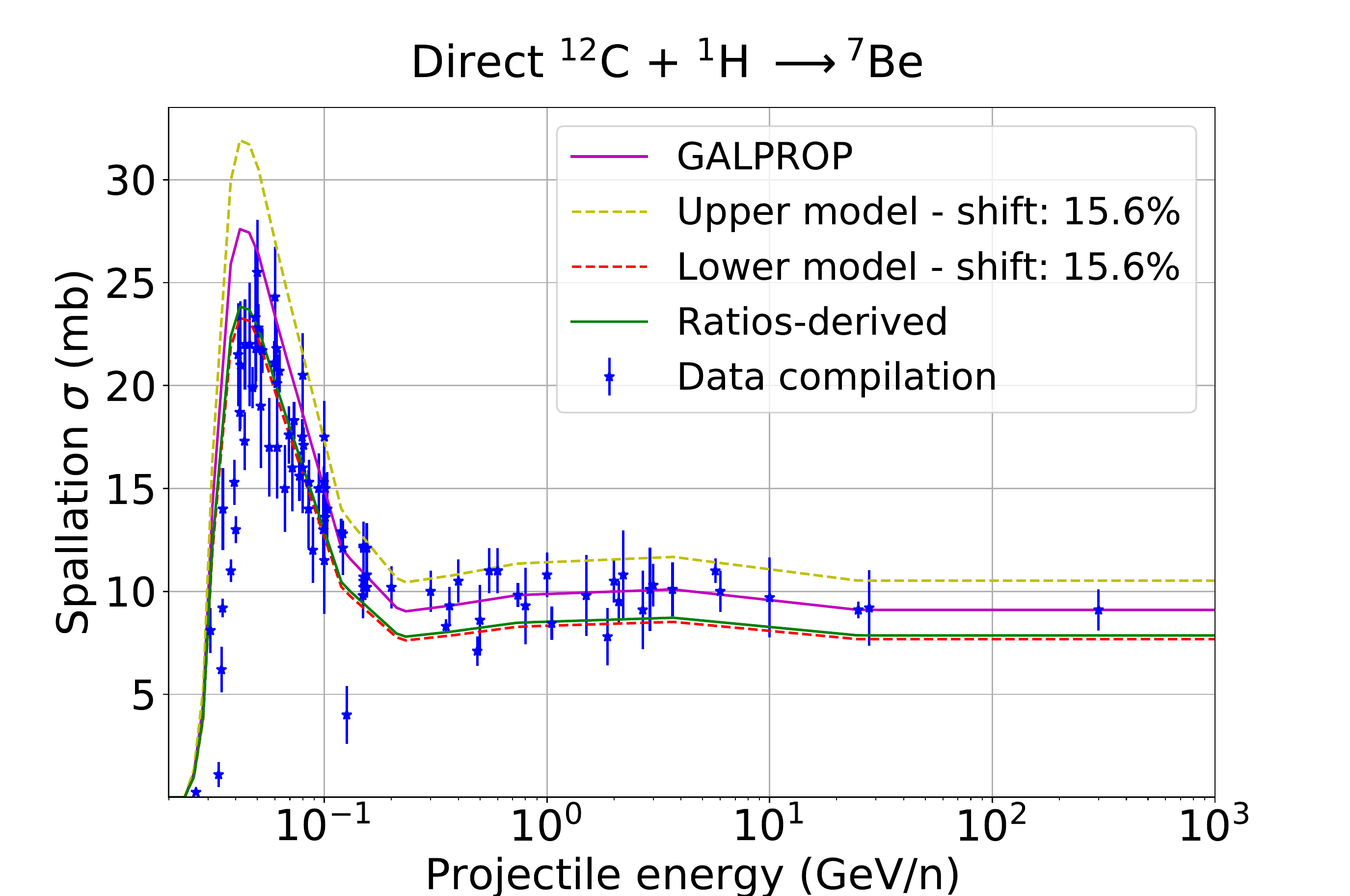}
\caption{As in Figure~\ref{fig:LUmodelsB} but for the cross sections describing the production of the $^{10}$Be, $^{9}$Be and $^{7}$Be isotope.}
\label{fig:LUmodelsBe}
\end{figure*}

\begin{figure*}[!bt]
\centering
\includegraphics[width=0.48\textwidth,height=0.23\textheight,clip] {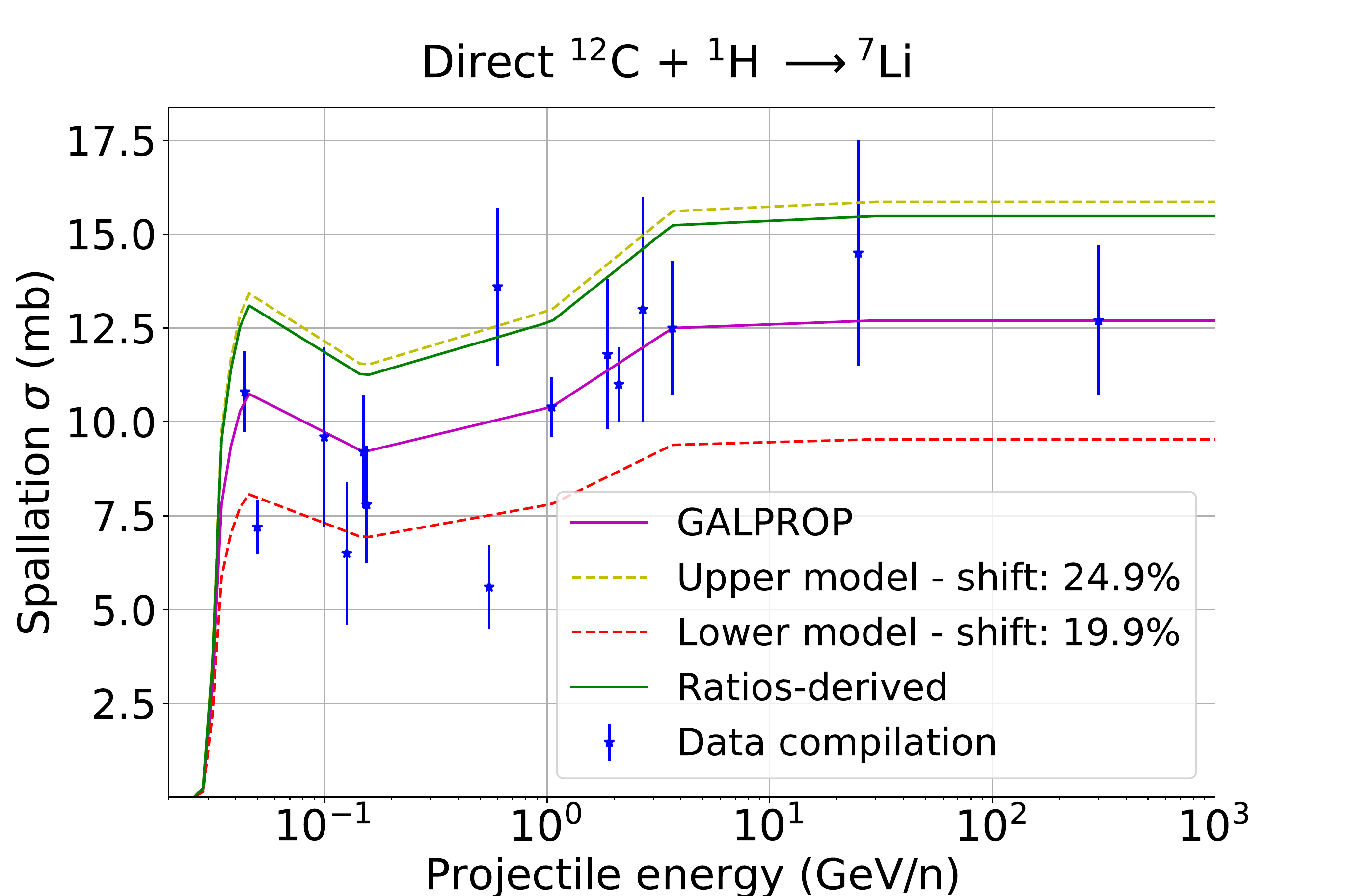} 
\includegraphics[width=0.48\textwidth,height=0.23\textheight,clip] {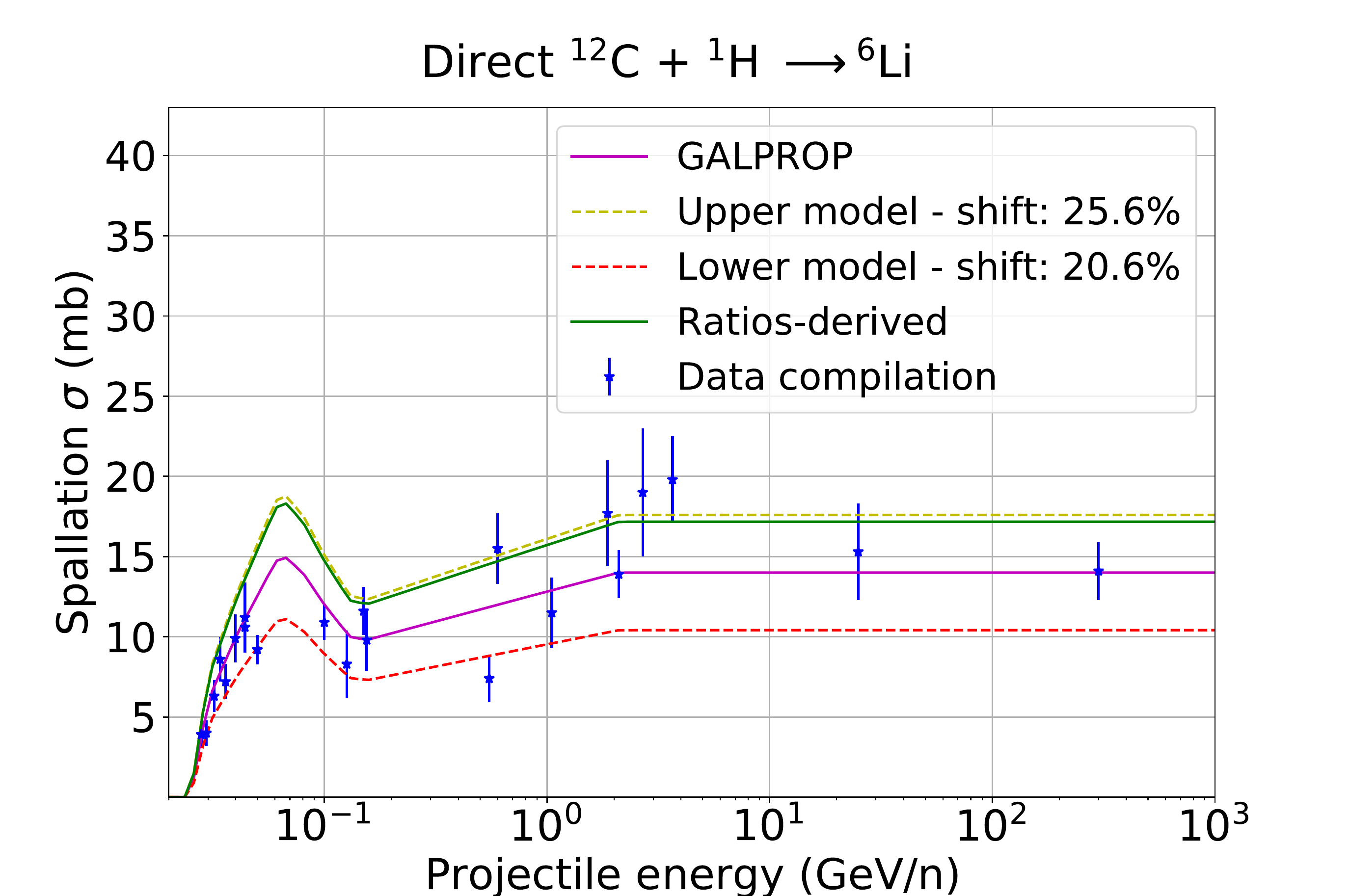} 

\includegraphics[width=0.48\textwidth,height=0.23\textheight,clip] {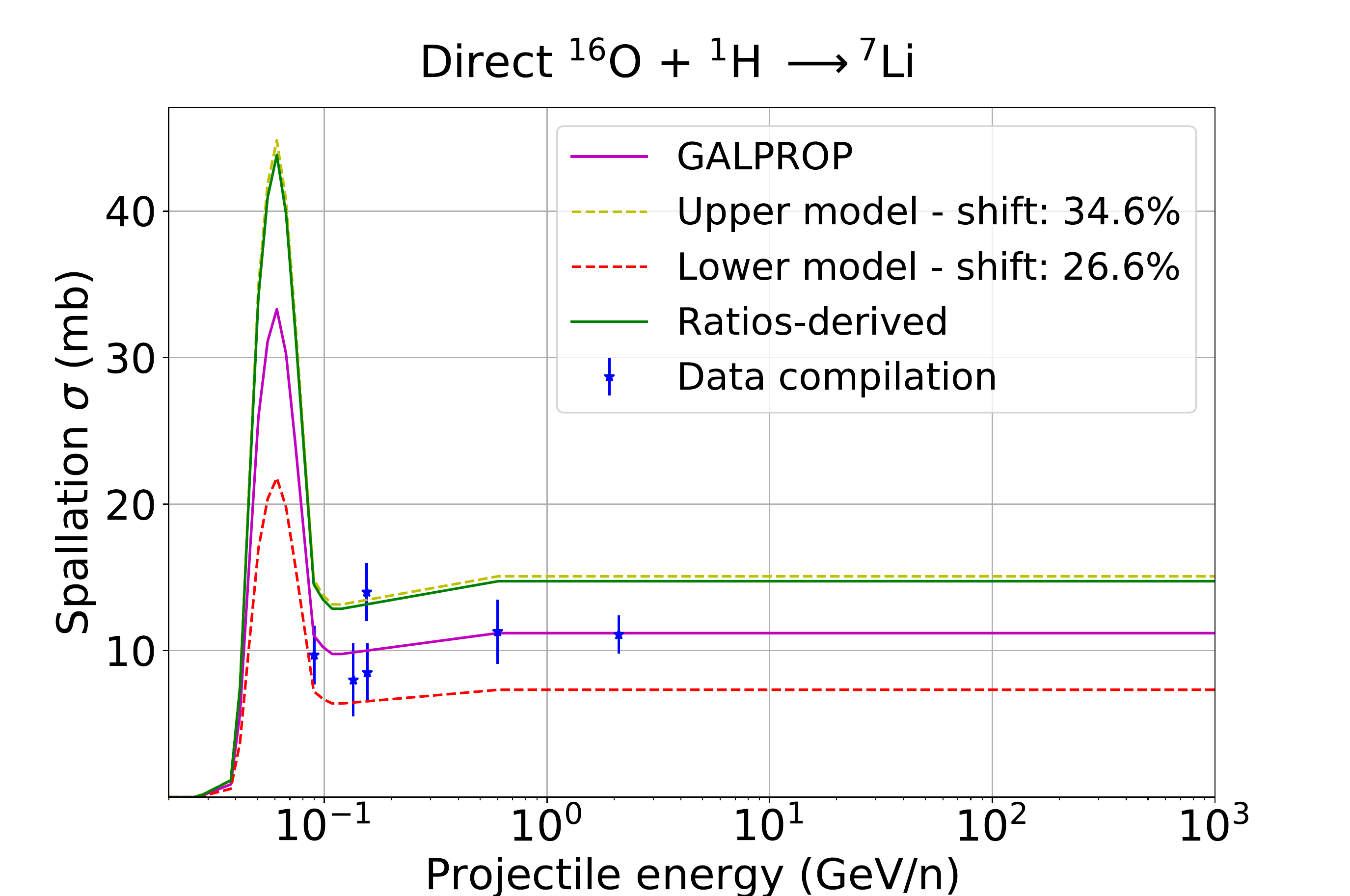} 
\includegraphics[width=0.48\textwidth,height=0.23\textheight,clip] {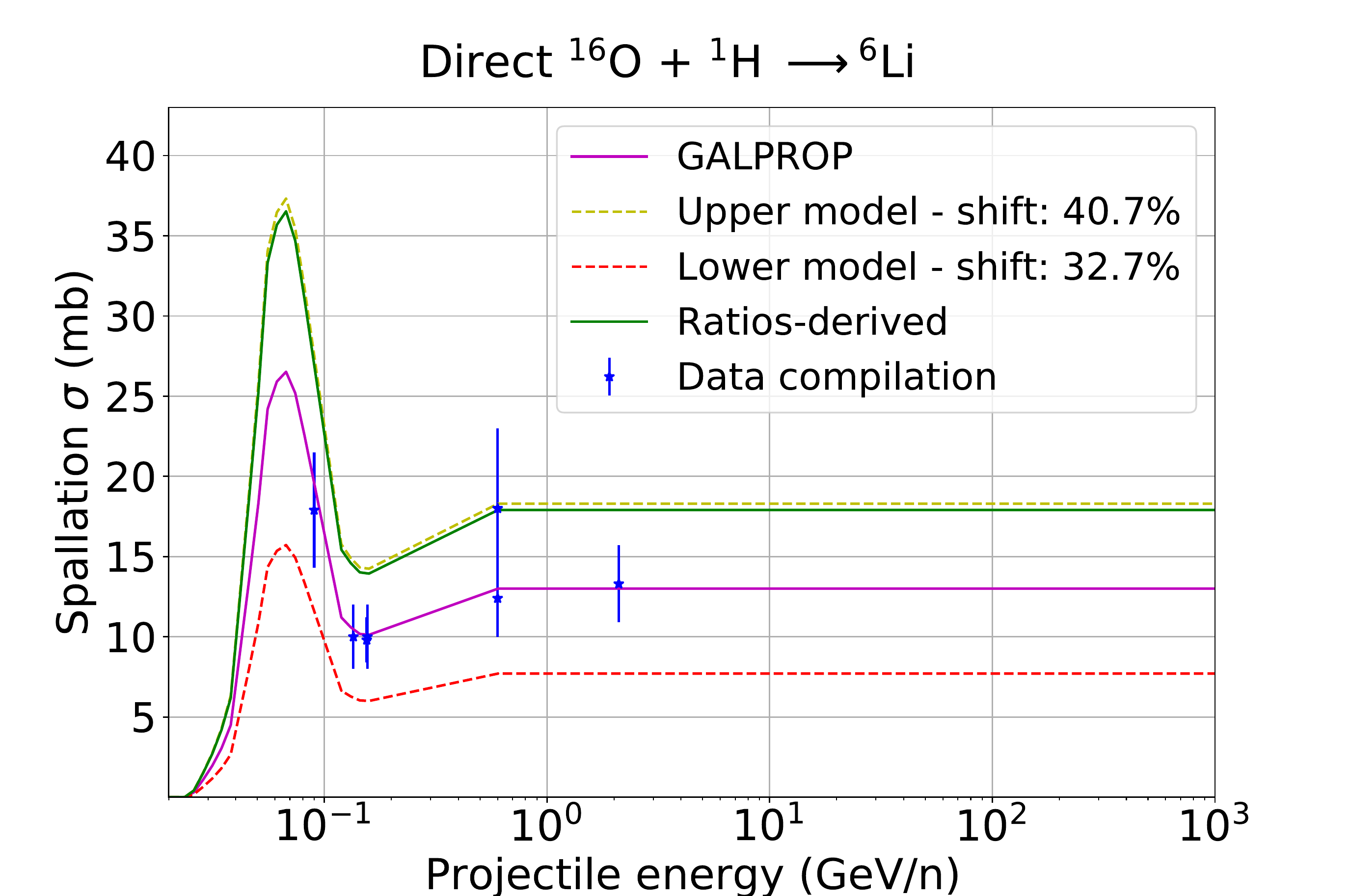}
\caption{As in Figures~\ref{fig:LUmodelsB} and~\ref{fig:LUmodelsBe} but for the cross sections describing the production of the $^{7}$Li and $^{6}$Li isotopes.}
\label{fig:LUmodelsLi}
\end{figure*}

The bracketing models are shown in figures~\ref{fig:LUmodelsB}, \ref{fig:LUmodelsBe} and~\ref{fig:LUmodelsLi}, where the shifts, typically corresponding to a $\pm 20\%$ variation from the cross sections normalization, are shown in the legends for each channel. We see that the Be channels are those with smaller uncertainties and more data points, while the Li channels are those with the higher uncertainties and less data points. We also see that the channels coming from the spallation of $^{16}$O exhibit larger experimental errors and less data points than those from the spallation of $^{12}$C. The original {\tt GALPROP} cross sections are also shown to illustrate the relative changes. These figures also show the cross sections obtained from the fit of the secondary-over-secondary flux ratios above $10 \units{GeV/n}$, which will be discussed in Section~\ref{sec:model_secratios} (see figure~\ref{fig:secsec_Uncert}). These cross sections are scaled from the original {\tt GALPROP} parametrisations and allow us to simultaneously reproduce the flux ratios involving Li, Be and B.

We point out here that we are not changing the cross sections from the spallation reactions of other nuclei than $^{12}$C and $^{16}$O (the main channels). Most of the channels with minor importance have no or very few data measurements, what originates most of the uncertainties on the total production cross sections of the isotopes we are studying. Although the individual contribution to the secondary CR fluxes from each of these channels is small, the sum of all their contributions is relevant ($\sim 25\%$ for B and $> 50\%$ for Li and Be). The next step here consists on the evaluation of the spectra of secondary CRs for these upper and lower (bracketing) models.

\subsection{Uncertainty effects and cross sections evaluation from the secondary-over-secondary ratios}
\label{sec:model_secratios}

\begin{figure*}[!t]
\centering
\hskip -0.25 cm \includegraphics[width=0.51\textwidth,height=0.235\textheight,clip] {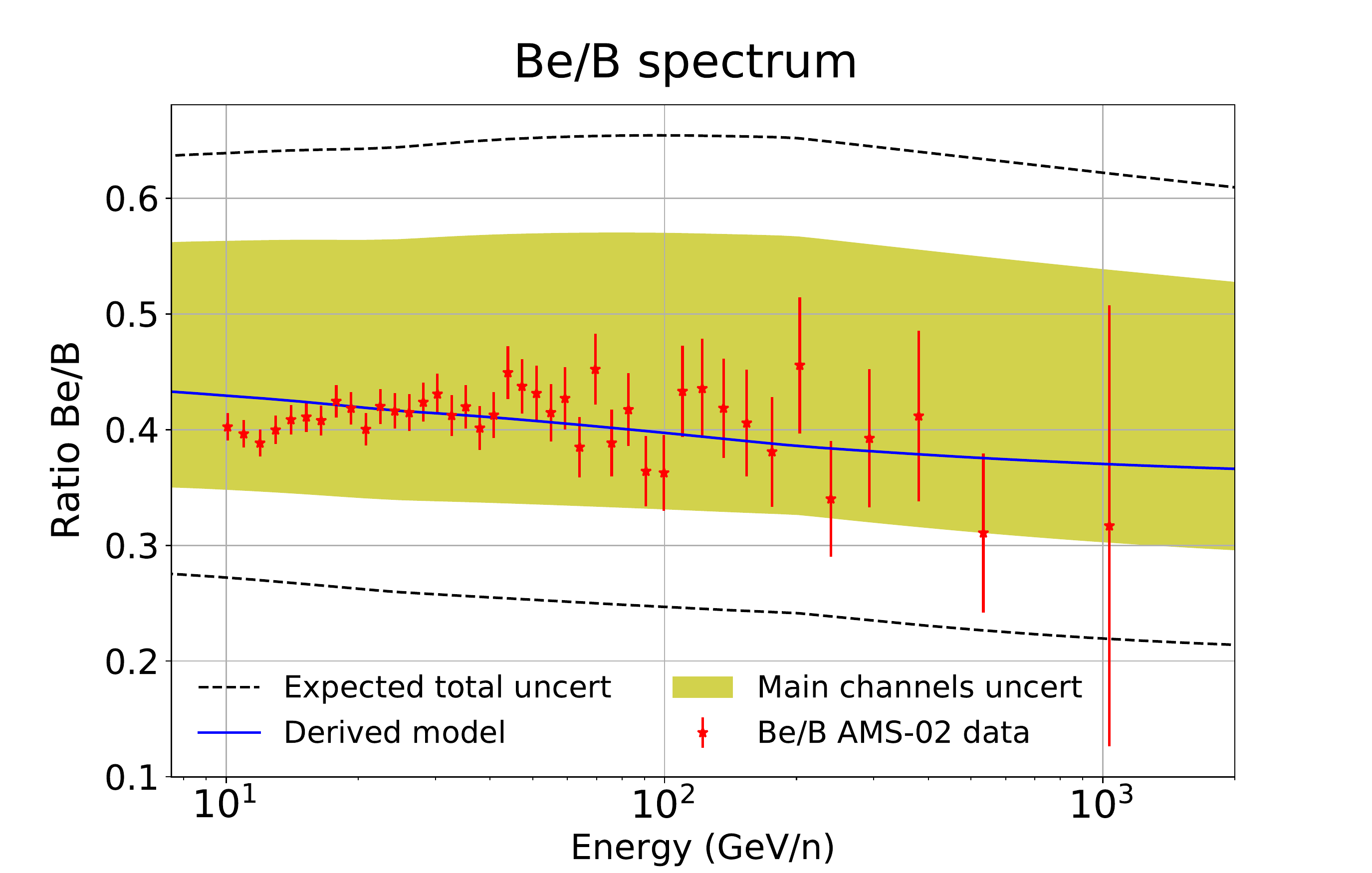} \hspace{-0.6cm}
\includegraphics[width=0.51\textwidth,height=0.235\textheight,clip] {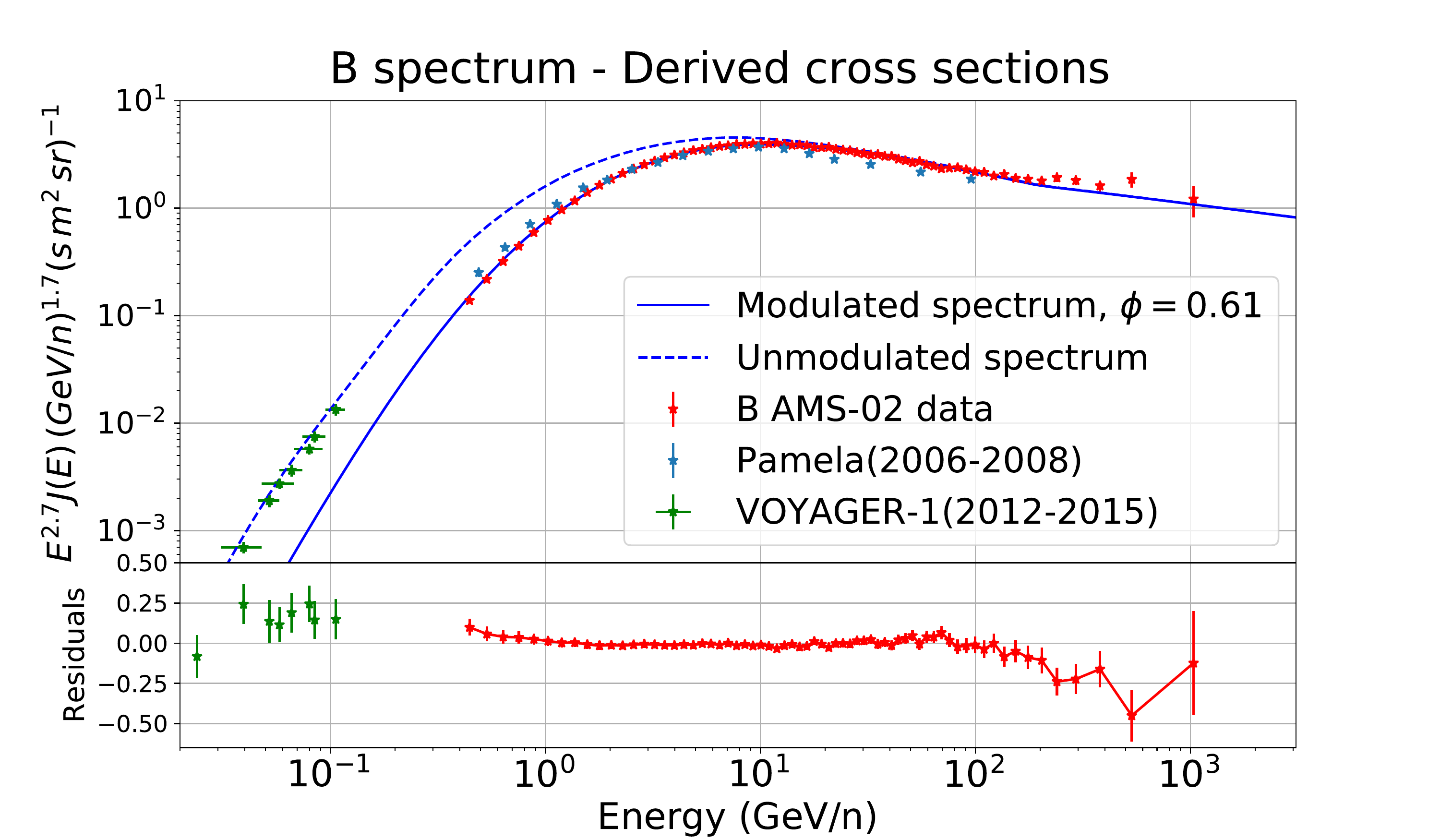} 

\hskip -0.25 cm \includegraphics[width=0.51\textwidth,height=0.235\textheight,clip] {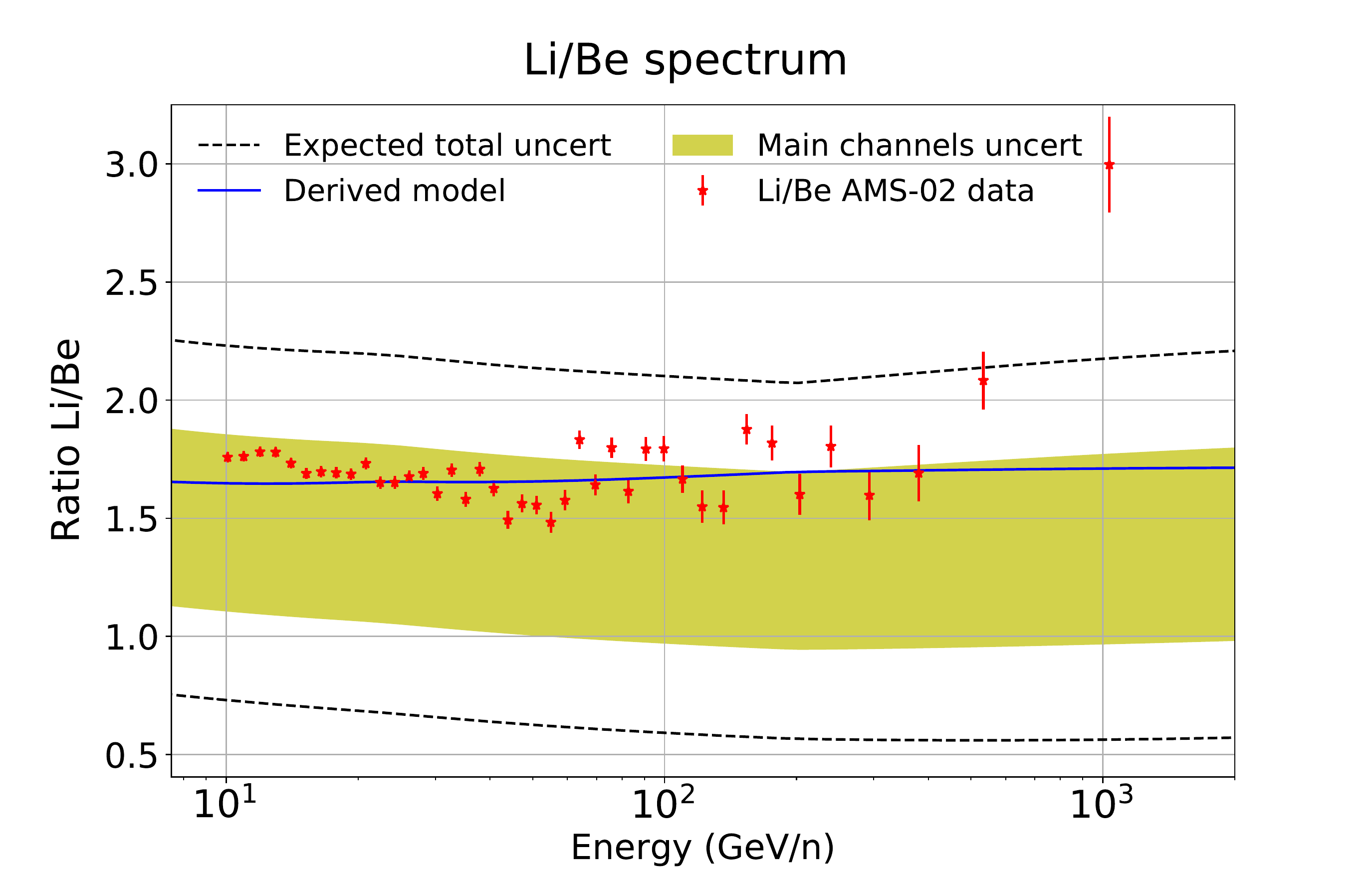} 
\hspace{-0.6cm}
\includegraphics[width=0.51\textwidth,height=0.235\textheight,clip] {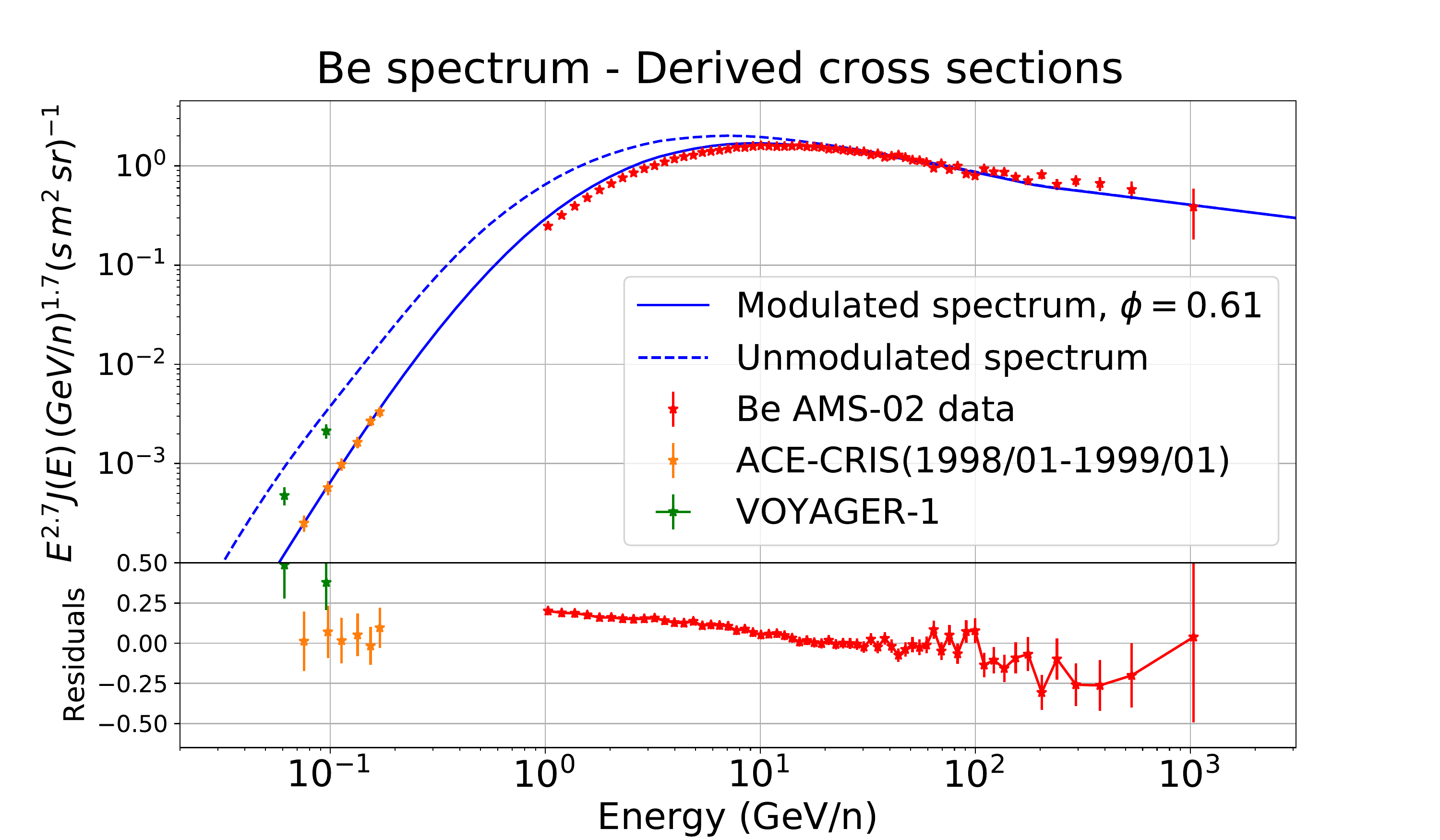}

\hskip -0.25 cm \includegraphics[width=0.51\textwidth,height=0.235\textheight,clip] {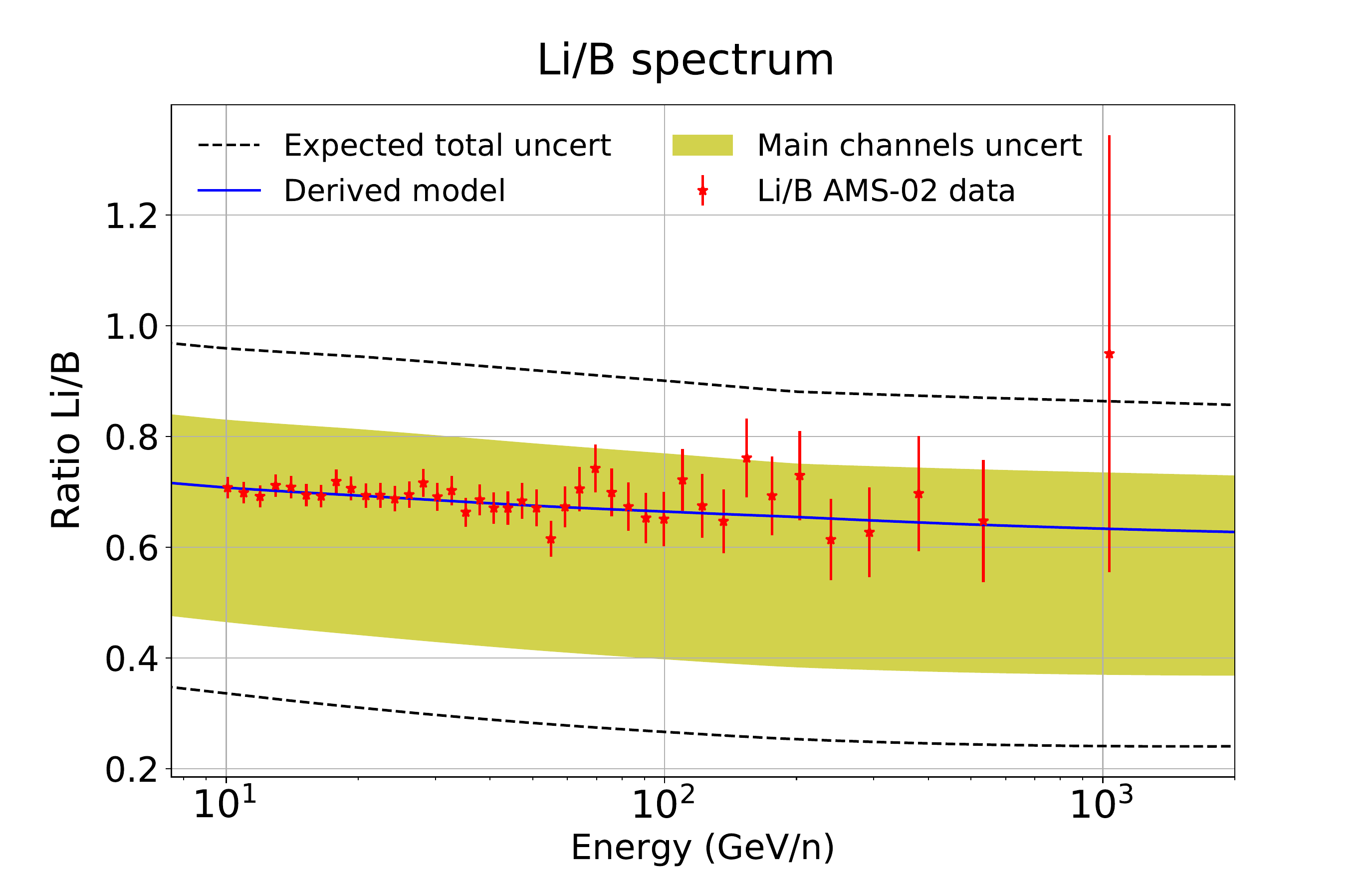}
\hspace{-0.6cm}
\includegraphics[width=0.51\textwidth,height=0.235\textheight,clip] {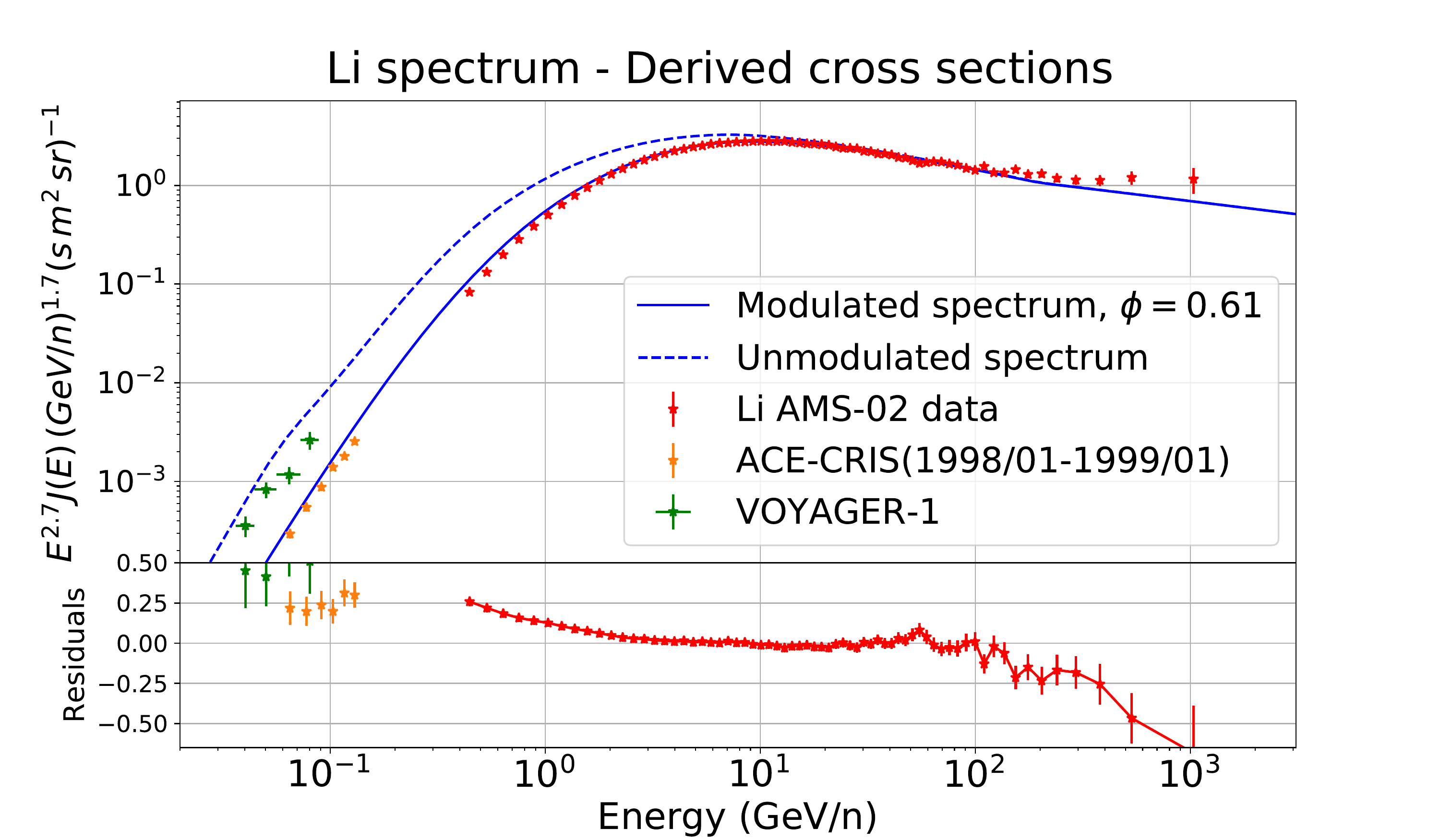} 

\caption{The secondary-over-secondary flux ratios involving Li, Be and B are shown in the left column. The yellow bands correspond to the uncertainties obtained using the bracketing cross section models for the main production channels. The bands between the black dashed lines correspond to the expected total uncertainties, obtained adding the contributions from all the minor production channels. The blue lines are obtained by simultaneously fitting the AMS-02 data, using the diffusion parameters obtained from the fit of the B/C ratio. 
In the right panels the B, Be and Li fluxes obtained with the cross sections derived from the fit procedure are compared with the AMS-02 data. The residuals are also shown. Data taken from \url{//https://lpsc.in2p3.fr/crdb/} \cite{Maurin_db1, Maurin_db2} and \url{https://tools.ssdc.asi.it/CosmicRays/} \cite{ssdc}.
}
\label{fig:secsec_Uncert}
\end{figure*}

To show the effects of the cross sections uncertainties on the secondary-over-secondary flux ratios, we have derived these ratios using the two bracketing models for the cross sections, demonstrating that the AMS-02 data lie between these two models. The results are shown in Figure~\ref{fig:secsec_Uncert}, where the yellow bands represent the values of the flux ratios within the two limiting models. The upper bound of each band corresponds to the situations in which the numerator is taken from the upper bracketing model and the denominator from the lower bracketing model (thus maximizing the ratio) and vice versa for the lower bound of the band. Furthermore, the expected full uncertainty bands on the flux ratios are also represented by black dashed lines. These bands are evaluated taking into account that the contribution from the $^{12}$C and $^{16}$O channels are about $77$\% of the total B flux and around $55$\% and $60$\% of the flux of Be and Li, respectively, according to tables IV, V and VI of~\cite{Genolini}, which implies to scale the fluxes in the bracketing models accordingly to have a rough estimation of the full bands. From this figure, we see that the full uncertainty band for the Li/Be ratio is nearly twice larger than that evaluated taking into account the uncertainty associated to their main production channels (i.e. with the two bracketing models), while for the other ratios it is around $70\%$ larger. In any case, we see that almost all the AMS-02 data lie within the bands obtained just varying the main production channels. These results confirms that primary components of B, Be or Li are not needed to explain the experimental data (although they could still be present).

Moreover, in this work, we have obtained a set of cross sections from a fit of the high energy part of the secondary-over-secondary flux ratios. In the analysis of the {\tt GALPROP} model, from Figure~\ref{fig:secsec_Webber}, it is obvious that an increase of Li cross sections is needed; however, just a change of this cross sections does not account for the discrepancies in the Be/B and Li/B flux ratios. This means that making variations just in the Be and Li cross sections (or whichever pair of secondaries) independently will never reproduce the three ratios at the same time, implying that a simultaneous adjustment of all the three fluxes is needed. Nevertheless, the simultaneous adjustment of the studied ratios does not have an exact solution (there can be degenerate solutions), so that finding the correct relation between the cross sections and the ratios is not straightforward. Therefore, we consider that the adjustment which implies a minimum rescaling from the original parametrisation should be favored.

The strategy followed here to simultaneously fit the high energy part of the secondary-over-secondary ratios consists of a progressive shift of the cross sections from the original parametrisation for each of the main channels,  until we find a configuration that matches the three ratios at the same time, thus searching for the configuration with less variations from the original parametrisations. Indeed, in this case we see that the fit convergence is reached when Li and Be cross sections are very close to touch the bracketing limits (Upper model for Li and Lower for Be), which means that there is very narrow margin to find other degenerate solutions. The fit yielded a renormalization of the cross sections of $\sim 5\%$ down for the B flux (constant for all the channels), of $\sim 18\%$ down for the Be flux ($\sim16\%$ in the $^{12}$C channels and $\sim 20\%$ in the $^{16}$O channels) and of $28\%$ up for the Li flux ($\sim 22\%$ for $^{12}$C channels and $\sim34\%$ $^{16}$O channels). The uncertainties on the scaling factors obtained for this derived model are of $4\%$ above $10 \units{GeV/n}$, as discussed in appendix~\ref{sec:appendixC}. These scaling factors can be compared to the post-fit nuisance parameters found in the combined LiBeB analysis of ref.~\cite{Weinrich_combined}, for their QUAINT model, taking into account that the scaling factors we compute are only related to the 
$^{12}$C and $^{16}$O channels. They found that the scaling parameters were of $\sim 4$\% down, $\sim 6$\% down and $\sim 12$\% up for B, Be and Li respectively. These are consistent with the values we found in our analysis, since we expect that the B scaling factor should be very similar to the value obtained when scaling all the channels and the scaling factors for Be and Li should be around half the values we obtained when scaling all the channels. As expected, the scaling needed in the cross sections of the B channels with respect to the original {\tt GALPROP} parametrisation is very small, while the Li and Be main channels need larger shifts, in agreement with the uncertainties that one would expect, as the impact of a change of these minor channels (very poorly known) for B production is very small. In turn, the impact of the main channels for Li and Be fluxes is approximately the same as that of the rest of the minor channels, which means larger level of uncertainty. The use of these scaling factors obtained from the combination of flux ratios among secondary CRs can help improving the description of the spallation cross sections used in CR propagation codes and may also be considered as a strategy to take care of the deficiencies on the description of the minor channels. This strategy has been already used for improving the estimation of the production of antiprotons from CR interactions with the interstellar gas in~\cite{ICPPA_Pedro}, obtaining good agreement with the AMS-02 antiprotons data.

As a final remark, from the lower plots of Figure~\ref{fig:secsec_Uncert} we see that the predicted fluxes of these three light secondary CRs 
reproduce the AMS-02 data at the same time 
in a broad energy region. The common discrepancy at high energies is due to the choice of the diffusion parameters, as already mentioned, which can be also the reason for the discrepancy of the Li flux below $2 \units{GeV/n}$ (although it is also related to the cross sections parametrisation). The influence of the halo size value in the low-energy part of the Be spectrum may explain its deviation from experimental data as commented in ref.~\cite{aguilar2018observation}.

In conclusion, we have demonstrated that we can reproduce the fluxes of B, Be and Li at the same time within the experimental cross sections uncertainties, with no need of including any primary extra source and that we can tune the spallation cross sections of production of secondary species to reproduce the secondary-over-secondary ratios and overcome the lack of knowledge we have in the normalization of the cross sections parametrisations. This is crucial in order to determine the diffusion parameters with better accuracy. In fact, with a correct balance of the secondary CRs, we could in principle also use Li and Be data in addition to B data to determine the diffusion parameters. This will be explored in a next paper.

\section{Implications of the cross sections on the halo size determination}
\label{sec:size}

The height of the galactic halo H plays a relevant role for the study of secondary-to-primary ratios involving leptons (the radiative energy loss rates are of the same order of magnitude as the reciprocal of the diffusion time), antiprotons and, as discussed above, unstable isotopes like $^{10}$Be. The usual way to constrain the halo size is by means of the study of the ratios of $^{10}$Be to the total Be flux or to the $^9$Be flux \cite{UlysesBe, moskaBe}. The spectrum of the isotope $^{10}$Be depends on an interplay between the diffusion time of the primary CRs ($\tau \propto E^{-\delta}$) and the decay time of this isotope (which is around $1.4 \units{My}$, as determined in ref.~\cite{chmeleff2010determination}). Other methods have been used to set constraints on H
from radio observations of lepton synchrotron emission \cite{bringmann2012radio}, from X-ray and gamma-ray studies \cite{biswas2018constraining}, from studies on CR leptons and other heavy nuclei \cite{moskalenko2000diffuse}.
\begin{figure*}[t]
\centering
\includegraphics[width=0.51\textwidth,height=0.25\textheight,clip] {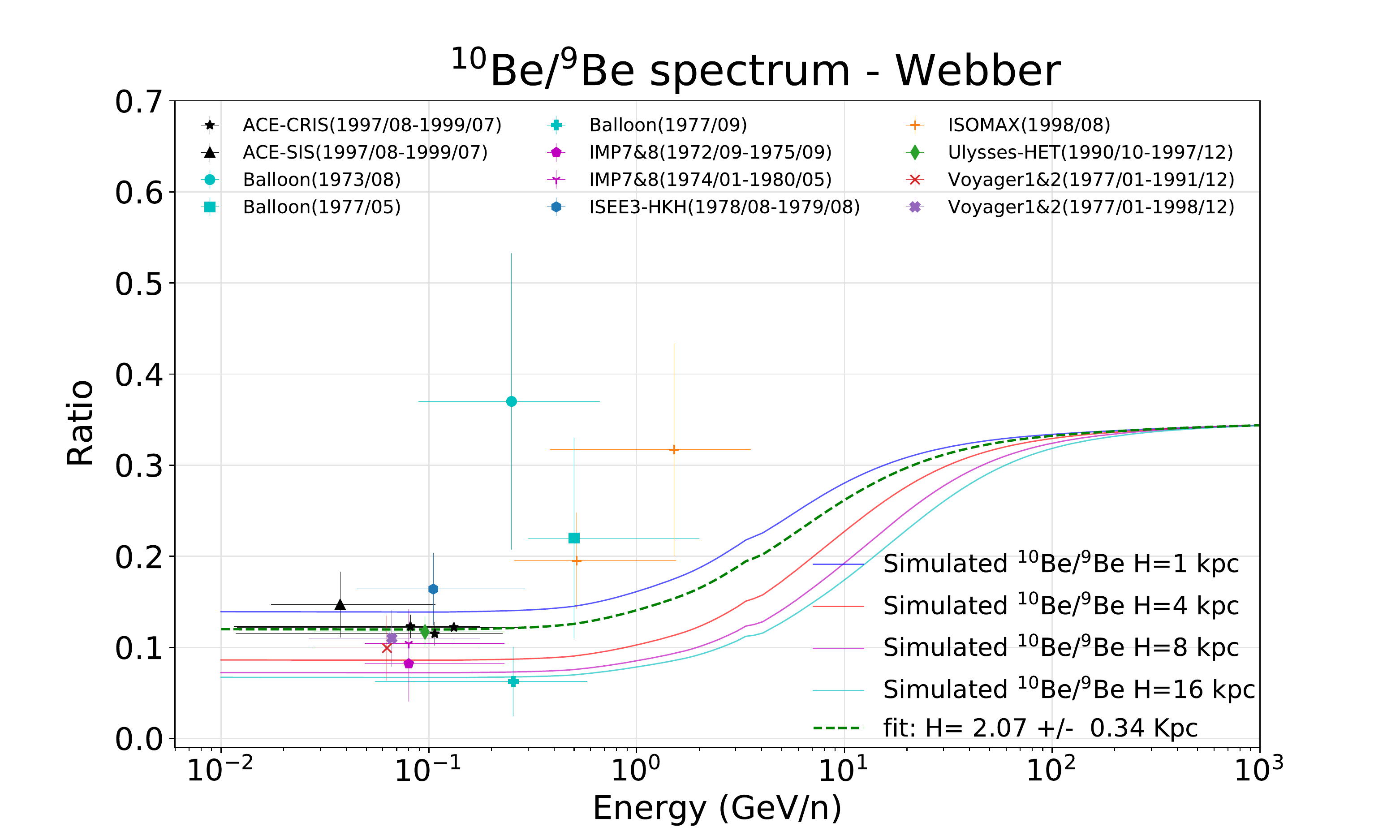} \hspace{-0.7cm}
\includegraphics[width=0.51\textwidth,height=0.25\textheight,clip] {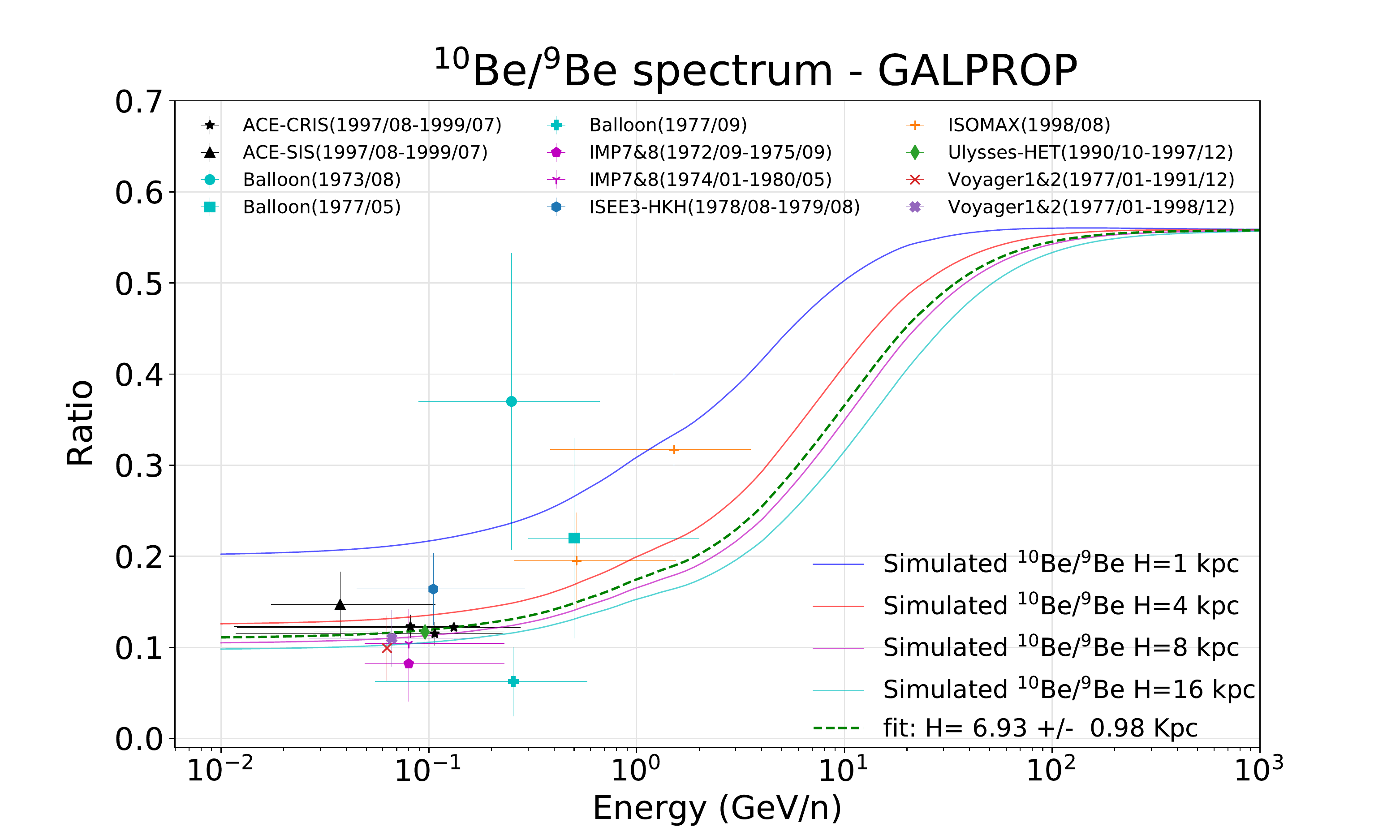} 

\includegraphics[width=0.51\textwidth,height=0.25\textheight,clip] {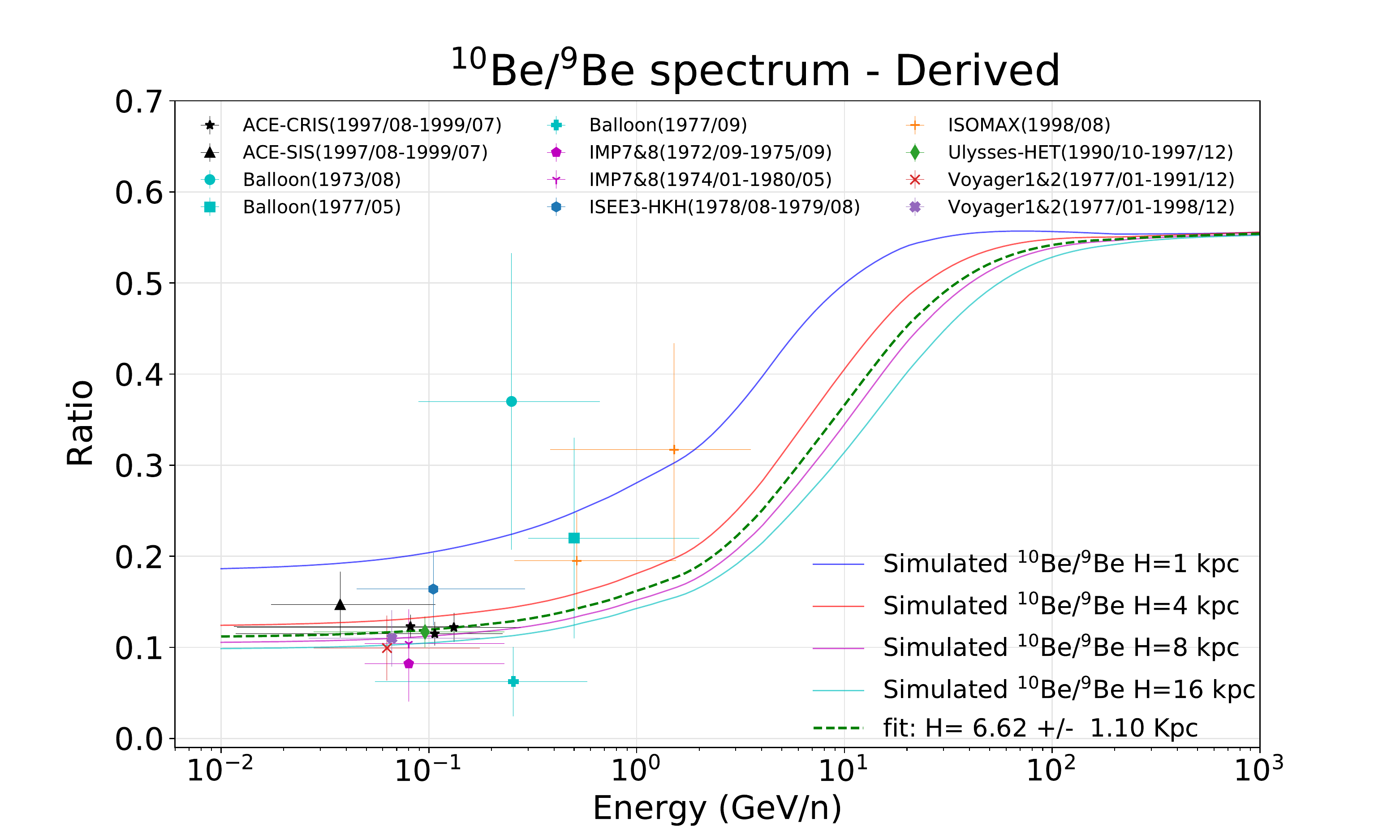} 
\hspace{-0.7cm}
\includegraphics[width=0.51\textwidth,height=0.25\textheight,clip] {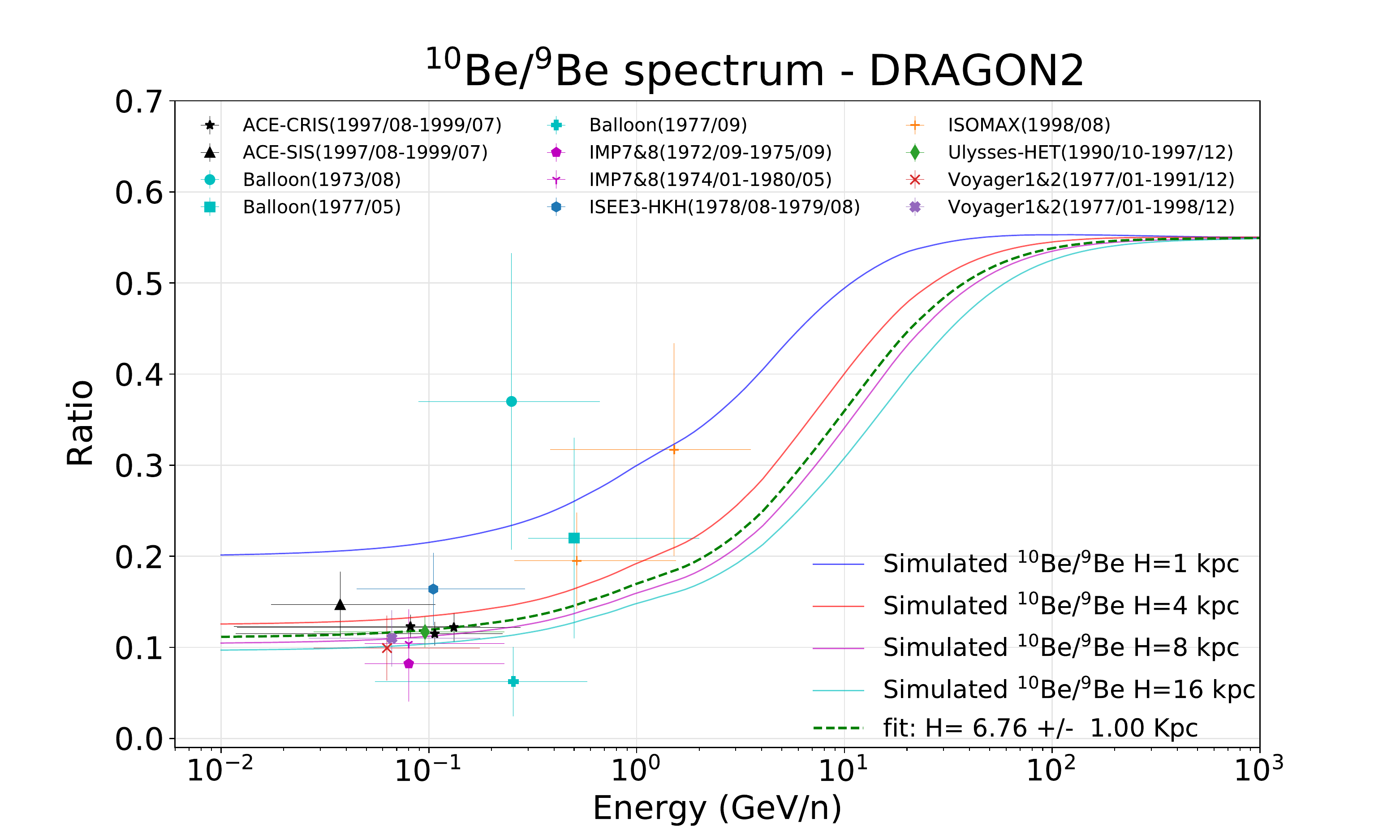}

\caption{$^{10}$Be/$^{9}$Be predicted flux ratios compared to all experimental data available and for every cross section parametrisation studied here. For each parametrisation, various simulations with different halo sizes are shown, along with the simulation yielding the best fit value. Data taken from \url{//https://lpsc.in2p3.fr/crdb/} \cite{Maurin_db1, Maurin_db2}.}
\label{fig:sizes}
\end{figure*} 

One of the obvious consequences of using secondary isotopes to determine any feature of the propagation is that the results will be highly influenced by the cross section model used. We have therefore decided to evaluate the halo size with four different cross section models: the {\tt DRAGON2}, {\tt Webber} and {\tt GALPROP} parametrisations and the cross sections derived in section \ref{sec:model_secratios}. For each model, we have implemented a fit of the halo size using the available data on the $^{10}$Be/$^{9}$Be flux ratios, from the ACE~\cite{ACEBe}, IMP~\cite{IMP1, IMP2}, ISEE~\cite{ISEE}, ISOMAX~\cite{Hams_2004}, Ulysses~\cite{UlysesBe} and Voyager~\cite{VoyagerMO} experiments. We have simulated different halo sizes from $1\units{kpc}$ to $16\units{kpc}$ and, for each halo size, we have evaluated the $^{10}$Be/$^{9}$Be flux ratios. As mentioned above, in each of simulation (i.e. for every halo size value) the diffusion parameters have been chosen to fit the B/C spectrum of the AMS-02 experiment. The flux ratios with halo sizes different from the tabulated values have been evaluated using a 2D interpolation with the tool \textit{RegularGridInterpolator} \footnote{\url{ https://docs.scipy.org/doc/scipy/reference/generated/ scipy.interpolate.RegularGridInterpolator.html}}. 
The error introduced by the interpolation is smaller than $1\%$ for every energy bin. The fit is performed with the \textit{$curve\_fit$} package from the \textit{$scipy.optimize$} library.

Figure~\ref{fig:sizes} shows the fit results for all the cross section models. For each cross section model the experimental data are compared with the predicted $^{10}$Be/$^{9}$Be flux ratios obtained for different halo sizes. As expected, the differences among the various predictions are larger at low energies, while the curves tend converge above $10\units{GeV/n}$. The curve corresponding to the halo sizes which yields the best fit is also shown.

\begin{figure}[!t]
\centering
\includegraphics[width=0.5\textwidth,height=0.25\textheight,clip] {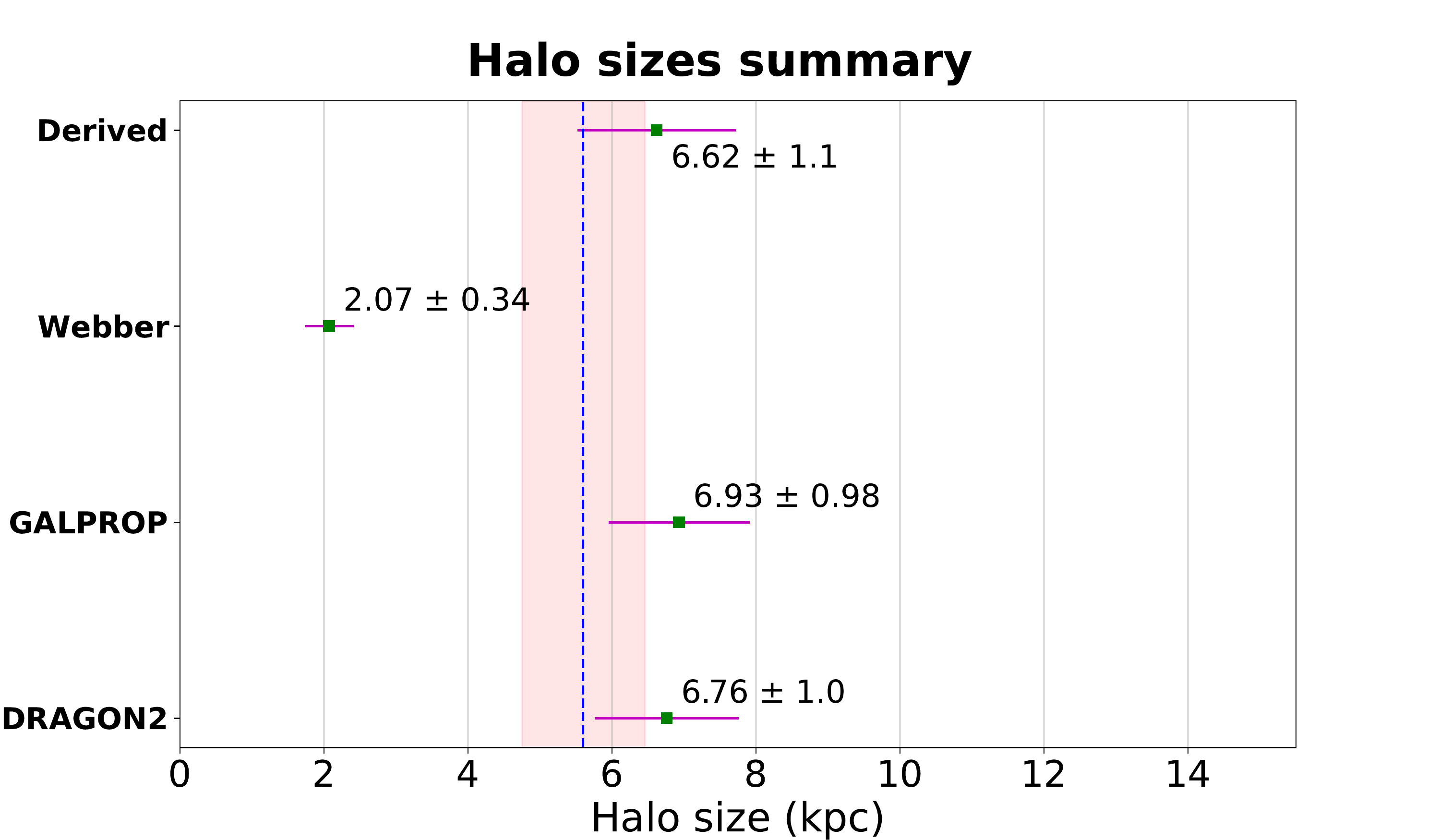} 
\vspace{0.1cm}
\caption{Summary of the results obtained for the fits of halo sizes to the $^{10}$Be/$^{9}$Be experimental data. Also the values obtained with the derived cross sections are shown for having a broader comparison. Error bars reflect only statistical uncertainties, as explained above. The dashed blue line represents the mean value of the halo best fit values and the red band the uncertainties of the mean ($1 \sigma$), calculated as the mean of the halo best fit values $\pm \sigma$. }
\label{fig:size_sum}
\end{figure}

The value of the halo size that best fits the experimental data for each model is accompanied by the uncertainties coming from its determination (i.e. the $1\sigma$ statistical uncertainties related to this fit). The fit results are summarized in Figure~\ref{fig:size_sum} together with a line indicating the mean halo size value. 
The error bar associated to the halo size fitted with the Webber cross section model is smaller than those associated to the halo sizes fitted with the other cross section models. This is due to the fact that, as can be seen from the curves in Figure~\ref{fig:sizes}, a change in the value of H leads to changes in the flux ratios which become smaller as H increases. Therefore, the smaller error bar found.

We see that the derived cross sections cast a value of the halo size very similar to the {\tt DRAGON2} and {\tt GALPROP} predictions, of around $6.6 \units{kpc}$ (very similar to the value found in refs.~\cite{CarmeloBeB, Weinrich_halo}, where the authors also employed the Be/B spectrum to make their predictions, getting a value of $\sim6\units{kpc}$ and $5^{+3}_{-2}\units{kpc}$, respectively). The small value found for the {\tt Webber} parametrisations can be explained having a look to the cross sections around hundreds $\units{MeV}$. For this parametrisation, the most important channels generally show a good agreement with cross sections data, but in the case of the $^{12}$C $\rightarrow$ $^{10}$Be channel, the discrepancy is considerable. Therefore, from this underestimation of the cross section, the halo size prediction is expected to be also underestimated (less $^{10}$Be production implies a lower halo size to reduce the fraction of decaying $^{10}$Be nuclei). 

In any case, the value obtained using the {\tt Webber} parametrisations seems to be out of the standard picture of a halo size between $3$ and $10 \units{kpc}$ height found in measurements of diffuse gamma-ray background \cite{zaharijas2012fermi} and with other radio observations \cite{di2013cosmic,beuermann1985radio,orlando2013galactic}. The mean of the best fit values obtained using the various cross sections shown in Fig.~\ref{fig:size_sum} (including the determination of the halo size using the {\tt Webber} parametrisation) is $5.6^{+0.86}_{-0.85} \units{kpc}$, and the value obtained using only the most updated cross sections is $6.8 \pm 1\units{kpc}$.

\section{Conclusions}
\label{sec:conc}
We are living in a very exciting time for CR physics thanks to the high quality data provided by the AMS-02 experiment, which allows making precise studies on the nature of CRs and their propagation throughout the Galaxy. However, while the astrophysical information seems to be accurate enough to face new physical challenges, the information about the spallation cross section, which is the other key element in the CR puzzle, is still largely incomplete. Since the study of secondary-over-primary CRs is the best tool to test the diffusion coefficient parametrisations, we need to have high precision on the secondary CRs fluxes and in their production (inclusive) cross sections.

The amount of channels that matter in the spallation network and the difficulty to perform the measurements lead to parametrisations that extend the energy range of the actual measurements and expand to all important nuclei (even those for which experimental data are missing). While we have some insight in the shape of cross sections as a function of energy, the normalization is still somewhat uncertain. 

In this paper we have discussed different cross sections parametrisations and we have investigated their direct impact in the fluxes of Li, Be and B. We have highlighted the immediate relation of the secondary-over-secondary ratios with the spallation cross sections used. We have shown that predictions from different cross sections models can differ in the amount of Be and Li fluxes at a level from $20\%$ to more than $40\%$. We also emphasize that the uncertainties in the determination of secondary CR fluxes are in general quite large, although those on the B flux seem to be remarkably smaller, since its production is essentially due to the contribution of C and O channels for which more experimental data are available. In addition, we have demonstrated the good performance achieved in the simulations when using the default {\tt DRAGON2} cross sections model, being able to reproduce every observable within very small errors. 

It has been shown that there is no need of invoking primary sources of secondary CRs, as the cross sections uncertainties can largely account for the discrepancies found with respect to data. To better show this, two models of cross sections have been considered as limiting cases. These bracketing models represent minimum and maximum credible renormalizations of the {\tt GALPROP} model of cross sections, maintaining the same energy dependence. 
We have investigated how the secondary-over-secondary flux ratios would behave under the combination of the bracketing models to test whether, between these two limiting cases, there is a space of cross sections that can reproduce all secondary CRs above $10 \units{GeV/n}$ at the same time. A set of cross sections has been derived by matching these flux ratios at high energy, demonstrating that they are an excellent tool to constrain and even adjust the cross sections parametrisations. This derived cross sections model balances the deficiencies on the description of those channels which are very poorly known and gives some insight on how much our cross sections network is biased for every secondary species. We showed that, while the B production channels should not be renormalized by more than $5\%$ to reproduce the ratios, the beryllium and lithium ones needed an overall renormalization of around $18\%$ and $28\%$ (in average), respectively. 

In this way we have been able to combine the information arising from secondary CRs to mitigate the systematic uncertainties related to spallation cross sections. We argue that this combined tuning of the normalization on a cross sections parametrisation to reproduce the secondary-over-secondary flux ratios serves to improve the determination of the diffusion coefficient parameters as well, and it is important  to obtain predictions which are consistent with all the observables at the same time.

Finally, a study of the effects on the halo size on the secondary-over-secondary flux ratios for each cross section model has been performed. In particular, the ratio $^{10}$Be/$^{9}$Be was studied for each of the models, as the $^{10}$Be isotope has a lifetime of the order of the diffusion time of CRs in the Galaxy. 
This allowed the discussion on the repercussion of the cross sections on the determination of the halo size too. The determination of the halo size from the most updated parametrisations gives a mean value around $6.8\pm 1.0 \units{kpc}$, which is in agreement with most of the values obtained in other works. 

In a next work, the diffusion parameters will be studied in a combined analysis of the secondary-over-primary and secondary-over-secondary flux ratios of B, Be and Li including nuisance factors to allow adjustments on the normalization of the cross sections parametrisations studied.

\vspace{0.5cm}
\acknowledgments
We remark the crucial help of Daniele Gaggero in the development of the paper and his examinations during the evolution of the ideas commented here and also in the implementation of the preliminary version of the {\tt DRAGON2} code. Special acknowledgements to Carmelo Evoli for his invaluable advice and comments during all the work process regarding the {\tt DRAGON2} preliminary version used here and the manuscript elaboration.
Many thanks to the instituto de física teórica (IFT) in Madrid for hosting Pedro De la Torre for a long stay there and specially to the DAMASCO (DArk MAtter AStroparticles and COsmology) group for their support and valuable conversations related to this work.
This work has been carried out using the RECAS computing infrastructure in Bari (\url{https://www.recas-bari.it/index.php/en/}). A particular acknowledgment goes to G. Donvito and A. Italiano for their valuable support. 

\bibliographystyle{apsrev4-1}
\bibliography{biblio}

\newpage
\appendix

\section{\large Cross sections for the main channels}
\label{sec:appendixA}

In this appendix we are showing a comparison between the different cross section models used in the study and the experimental data in the most important reaction channels for the light secondary CRs Li, B, Be. Experimental data are taken from various experiments and authors: some can be found in EXFOR (Experimental Nuclear Reaction Data)\footnote{\url{https://www-nds.iaea.org/exfor/exfor.dhtm}} others in the {\tt GALPROP} database of cross sections ($isotope\_cs.dat$) and the rest come from various publications and experiments (Bodemann1993, Davids1970, Fontes1977, Korejwo1999, Korejwo2002, Moyle1979, Olson1983, Radin1979, Read1984, Roche1976, W90, W98a and Zeitlin2011). This data is available upon request. More information about the references used can be found in section 5 of ref.~\cite{DRAGON2-2} and mainly in the appendix of ref.~\cite{CarmeloBlasi}.

\vskip 0.8cm
\textbf{Production of B isotopes from $^{12}$C and $^{16}$O}

\begin{figure*}[!ht]
\label{fig:XSB}
\centering
\includegraphics[width=0.48\textwidth,height=0.23\textheight,clip] {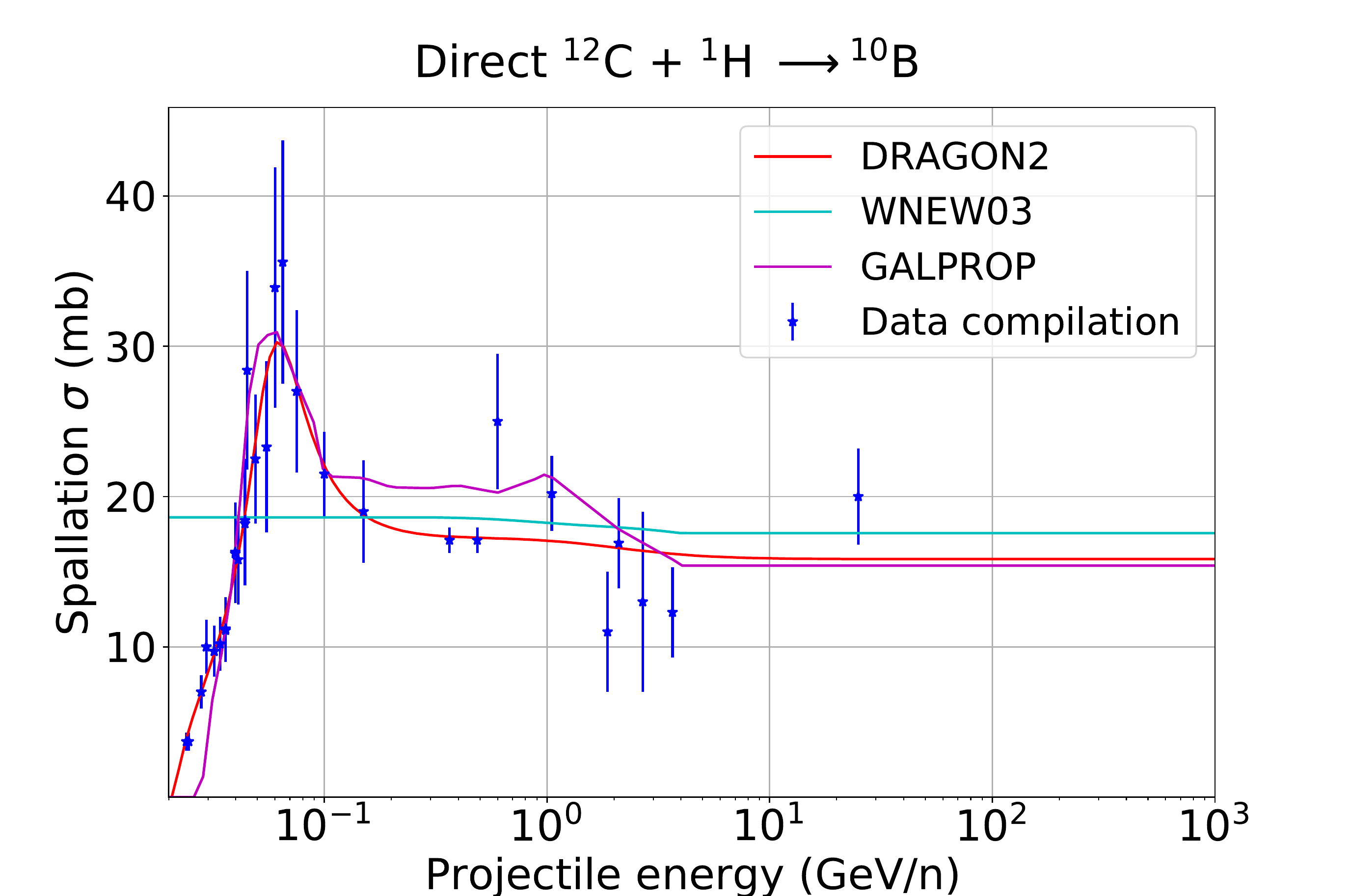} 
\includegraphics[width=0.48\textwidth,height=0.23\textheight,clip] {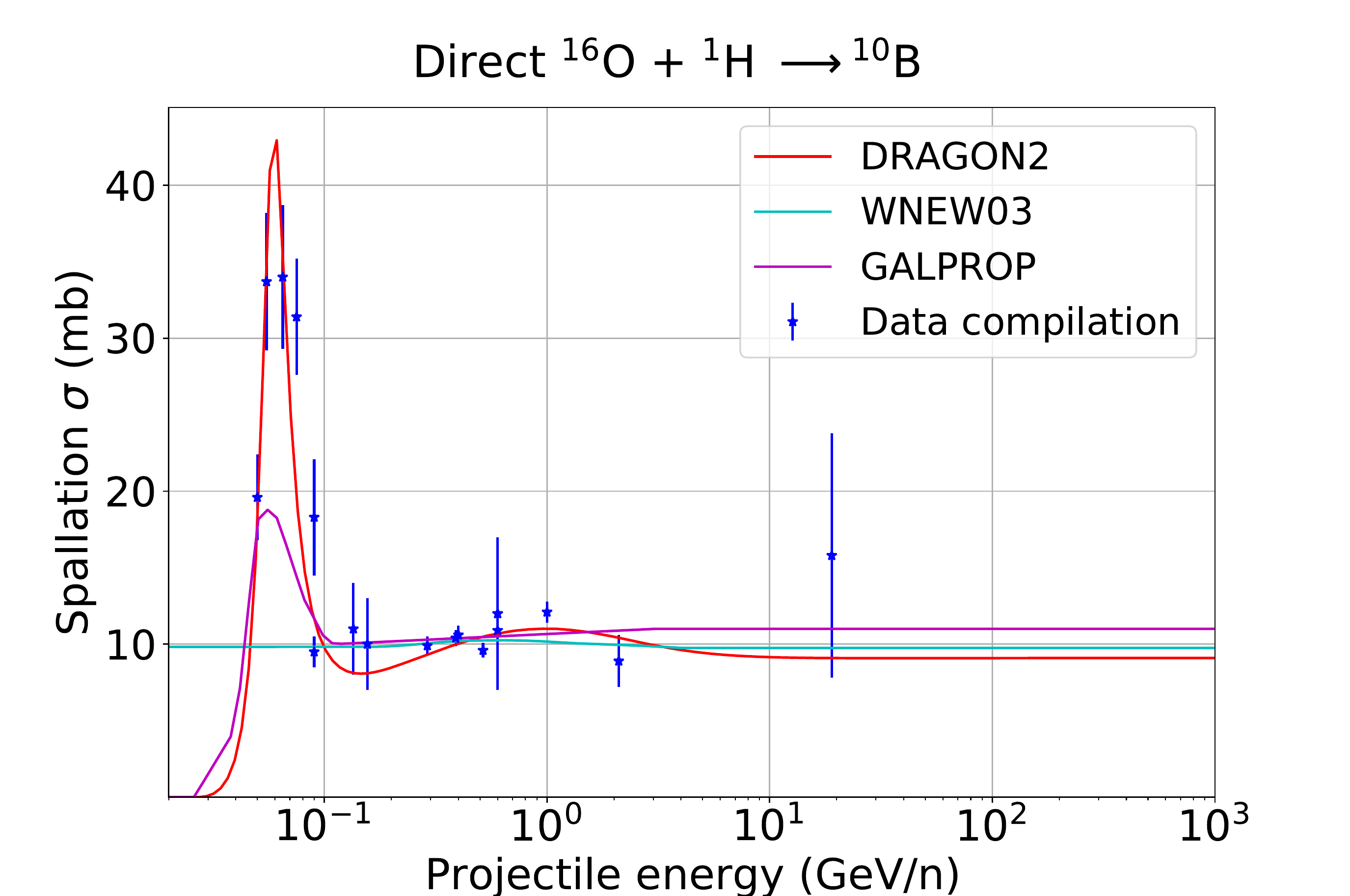} 
\includegraphics[width=0.48\textwidth,height=0.23\textheight,clip] {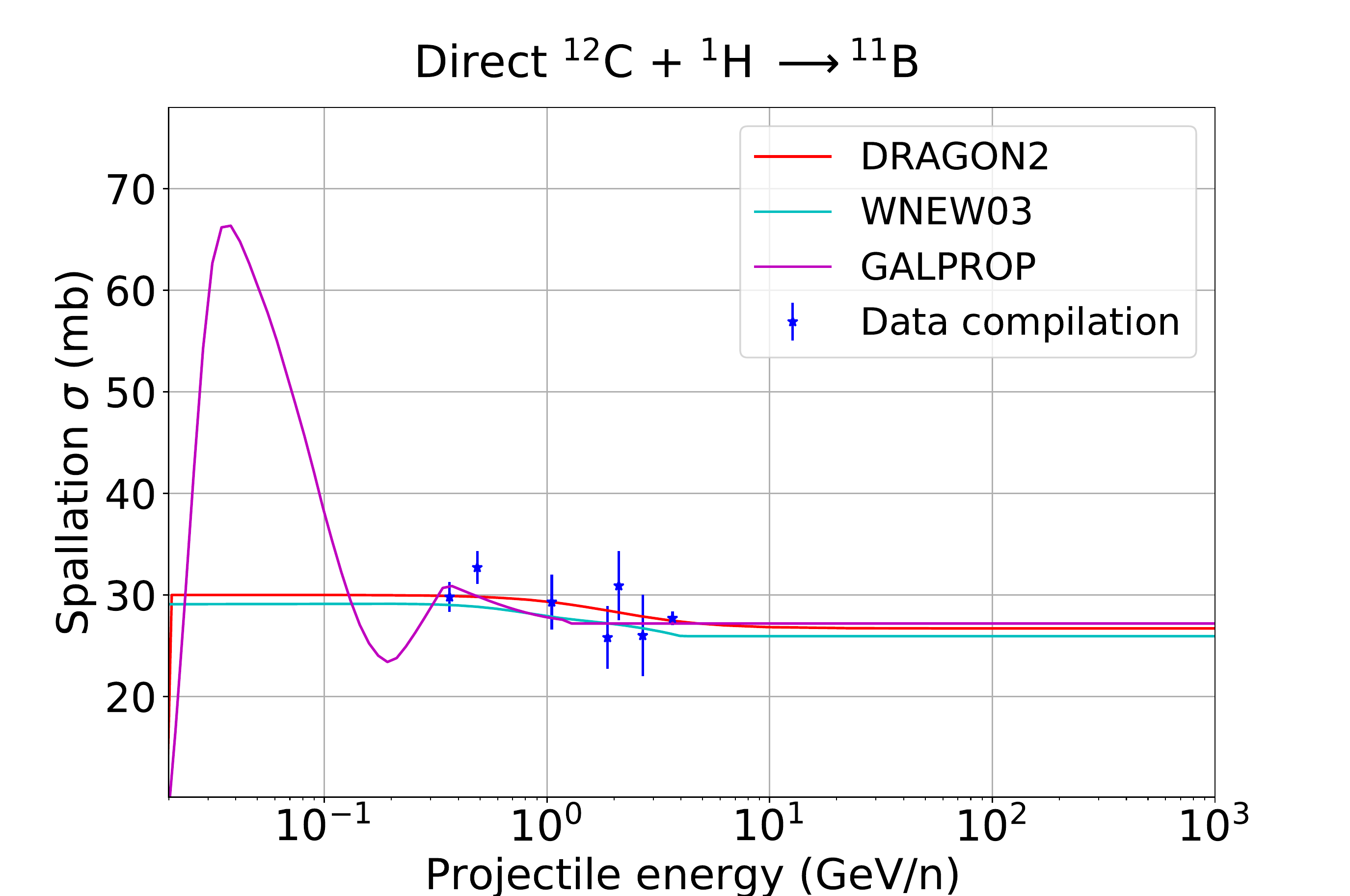} 
\includegraphics[width=0.48\textwidth,height=0.23\textheight,clip] {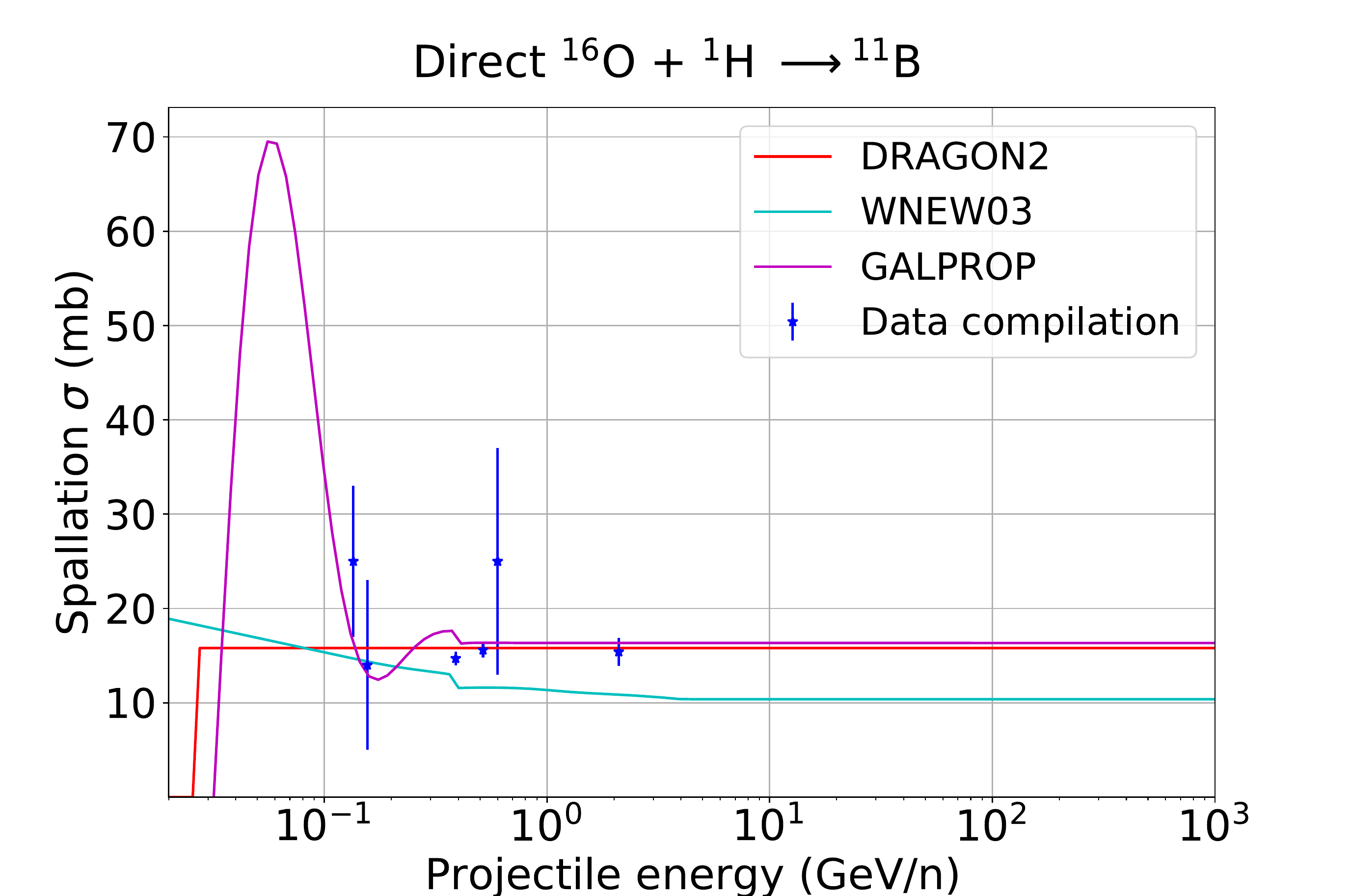} 
\caption{Cross sections compared to available experimental data for the production of $^{10}$B (top) and $^{11}$B (bottom) coming from $^{12}$C (left) and $^{16}$O (right).}
\end{figure*} 

\newpage
\textbf{Production of Be isotopes from $^{12}$C and $^{16}$O}

\begin{figure*}[!th]
\label{fig:XSBe}
\centering
\includegraphics[width=0.48\textwidth,height=0.22\textheight,clip] {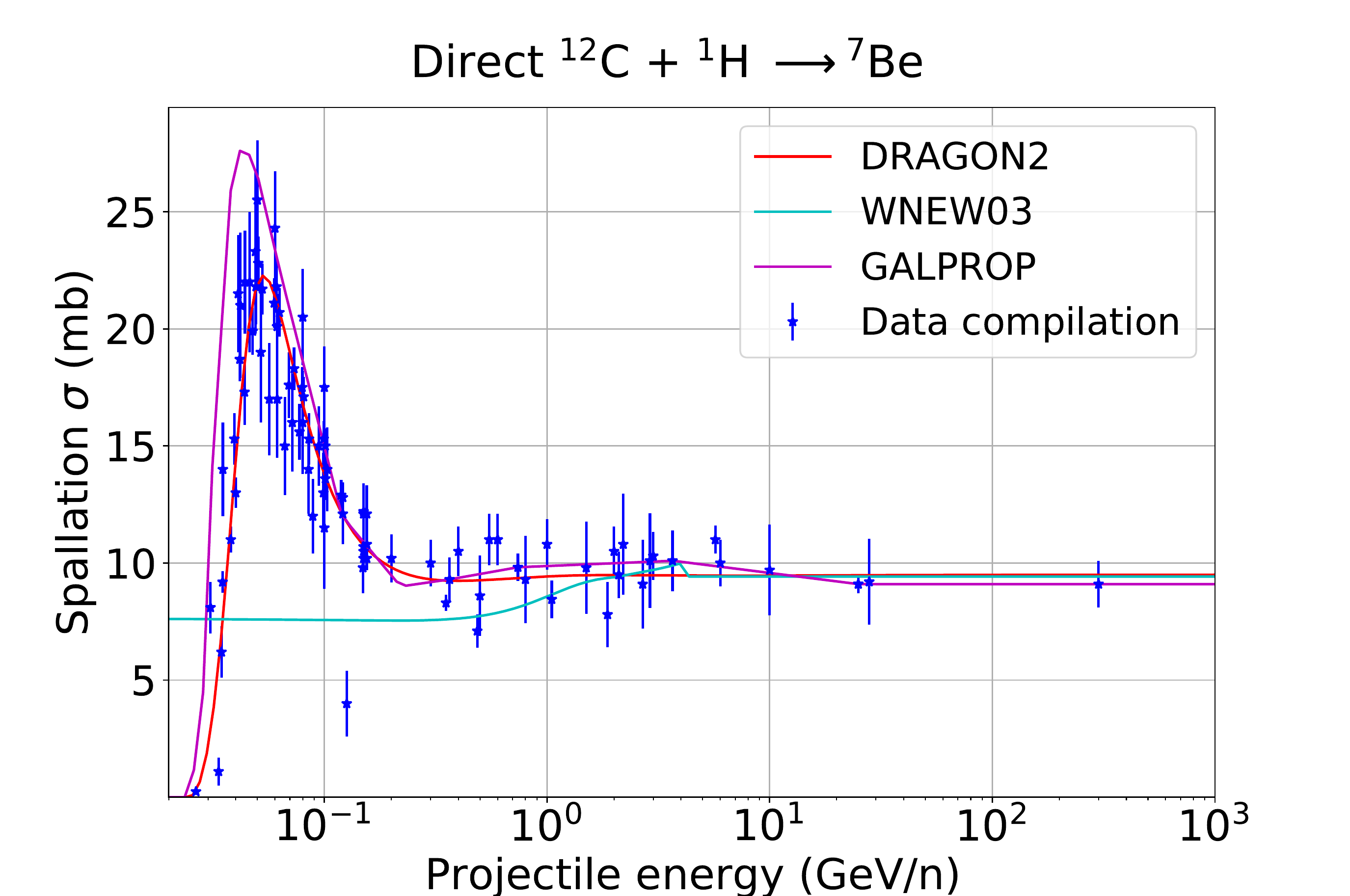}
\includegraphics[width=0.48\textwidth,height=0.22\textheight,clip] {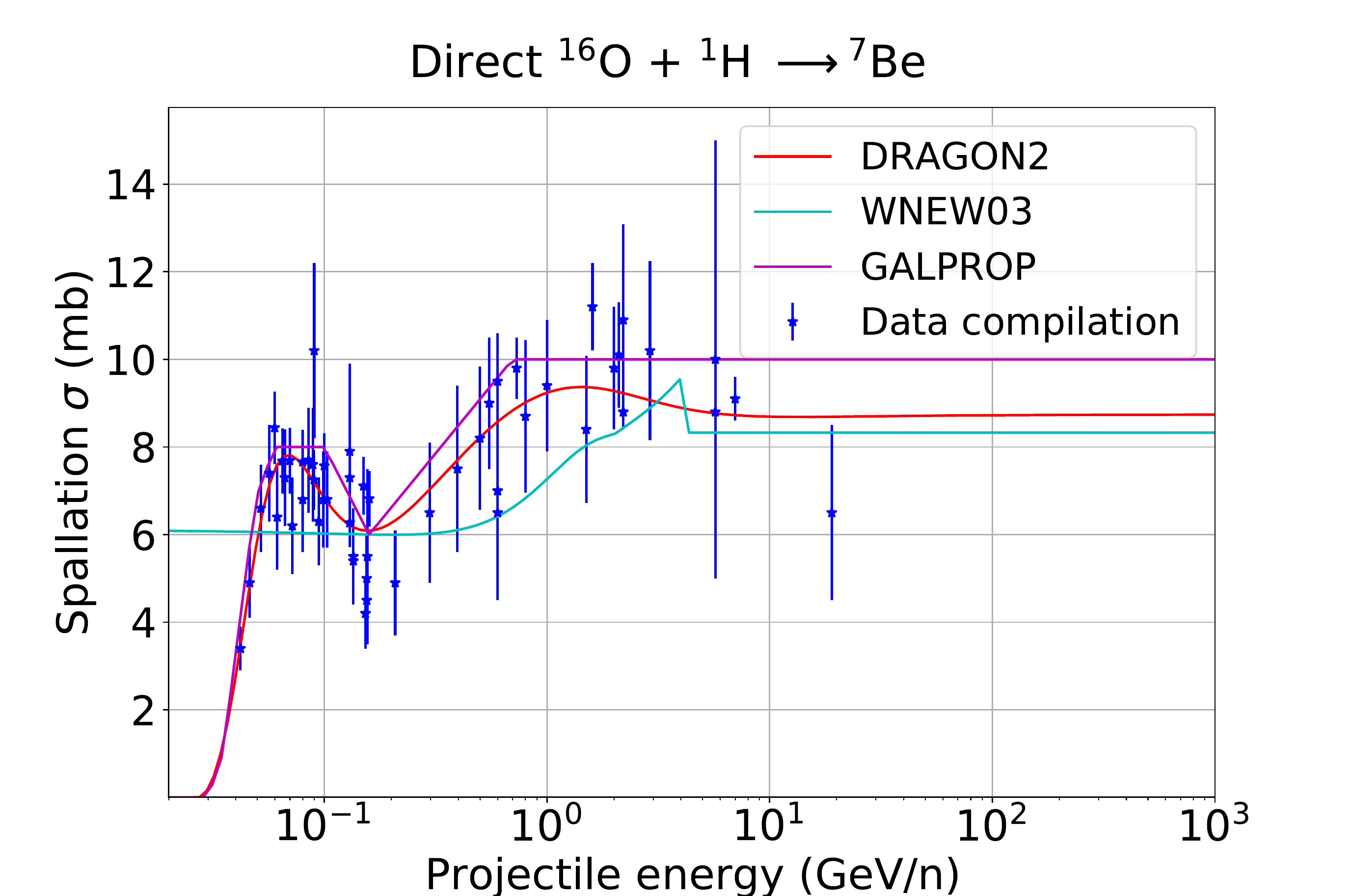}

\includegraphics[width=0.48\textwidth,height=0.22\textheight,clip] {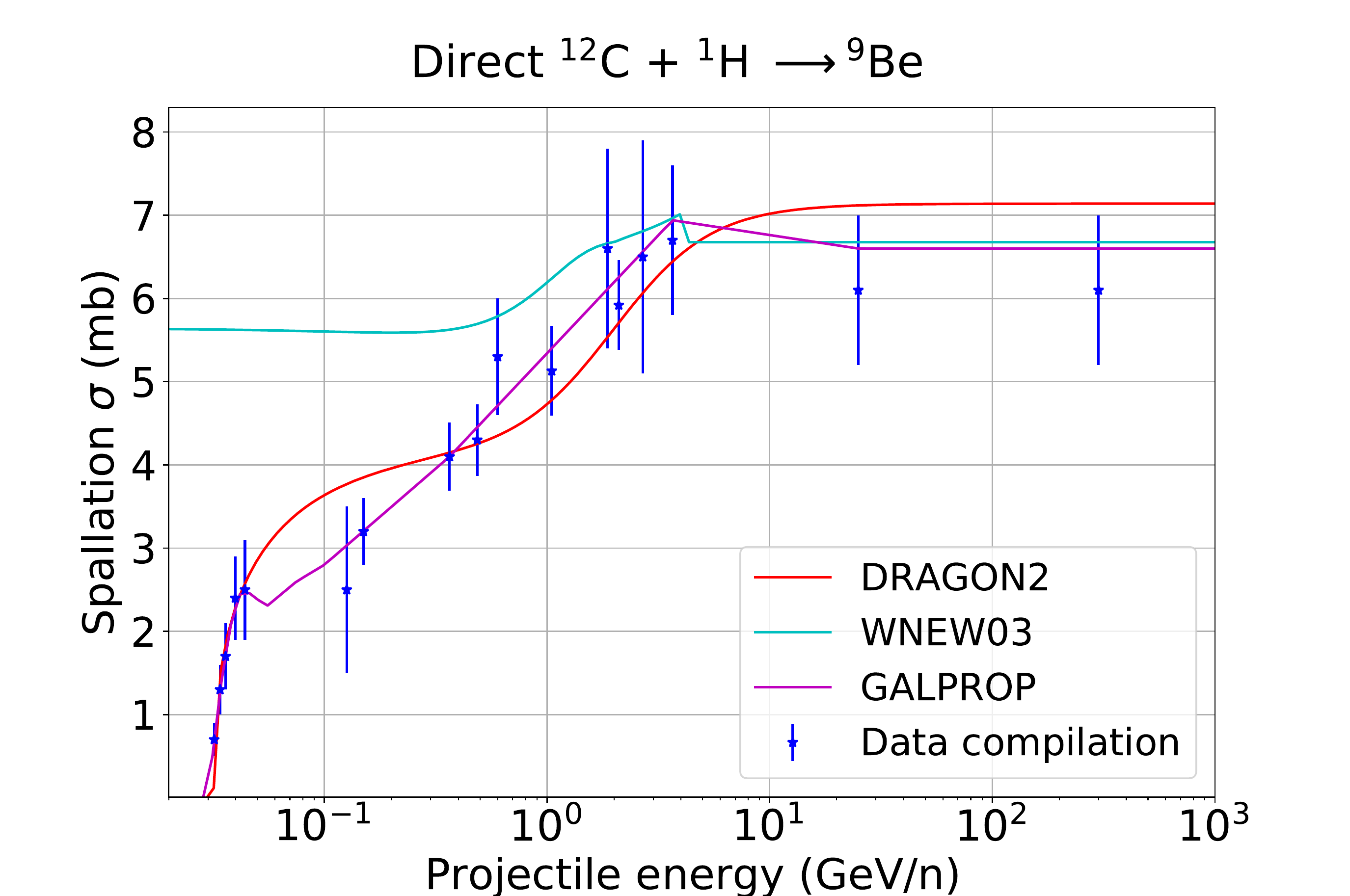}
\includegraphics[width=0.48\textwidth,height=0.22\textheight,clip] {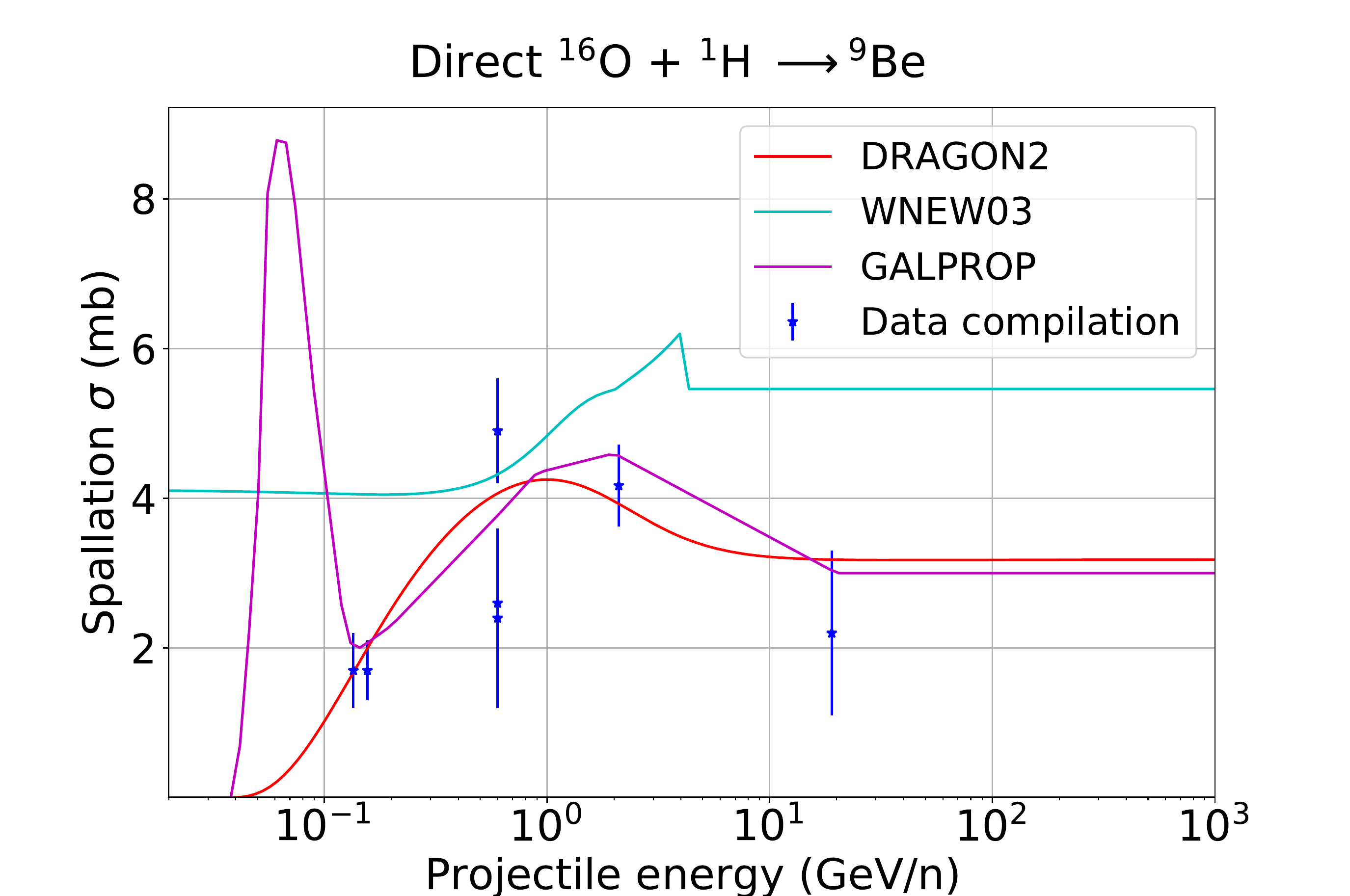}

\includegraphics[width=0.48\textwidth,height=0.22\textheight,clip] {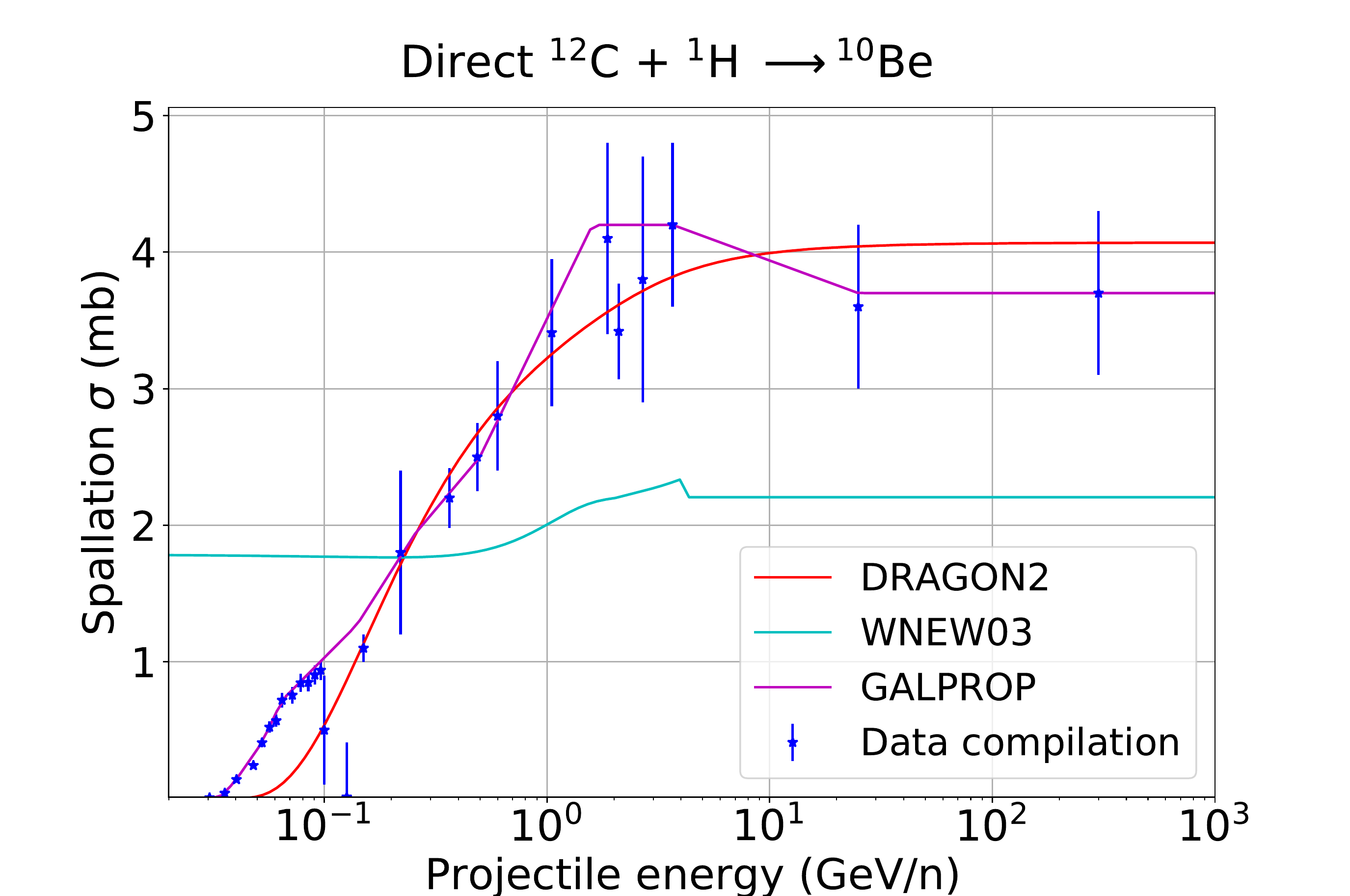} 
\includegraphics[width=0.48\textwidth,height=0.22\textheight,clip] {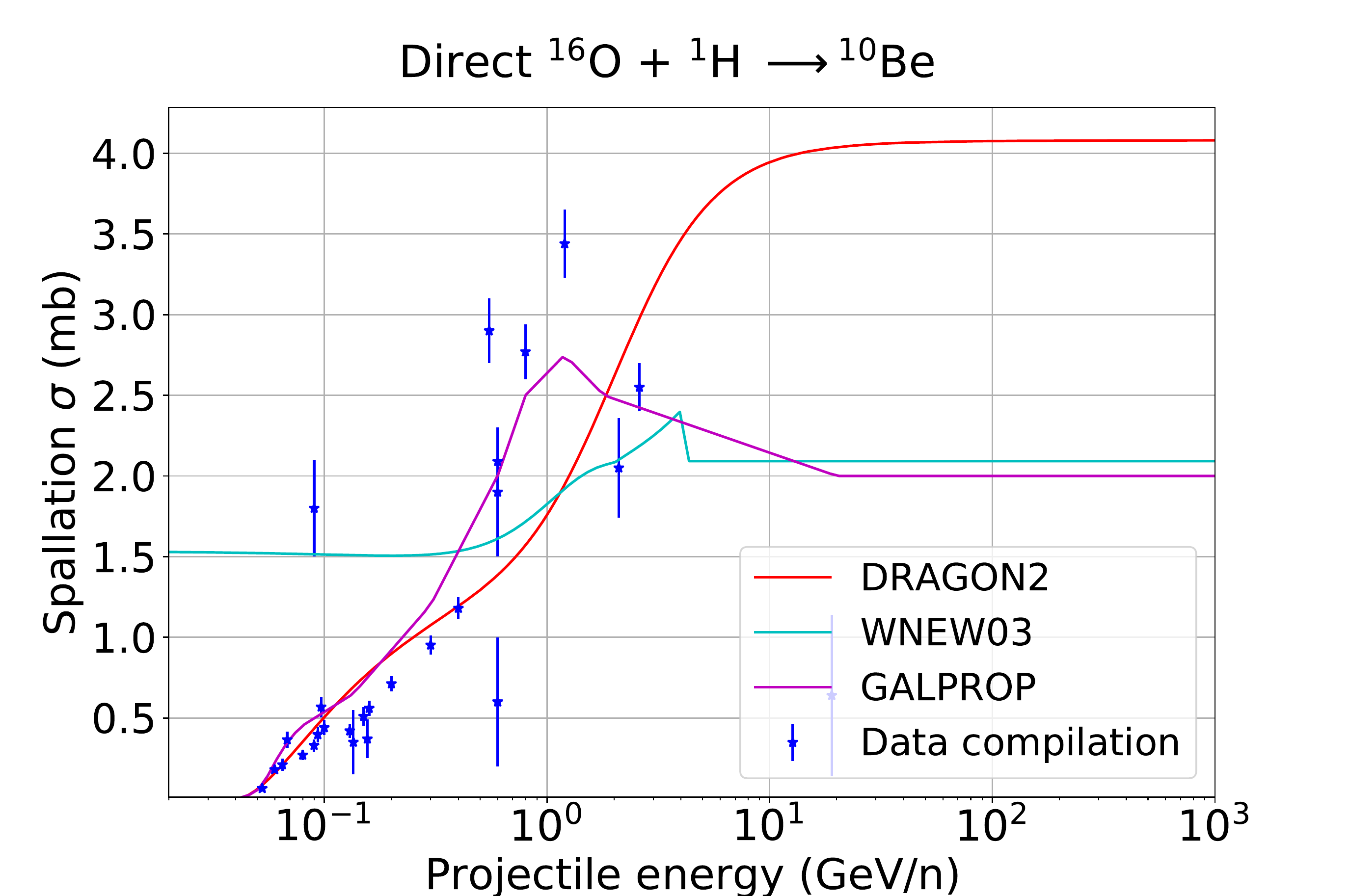} 
\caption{Cross sections compared to available experimental data for the production of $^{7}$Be (top), $^{9}$Be (middle) and $^{10}$Be (bottom) coming from $^{12}$C (left) and $^{16}$O (right).}
\end{figure*} 

\newpage
\textbf{Production of Li isotopes from $^{12}$C and $^{16}$O}

\begin{figure*}[!th]
\label{fig:XSLi}
\centering
\includegraphics[width=0.48\textwidth,height=0.22\textheight,clip] {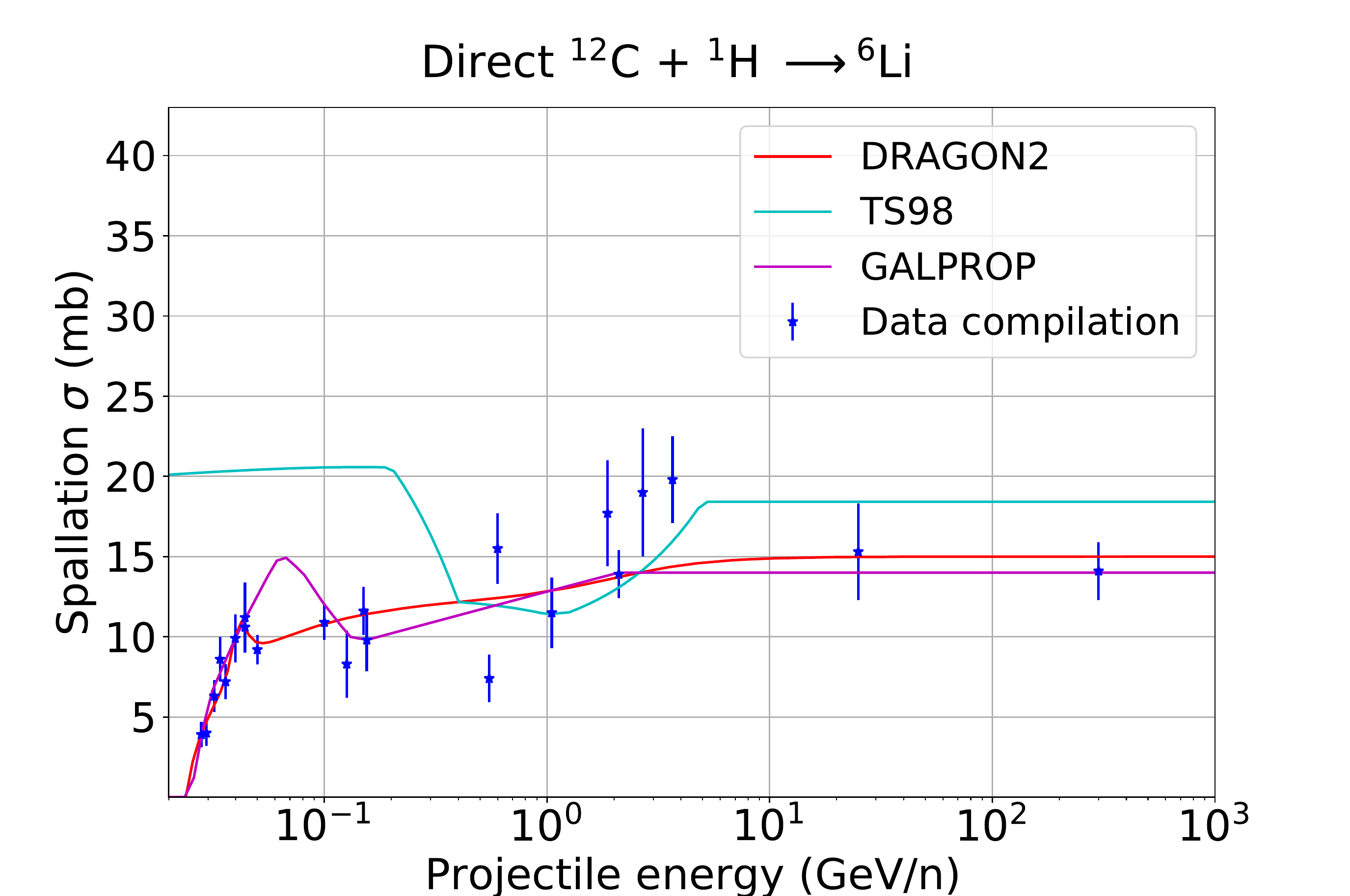}
\includegraphics[width=0.48\textwidth,height=0.22\textheight,clip] {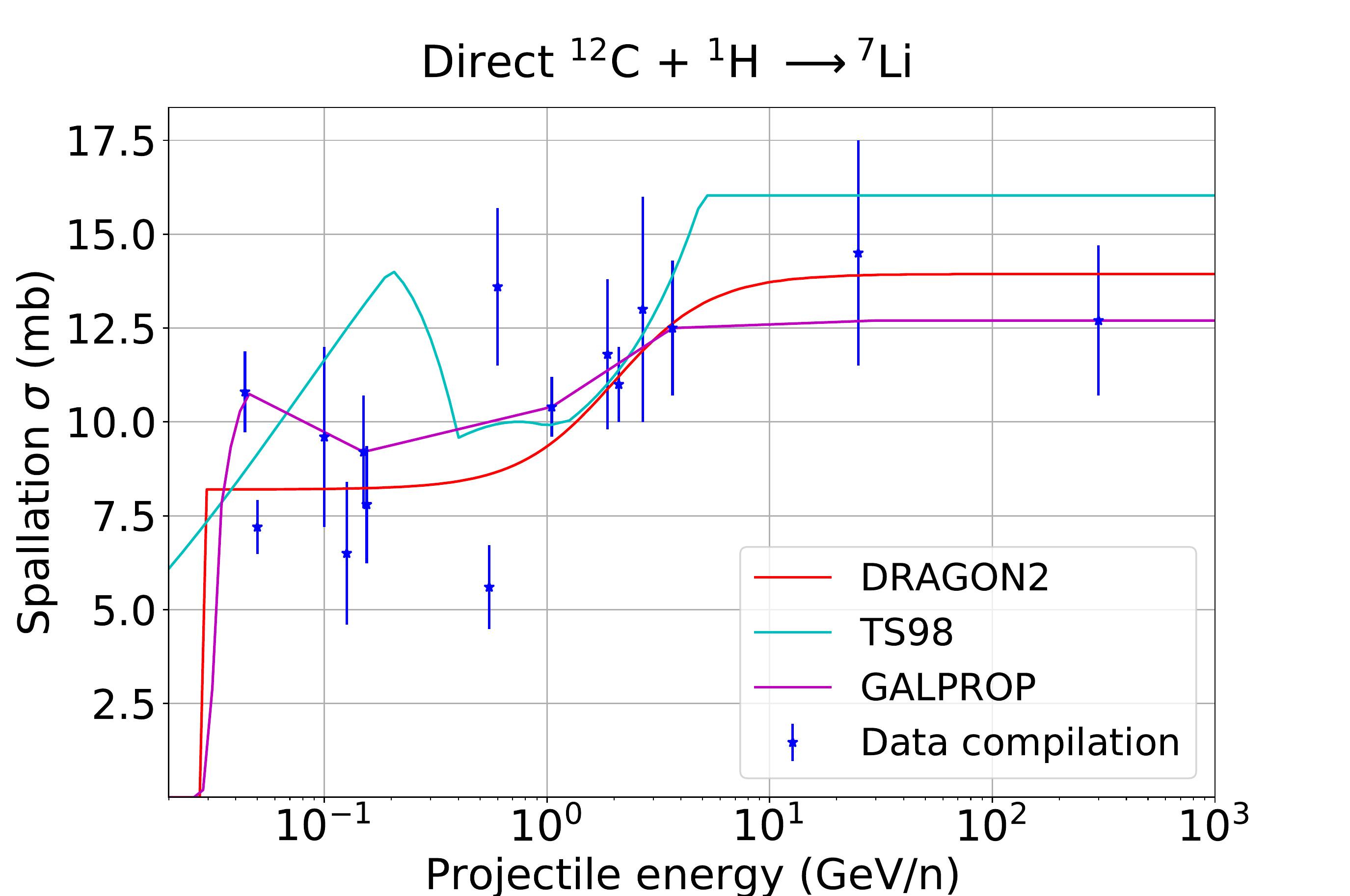} 
\includegraphics[width=0.48\textwidth,height=0.22\textheight,clip] {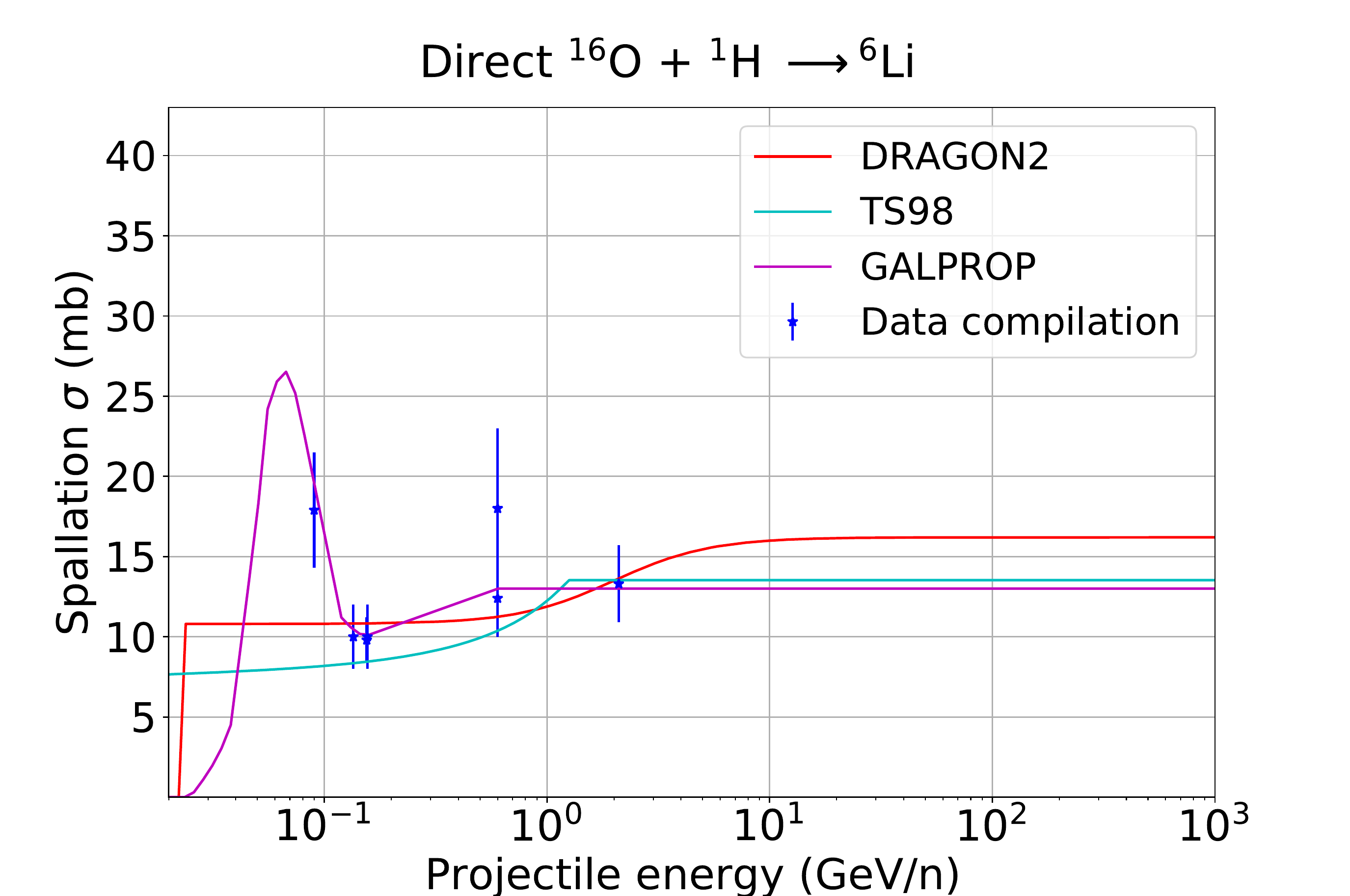}
\includegraphics[width=0.48\textwidth,height=0.22\textheight,clip] {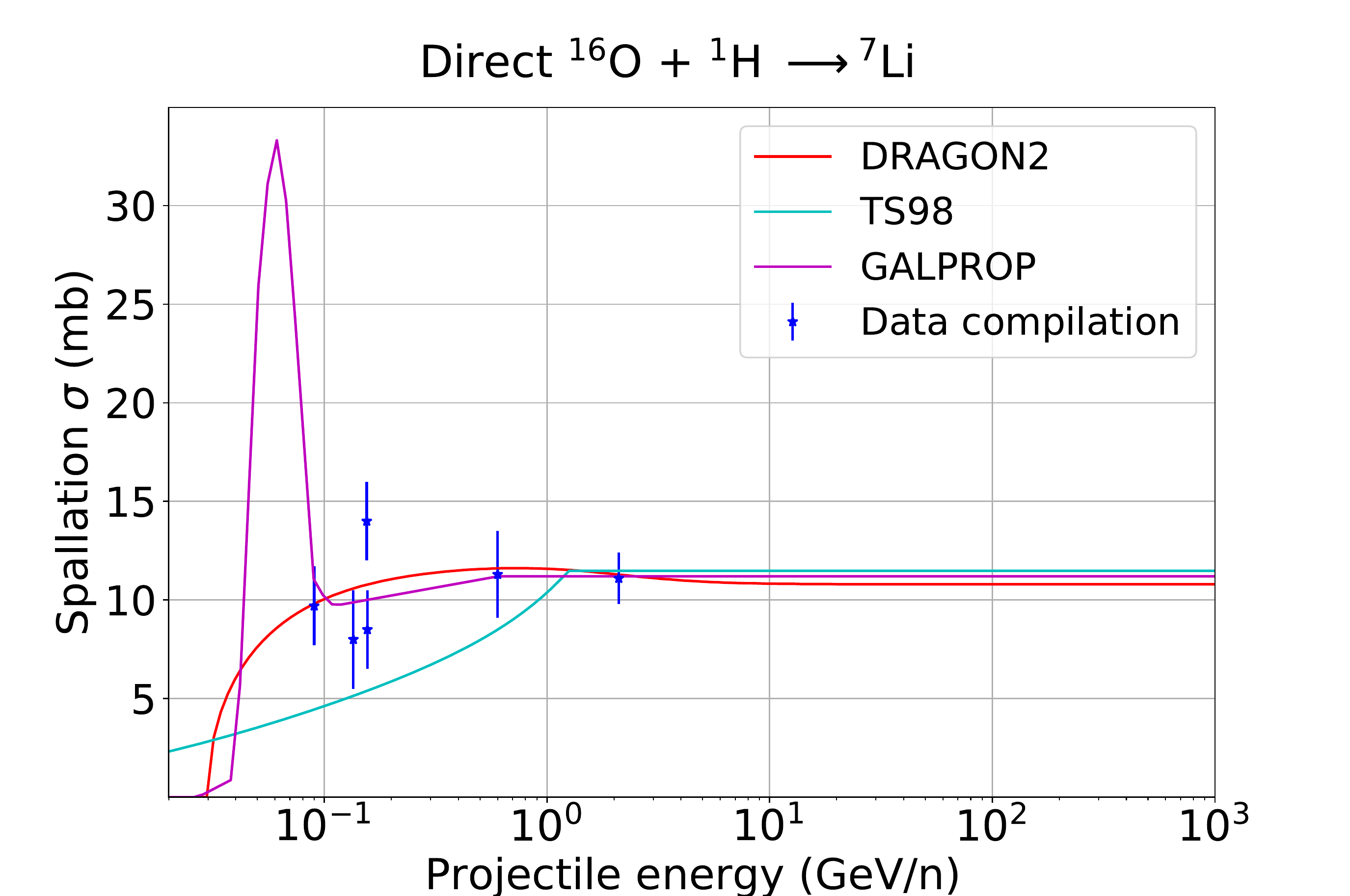} 
\caption{Cross sections compared to available experimental data for the production of $^{6}$Li (top) and $^{7}$Li (bottom) coming from $^{12}$C (left) and $^{16}$O (right).}
\end{figure*} 

\newpage
\section{\large Primary spectra}
\label{sec:appendixB}

In this section we show the spectra of the main primary CRs involved in the generation of the secondary CRs Li, Be and B. 

\begin{figure*}[hb]
\centering
\includegraphics[width=0.49\textwidth,height=0.267\textheight,clip] {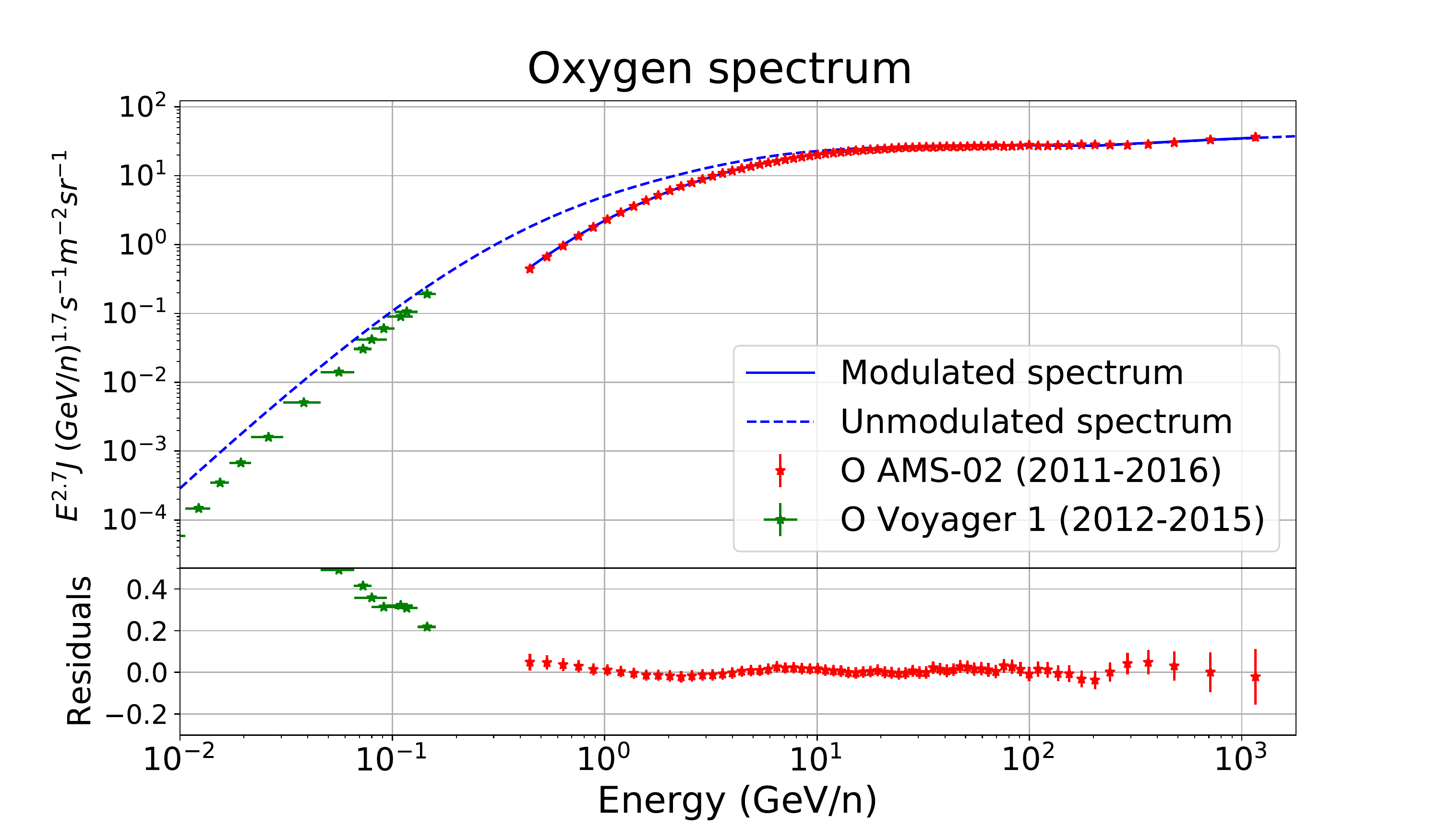} 
\includegraphics[width=0.49\textwidth,height=0.267\textheight,clip] {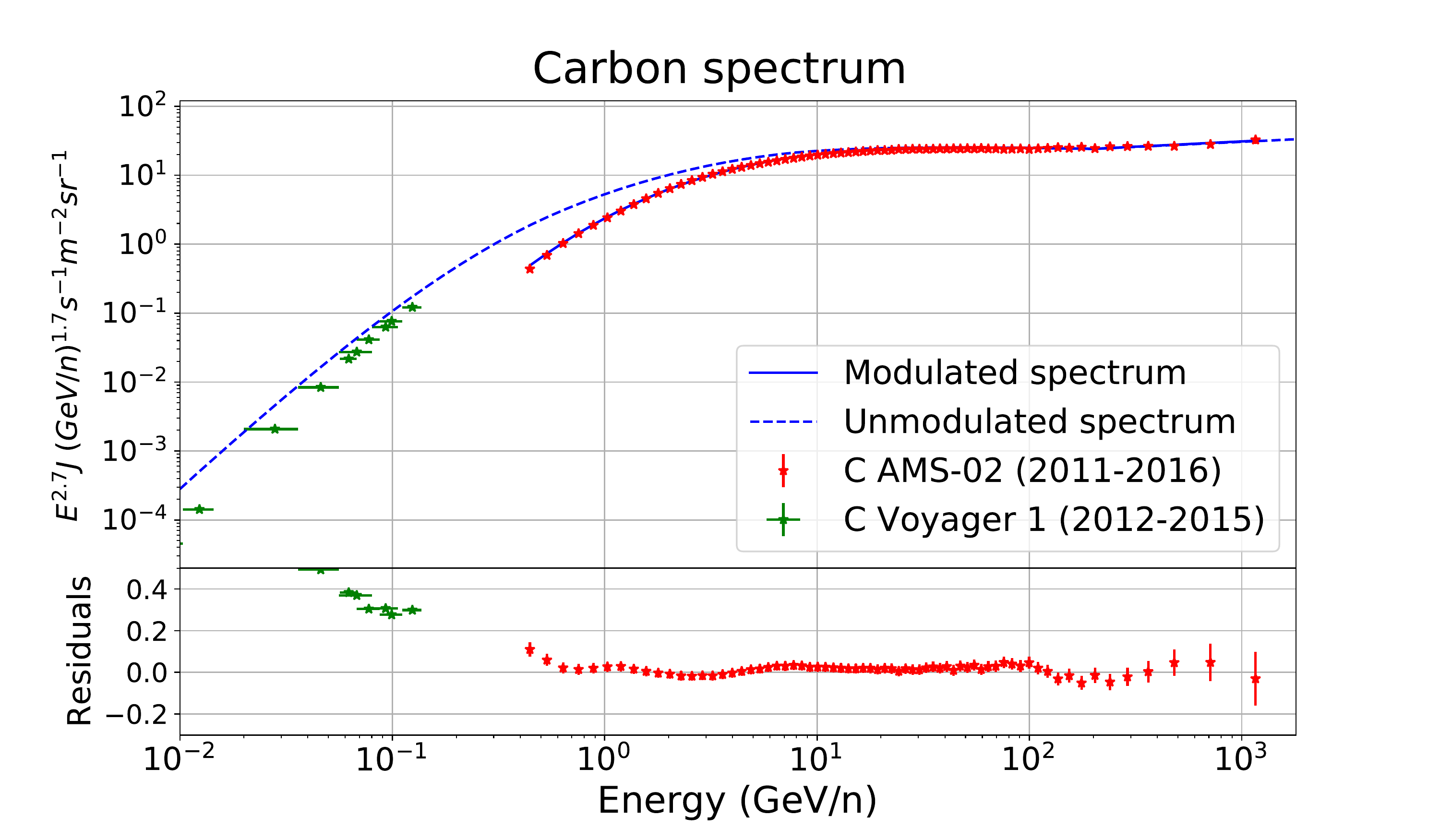} 
\vspace{-0.2cm}

\includegraphics[width=0.49\textwidth,height=0.267\textheight,clip] {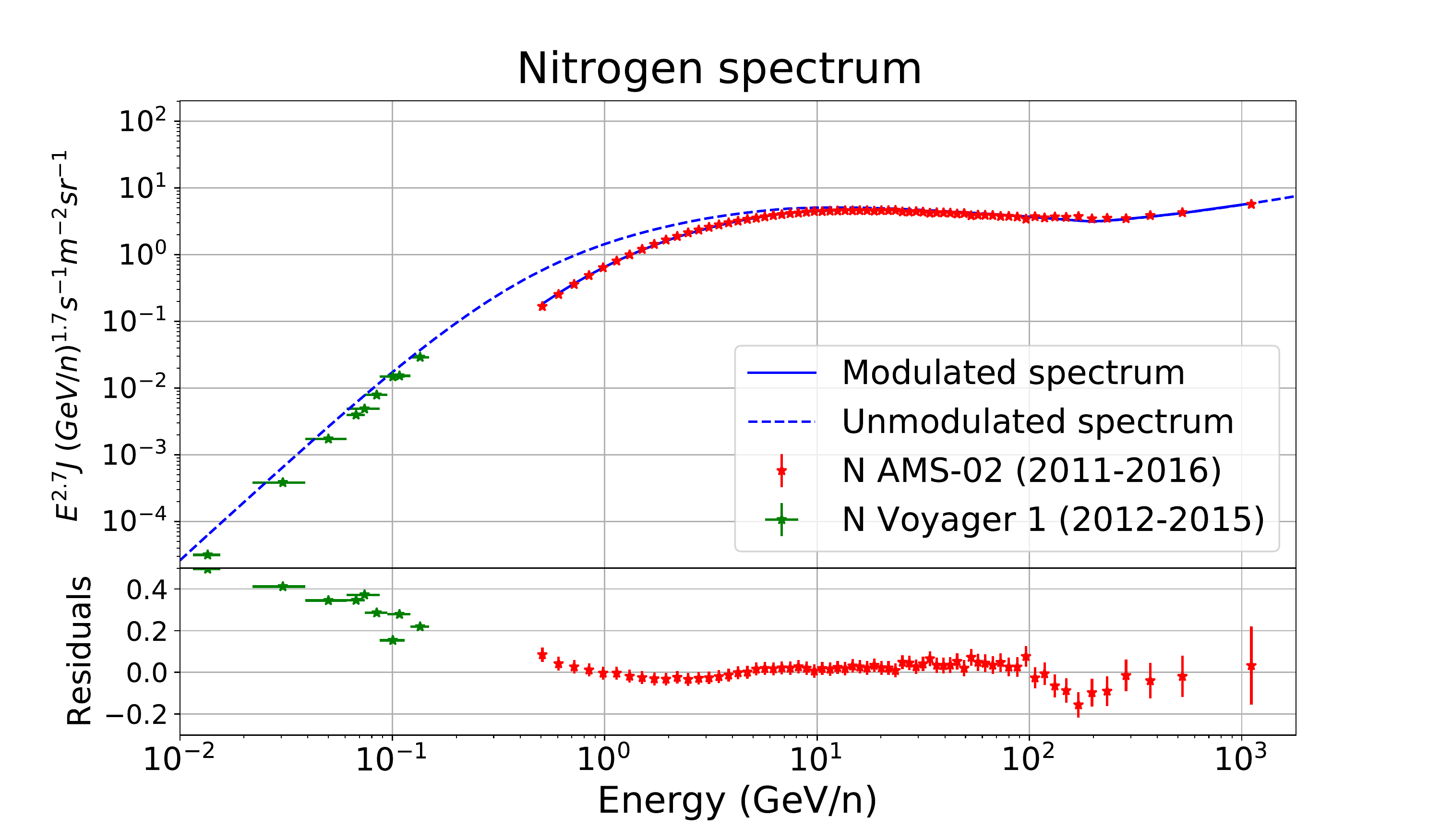} 
\includegraphics[width=0.49\textwidth,height=0.267\textheight,clip] {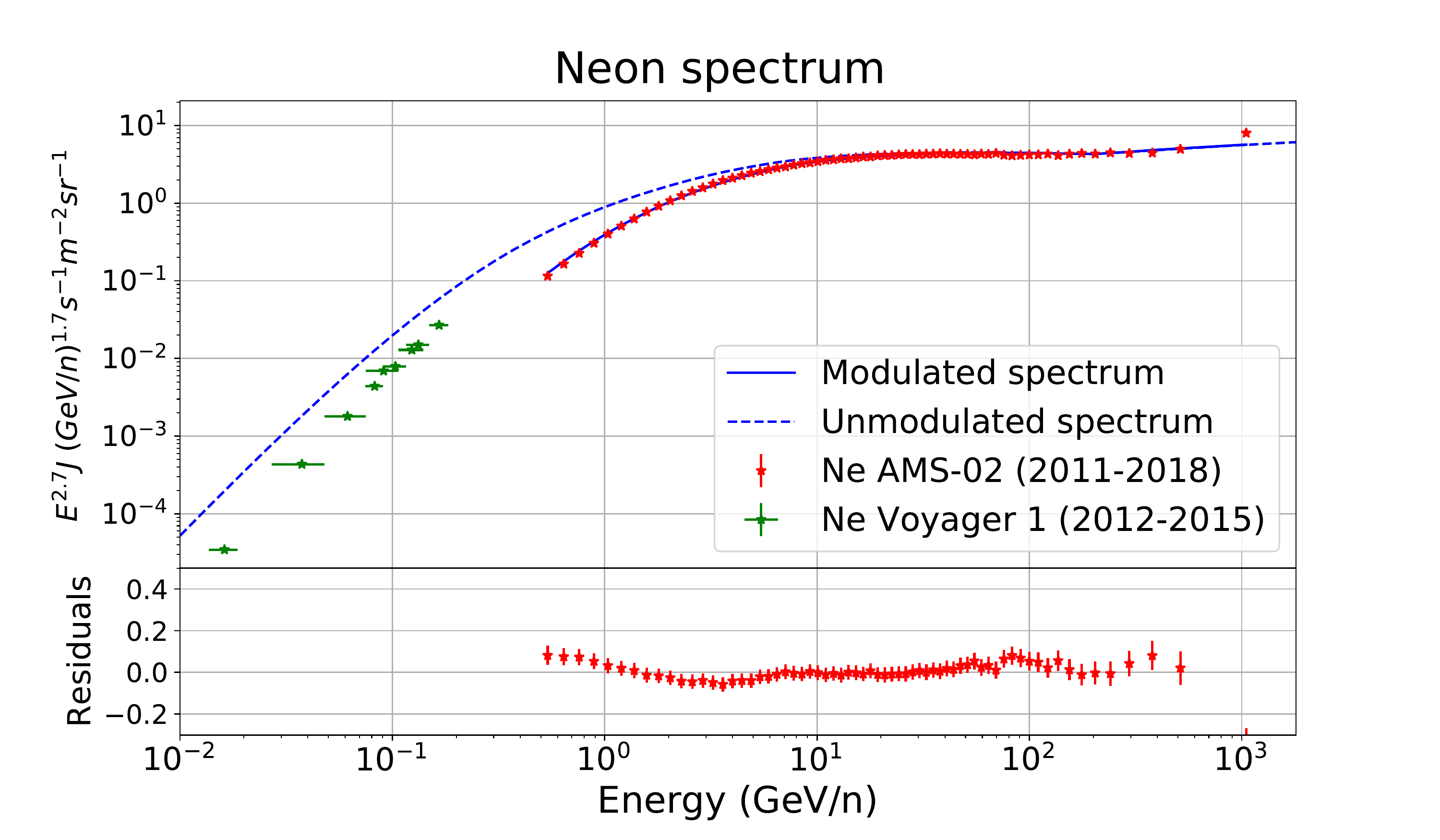}
\vspace{-0.2cm}

\includegraphics[width=0.49\textwidth,height=0.267\textheight,clip] {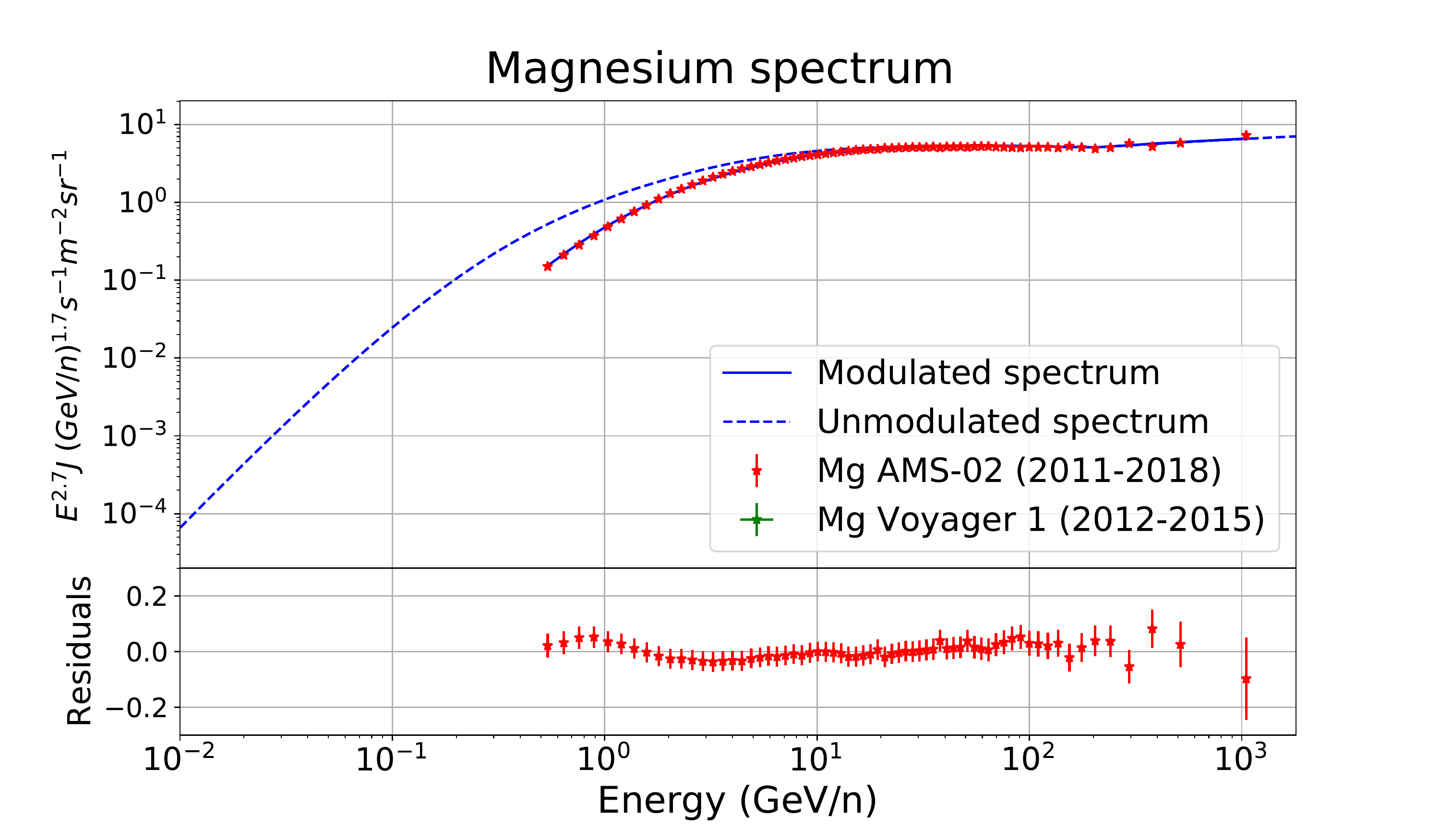}
\includegraphics[width=0.49\textwidth,height=0.267\textheight,clip] {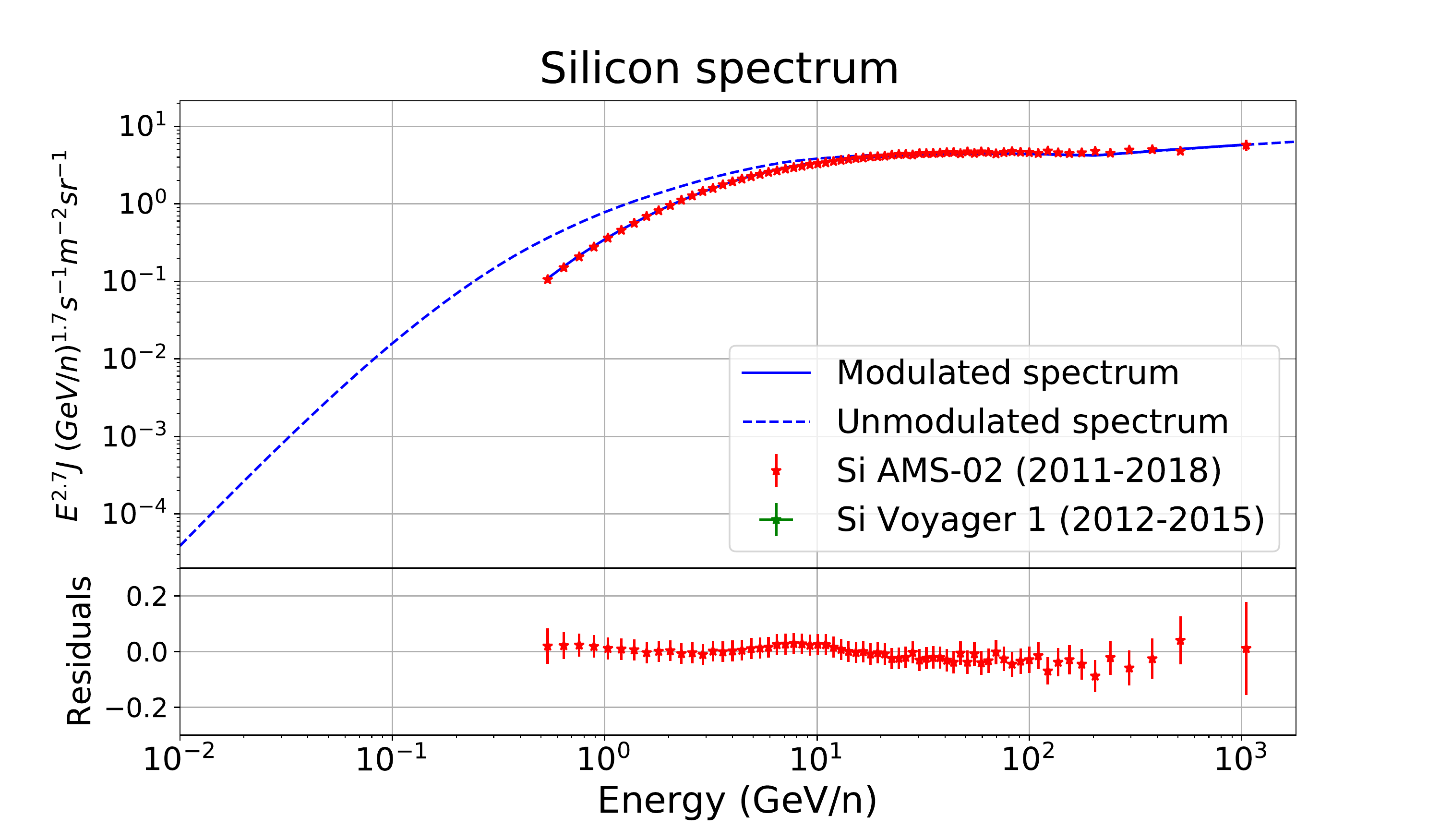}
\caption{Fits of the primary CR spectra used in the simulations, using the {\tt DRAGON2} cross sections. Data from the Voyager-1 mission  and AMS-02 are plotted, including the very recent Ne, Mg and Si data from AMS-02. Voyager-1 is outside the heliosphere, which means no solar modulation effects, while the Fisk potential value used for to fit AMS-02 data is $\phi = 0.61 \units{GV}$. Data taken from \url{//https://lpsc.in2p3.fr/crdb/} \cite{Maurin_db1, Maurin_db2} and \url{https://tools.ssdc.asi.it/CosmicRays/} \cite{ssdc}.}
\label{fig:primFit}
\end{figure*}

\begin{figure*}[!htb]
\begin{tabular}{c c c}
\hskip 0.6cm \textbf{\underline{Webber}} & \hskip 1.3cm \textbf{\underline{GALPROP}} & \textbf{\underline{DRAGON2}} \\
\hskip -0.63cm
\includegraphics[width=0.4025\textwidth,height=0.2\textheight,clip] {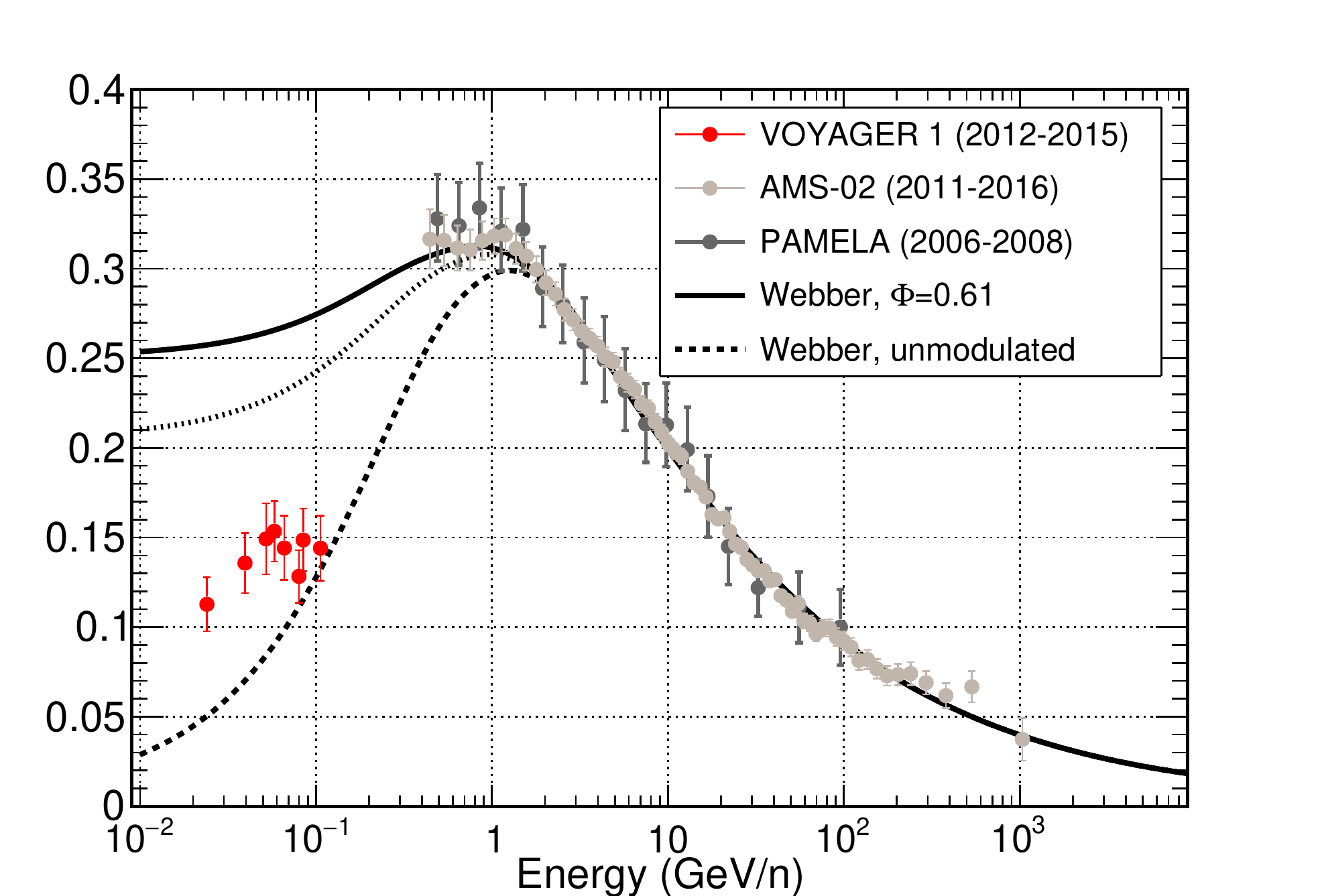} 
\hspace{-1.7cm} &
\includegraphics[width=0.4025\textwidth,height=0.2\textheight,clip] {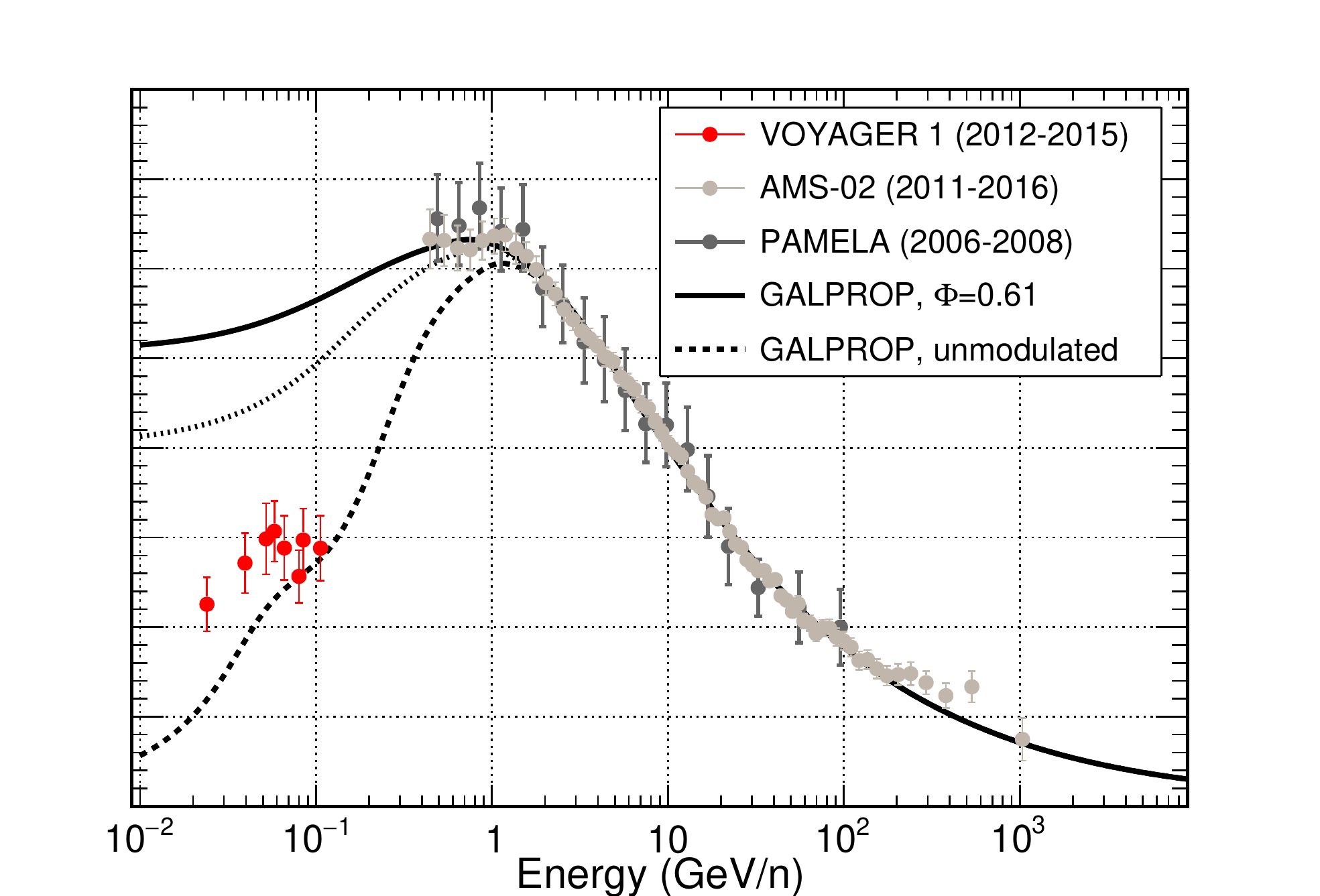} \hspace{-1.7cm} &
\includegraphics[width=0.4025\textwidth,height=0.2\textheight,clip] {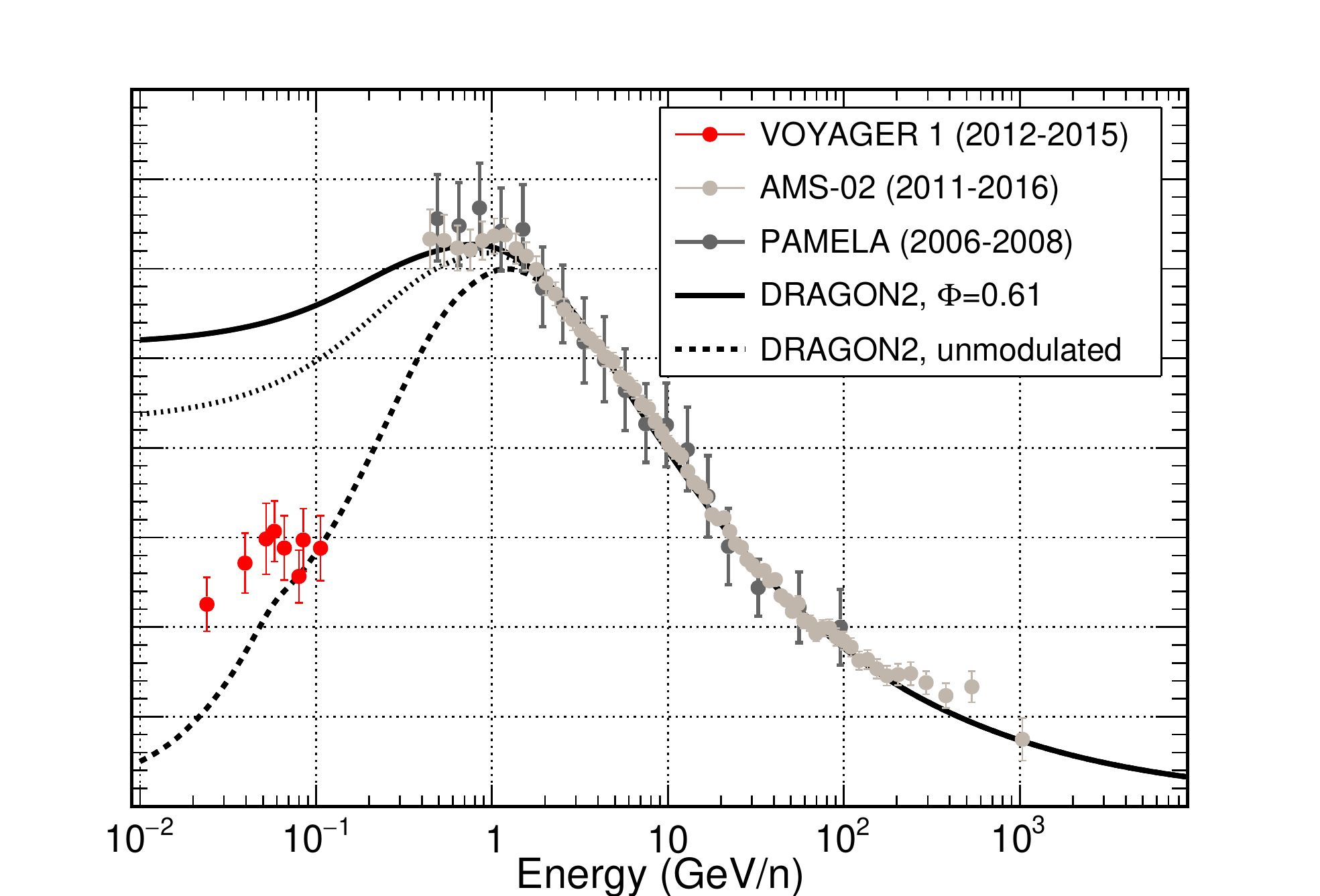}
    \end{tabular}
\caption{Fits of the boron-over-carbon ratios for the three compared cross sections models. These models are set to a Fisk potential of $\phi = 0.61 \units{GV}$ in the time period corresponding to the AMS-02 data taking. A line with a solar modulation $\phi = 0.4 \units{GV}$ is added for completeness. Data taken from \url{//https://lpsc.in2p3.fr/crdb/} \cite{Maurin_db1, Maurin_db2} and \url{https://tools.ssdc.asi.it/CosmicRays/} \cite{ssdc}.}
\label{fig:BCComp}
\end{figure*}

\vspace{0.3cm}
In addition, the fits of the boron-over-carbon spectra with the three tested cross sections are shown in fig. \ref{fig:BCComp}, for the unmodulated and modulated predictions, in comparison to the experimental data of the Voyager-1, PAMELA and AMS-02 experiments.

\vspace{0.5cm}

\section{\large Theory on secondary-over-secondary flux ratios}
\label{sec:appendixC}

In this appendix, the basic theory about the secondary-over-secondary flux ratios is illustrated.
Here we demonstrate that they are mainly dependent on the cross sections of secondary CR production and on the injection spectra used to describe primary CRs. 
Roughly speaking, if one considers that the flux of a primary CR species at Earth is $J_{\alpha} \propto q_{\alpha}\times\tau_{prop}(E)$ and $\tau_{prop}(E) = H^2/D(E)$, it will take the form $J_{\alpha} \sim C_{\alpha}E^{-\gamma_{\alpha} - \delta}$ at high energies (more precisely, around few GeVs, when $1/\sigma_{\alpha}^{inel}(E) \gg n_g\frac{h}{H}c \tau_{prop}(E)$). The secondary CR fluxes take the form $J_{i} \propto \sum^{\alpha} J_{\alpha}cn_g\sigma_{\alpha
\rightarrow
i}\times\tau_{prop}(E)$, whose ratio leaves the expression:

\begin{equation}
\label{eq:sectosec}
\begin{split}
\frac{J_k}{J_j}\Bigl(E\Bigr) \propto \frac{\sum^{\alpha \rightarrow k}J_{\alpha}(E)\sigma_{\alpha \rightarrow k}(E)}{\sum^{\alpha \rightarrow j}J_{\alpha}(E)\sigma_{\alpha \rightarrow j}(E)} \hspace{0.8cm} {\xrightarrow{\makebox[1cm]{high energies}}} \hspace{0.6cm} \sim \frac{\sum^{\alpha \rightarrow k}C_{\alpha}E^{-\gamma_{\alpha}}\sigma_{\alpha \rightarrow k}(E)}{\sum^{\alpha \rightarrow j}C_{\alpha}E^{-\gamma_{\alpha}}\sigma_{\alpha \rightarrow j}(E)}
\end{split}
\end{equation}

Notice that the term $E^{-\delta}$ is common in the summation terms and therefore is common in numerator and denominator, cancelling out. With the equation~\ref{eq:sectosec} one appreciates that these ratios have direct dependence on the local spectrum of primary nuclei ($\alpha$) and on the overall spallation cross sections. Nonetheless, the local primary spectrum may have some dependence on the diffusion parameters at low energies and reacceleration can slightly contribute. The total repercussion of these parameters in the low energy region can be estimated to be at level of $10\%$. Then, due to the presence of the radioactive $^{10}$Be, the most important contributor to the low energy uncertainties is the halo size being able to introduce variations of more than $10\%$, respectively. At the end, these ratios are mainly dependent on the source term of CRs and on their spallation cross sections at high energies. The fact that these ratios do not hold exactly for different diffusion parameters is caused by multi-step reactions, which introduce more non-linearities difficult to predict. Despite this fact, as AMS-02 data for the local spectrum of primary CRs are highly precise, the largest uncertainty at high energy lies on the spallation cross sections, and this is what makes them so suitable for adjusting the overall cross sections with high precision (AMS-02 precision). The next figure shows the effect of these changes on the Li/B ratio (since the importance of the halo size and gas profile is negligible):

\begin{figure*}[!hbt]
\label{fig:Rat_mod}
\centering
\includegraphics[width=0.65\textwidth,height=0.3\textheight,clip] {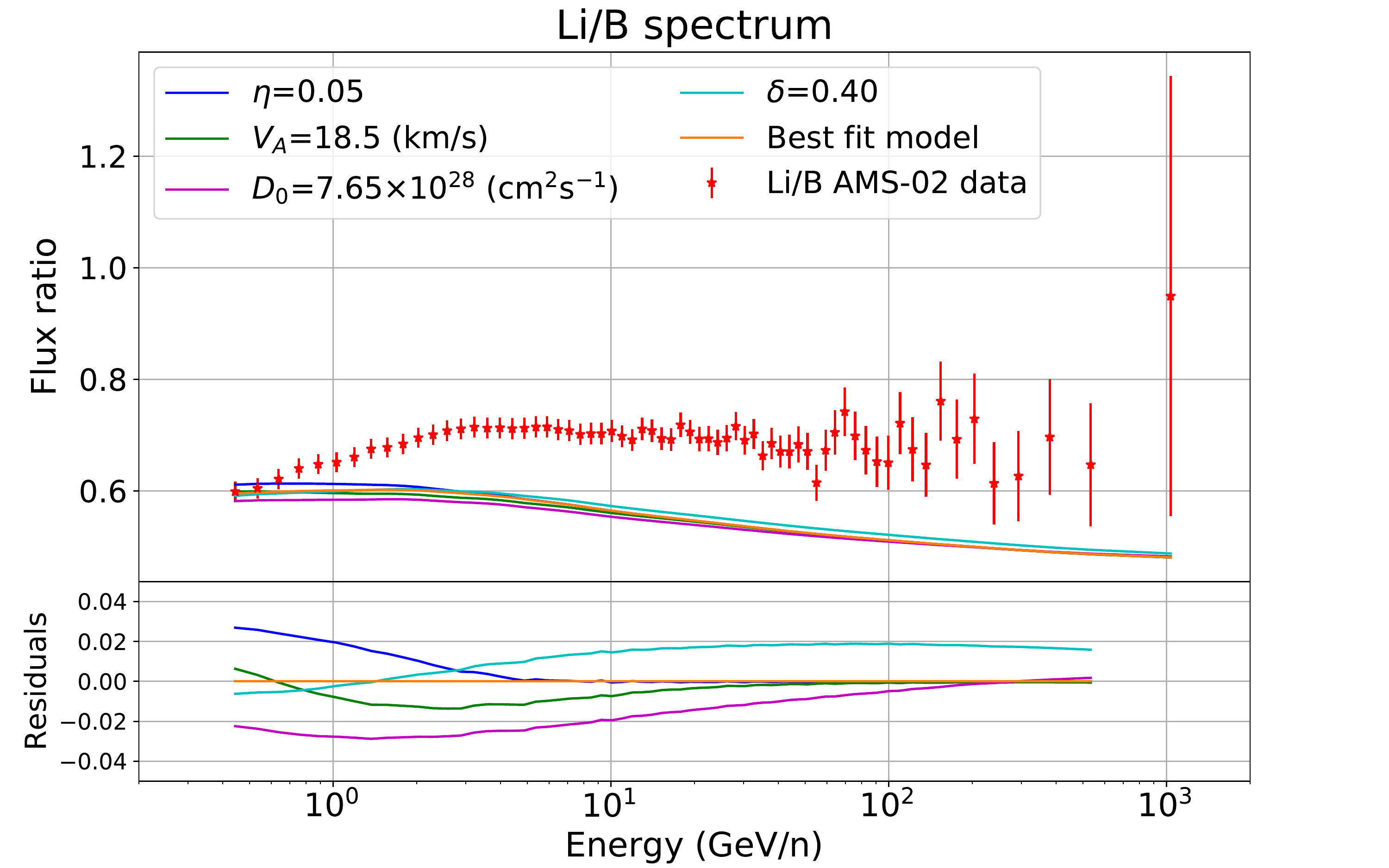} 
\caption{Simulated ratios with same primary source term (except for the simulation with change in the diffusion spectral index, $\delta$, in which the source term was changed to hold same $\alpha + \delta$ value) and changing diffusion parameters. The best fit model has been shown in table~\ref{tab:diff_params} and the simulations change by $\Delta\eta \sim 0.5$, $\Delta V_A \sim 7 \units{km/s}$, $\Delta D_0 \sim 1\times 10^{28} \units{cm^2/s}$, $\Delta \delta\sim0.04$, respectively, much higher than the usual uncertainties quoted in the best fits of these parameters~\cite{Niu:2017qfv, Niu:2018waj, Boschini:2019gow}. The residuals are shown with respect to the best fit model (orange line).}
\end{figure*} 

As we see, at $10 \units{GeV/n}$ the maximum difference between the predictions, when changing the diffusion parameters as described, is around $4\%$, while at $30 \units{GeV/n}$ it goes down to $3\%$ for variations of the diffusion parameters larger than the usual uncertainties related to their determination. 
Nonetheless, the differences at high energy are mainly due to multi-step reactions and the subtle differences in the primary fluxes because of the change in the amount of secondary C (roughly $1/5$ of its flux is secondary at low energies) and N ($\sim 73\%$ of it is secondary at $10 \units{GeV/n}$), which leads to sizeable changes in the secondary CRs. 

\end{document}